\documentclass[a4paper,11pt]{article}

\usepackage[utf8]{inputenc}
\usepackage[english]{babel}
\usepackage{physics}
\usepackage{graphicx}
\usepackage{float}
\usepackage[rightcaption]{sidecap}
\graphicspath{ {images/} }
\usepackage{appendix}
\usepackage{blindtext}
\usepackage{authblk}
\usepackage{multicol}
\usepackage{multirow}
\usepackage{geometry}
\usepackage{enumerate}
\usepackage{mathtools}

\usepackage{vmargin}
\usepackage{amsmath, amsthm, amssymb}
\usepackage{titlesec}
\usepackage{setspace}

\usepackage{tensor}
\usepackage[version=3]{mhchem}
\usepackage{simpler-wick}

\usepackage[shortlabels]{enumitem}
\usepackage{pdflscape}
\usepackage{afterpage}
\usepackage{capt-of}
\usepackage{changepage}

\usepackage{booktabs}
\usepackage{caption}
\usepackage{subcaption}
\usepackage{adjustbox}
\usepackage{xcolor}

\usepackage{csquotes}
\usepackage{cite}

\usepackage{dsfont}

\usepackage{rawfonts}

\usepackage{tikz-feynman}
\usetikzlibrary{arrows.meta, positioning, decorations.pathmorphing}

\usepackage{fancyhdr}
\newcommand{\institutionalpreprintMainz}{MITP-26-006}
\newcommand{\institutionalpreprintCERN}{CERN-TH-2026-035}

\usepackage[colorlinks=true,
            linkcolor=blue,
            citecolor=blue,
            urlcolor=blue,
            linktoc=page
            ]{hyperref}

\fancypagestyle{firstpage}{%
  \fancyhf{}
  \fancyhead[R]{\small\institutionalpreprintMainz\\ \institutionalpreprintCERN}
  \fancyfoot[C]{\thepage}
  
}

\makeatletter

\renewcommand\AB@affilsepx{\protect\\[\affilsep]}
\makeatother

\title{\textbf{Higher-order hadronic vacuum polarization contribution to the muon \boldmath{$g{-}2$} from lattice QCD}}
\date{} 
\author[1]{Arnau Beltran}
\author[2,3]{Alessandro Conigli}
\author[4]{Simon Kuberski}
\author[1,2,4]{Harvey B. Meyer}
\author[1,5]{Konstantin Ottnad}
\author[1,2,3]{Hartmut Wittig}

\affil[1]{\small \textit{PRISMA}$^{++}$ Cluster of Excellence and Institut f\"{u}r Kernphysik, Johannes Gutenberg-Universit\"{a}t Mainz, 55099 Mainz, Germany}
\affil[2]{\small Helmholtz-Institut Mainz, Johannes Gutenberg-Universit\"{a}t Mainz, 55099 Mainz, Germany}
\affil[3]{\small GSI Helmholtz Centre for Heavy Ion Research, 64291 Darmstadt, Germany}
\affil[4]{\small Theoretical Physics Department, CERN, 1211 Geneva 23, Switzerland}
\affil[5]{\small Helmholtz-Institut f\"ur Strahlen und Kernphysik (Theory), Rheinische Friedrich-Wilhelms-Universit\"at Bonn, Nussallee 14-16, 53115 Bonn, Germany}
\makeatletter
\g@addto@macro\@author{%
  \protect\\[2.0em]\protect\Affilfont\small
  \href{mailto:abeltran@uni-mainz.de}{\texttt{abeltran@uni-mainz.de}},
  \href{mailto:aconigli@uni-mainz.de}{\texttt{aconigli@uni-mainz.de}},
  \href{mailto:simon.kuberski@cern.ch}{\texttt{simon.kuberski@cern.ch}},
  \href{mailto:h.b.meyer@cern.ch}{\texttt{h.b.meyer@cern.ch}},
  \href{mailto:ottnad@hiskp.uni-bonn.de}{\texttt{ottnad@hiskp.uni-bonn.de}},
  \href{mailto:hartmut.wittig@uni-mainz.de}{\texttt{hartmut.wittig@uni-mainz.de}}}
\makeatother

\begin{document}

\maketitle
\thispagestyle{firstpage}

\begin{abstract}
    \normalsize
    We present the first lattice QCD calculation of the next-to-leading order hadronic vacuum polarization contribution to the muon anomalous magnetic moment with sub-percent precision. We employ the time-momentum representation for the space-like kernel, which is combined with the spatially summed vector correlator computed on CLS ensembles with $N_{\mathrm{f}}=2+1$ flavors of $\mathrm{O}(a)$-improved Wilson fermions, covering six lattice spacings between $0.039$ and $0.097$\,fm and a range of pion masses including the physical value. After accounting for finite-size corrections and isospin-breaking effects, we obtain as our final, continuum-extrapolated result $a_\mu^{\mathrm{hvp,\,nlo}}=-101.57{(26)}_{\mathrm{stat}}{(54)}_{\mathrm{syst}}\times10^{-11}$. It lies below the estimate provided by the 2025 White Paper of the Muon $(g-2)$ Theory Initiative by 1.4$\sigma$ but is two times more precise. It also exhibits a strong tension of 4.6$\sigma$ with data-driven evaluations based on hadronic cross section measurements excluding the recent result by CMD-3.
\end{abstract}

\newpage 

\tableofcontents

\newpage

\section{Introduction}
\label{sec:introduction}
The muon anomalous magnetic moment, $a_\mu$, is a precision observable that plays a crucial role for testing the validity of the Standard Model (SM)~\cite{Jegerlehner:2009ry,Jegerlehner:2017gek,Wittig:2025azm}. A significant deviation between the direct experimental measurement of $a_\mu$ and its SM-based theoretical prediction would, if observed, constitute incontrovertible evidence of a quantitative failure of the SM. The final result released by the E989 experiment at Fermilab~\cite{Muong-2:2021ojo, Muong-2:2023cdq, Muong-2:2025xyk} has pushed the precision of the direct measurement to an impressive 124 ppb. The error on the SM prediction, according to the 2025 White Paper by the Muon $g-2$ Theory Initiative (WP25~\cite{WP25}), is about four times larger and dominated by the uncertainty in the leading-order (LO) hadronic vacuum polarization (HVP) contribution, $a_\mu^{\mathrm{hvp,\,lo}}$, which arises at order~$\alpha^2$ in the fine-structure constant. Given the worldwide efforts to improve the precision of the SM prediction, it is important to consider also the sub-dominant contributions to the HVP of order~$\alpha^3$ and beyond. This is particularly relevant given that results for $a_\mu^{\mathrm{hvp,\,lo}}$ obtained using the traditional data-driven dispersive approach depend strongly on the experiment that provides the cross section data for the dominant two-pion contribution.

The SM prediction for $a_\mu$ quoted in WP25 is based on an average over many lattice calculations of $a_\mu^{\mathrm{hvp,\,lo}}$ employing different discretizations of the QCD action~\cite{RBC:2018dos, Giusti:2019xct, Borsanyi:2020mff, Lehner:2020crt, Wang:2022lkq, Aubin:2022hgm, Ce:2022kxy, ExtendedTwistedMass:2022jpw, RBC:2023pvn, Kuberski:2024bcj, Boccaletti:2024guq, Spiegel:2024dec, RBC:2024fic, Djukanovic:2024cmq, ExtendedTwistedMass:2024nyi, MILC:2024ryz, FermilabLatticeHPQCD:2024ppc}. By contrast, the WP25 estimate for the next-to-leading order (NLO) HVP contribution is still based on the data-driven dispersive method~\cite{Keshavarzi:2019abf, DiLuzio:2024sps}, accounting for the spread of results obtained using different experimental input. Clearly, for the sake of consistency, estimates for both the LO and NLO HVP contributions ought to originate from the same methodology as long as there are unexplained tensions between different approaches. A first exploratory lattice calculation of the NLO HVP contribution was performed in Ref.~\cite{Chakraborty:2018iyb}, achieving a precision of 13\%.

In this paper we present the first lattice QCD calculation of the NLO HVP contribution, $a_\mu^{\mathrm{hvp,\,nlo}}$, with sub-percent precision. Our work extends our previous determinations of the window observables that add up to the total LO HVP contribution~\cite{Ce:2022kxy, Kuberski:2024bcj, Djukanovic:2024cmq}, as well as the closely related hadronic contributions to the running of electroweak couplings~\cite{Ce:2022eix, Conigli:2025qvh, Conigli:2026stl}. We employ the space-like representation of the relevant NLO kernel functions~\cite{Nesterenko:2021byp, Balzani:2021del} in the time-momentum representation (TMR)~\cite{Bernecker:2011gh, Balzani:2024gmu} in combination with the spatially summed vector correlator. The latter has been computed on 32 gauge ensembles generated with $2+1$ flavors of O($a$)-improved Wilson fermions by the Coordinated Lattice Simulations (CLS) effort.

After extrapolating to the continuum limit and correcting for finite-size effects and isospin breaking, we obtain
\begin{equation}
  a_\mu^{\mathrm{hvp,\,nlo}}=
  (-101.57\pm0.26_{\mathrm{stat}}\pm0.54_{\mathrm{syst}})\times10^{-11}\,.
\end{equation}
Our result has a total relative error of 0.6\%, which corresponds to a similar level of precision as that reached for data-driven dispersive evaluations. It is slightly below the estimate of $(-99.6\pm1.3)\times10^{-11}$ quoted in the 2025 White Paper update~\cite{WP25}, which accounts for the tension encountered among different experimental measurements of the two-pion contribution. When compared with the 2019 data-driven determination of Keshavarzi et al.~\cite{Keshavarzi:2019abf} (which does not include the more recent measurement by CMD-3~\cite{CMD-3:2023alj, CMD-3:2023rfe}), one observes a tension of 4.6~standard deviations. This is consistent with the pattern of deviations observed when comparing lattice and data-driven estimates of the LO HVP contribution.

This paper is organized as follows. In Sec.~\ref{sec:setup}, we establish the framework for our calculation, introducing the notation for the diagram sets and the time-momentum representation formalism. We also motivate the use of time windows and describe the lattice ensembles employed in our global chiral-continuum extrapolation. Section~\ref{sec:strategy} presents the strategies adopted for the short- and long-distance windows, the application of finite-volume corrections, and the treatment of diagram NLOc. In Sec.~\ref{sec:results}, we present our results: first the isospin-symmetric results decomposed by window and channel, followed by the isospin-breaking corrections and other subleading contributions, and finally our complete NLO HVP determination. We conclude in Sec.~\ref{sec:conclusions}.

\section{Setup}
\label{sec:setup}

\begin{figure}[t]
    \centering
    \includegraphics[width=0.8\textwidth]{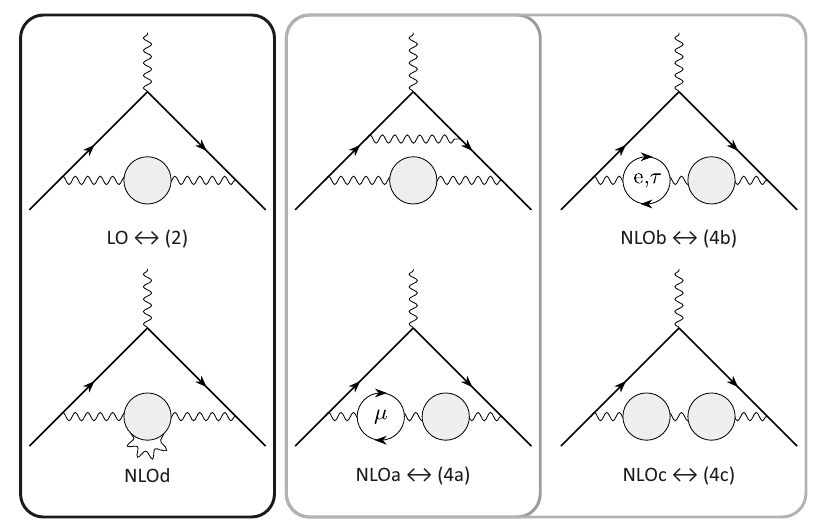}
    \caption{Feynman diagrams contributing to the hadronic-vacuum polarization to the muon $g-2$ up to $\mathrm{O}(\alpha^3)$. On the left side, we show the LO diagram, contributing at order $\alpha^2$, along with its electromagnetic isospin breaking correction. On the right side we show all the higher-order diagrams to be studied in this work. The diagram set NLOa considers all possible corrections coming from photon-lines and muon-loops. The set NLOb consists of corrections coming from electron- and tau-loops. Finally, diagram NLOc is composed of two QCD insertions.}
    \label{fig:feynman_diag}
\end{figure}

\subsection{Higher-order diagrams}
\label{sec:ho_diag}

Figure~\ref{fig:feynman_diag} shows all Feynman diagrams contributing to the HVP up to $\mathrm{O}(\alpha^3)$. In this work, we focus on the non-perturbative computation of the NLO HVP diagrams. These are commonly categorized into three groups—NLOa, NLOb, and NLOc—based on their topological structure. 
\begin{itemize}
    \item Diagram set NLOa accounts for corrections from photon lines and muon loops, yielding a total of 14 diagrams with competing signs: photon-line contributions are negative while muon-loop contributions are positive. After partial cancellation, data-driven determinations~\cite{Keshavarzi:2019abf} indicate that NLOa provides the dominant negative contribution to the NLO HVP. 
    \item Diagram set NLOb includes the remaining lepton loops (electron and tau), consisting of four diagrams—two for each lepton flavor. The electron loop contributes a large positive correction that partially cancels the negative NLOa contribution. 
    \item Finally, NLOc consists of a single diagram with two QCD insertions and is subleading compared to the other two contributions.
\end{itemize}
A fourth set of diagrams, usually labeled NLOd, is often found in the literature. These corrections account for electromagnetic effects within the QCD insertions and are, by convention, regarded as the electromagnetic component of the isospin-breaking corrections to the LO diagram. These contributions have been determined as part of our calculations of the LO HVP, published in~\cite{Ce:2022kxy,Kuberski:2024bcj,Djukanovic:2024cmq}.

Regarding notation, we refer to the different contributions either by their electromagnetic ordering, LO and NLO, or by a condensed notation introduced in Ref.~\cite{Balzani:2021del}, where the LO and NLO diagrams are labeled $(2)$ and $(4)$ respectively, and the NLO sub-contributions are labeled as $(4a)$, $(4b)$, and $(4c)$, for the diagram sets NLOa, NLOb, and NLOc respectively (see Fig.~\ref{fig:feynman_diag}).

\subsection{Definitions in the TMR}
\label{sec:def}

In this work, we make use of the time-momentum representation (TMR)~\cite{Bernecker:2011gh} to express all HVP contributions as one or multiple integrals over the convolution of a time-kernel function $\tilde{f}$ and the zero-momentum projection of the electromagnetic correlator $G$. Without loss of generality, any HVP observable can be written as
\begin{equation}
    a_\mu^{\mathrm{hvp,\,(i)}} = {\left(\frac{\alpha}{\pi}\right)}^{2+N_i}\int dt_1\dots dt_{m_i} \tilde{f}^{(i)}(\hat{t}_1,\dots,\hat{t}_{m_i})\, G(t_1)\times\dots\times G(t_{m_i})\ ,
    \label{eq:general_def_hvp}
\end{equation}
where $m_i$ is the number of QCD insertions in the internal photon propagators, $N_i$ denotes the electromagnetic perturbative order (i.e. $N_i=0$ for the LO, $N_i=1$ for the NLO diagrams, etc) of diagram set $(i)$, and we have introduced the dimensionless variable $\hat{t}=m_\mu t$. The function $G(t)$ is related to the two-point function of the electromagnetic currents through
\begin{equation}
    \delta_{kl} \, G(t) = - \int d^3x \, \langle j^\gamma_k(t,\mathbf{x})\, j^\gamma_l(0)\rangle\ ,\qquad 
    j^\gamma_\mu = \frac{2}{3}\overline{u}\gamma_\mu u - \frac{1}{3}\overline{d}\gamma_\mu d - \frac{1}{3}\overline{s}\gamma_\mu s + \frac{2}{3}\overline{c}\gamma_\mu c + \dots\ .
    \label{eq:G_def}
\end{equation}
We make use of the Gell-Mann matrices $\lambda^{m=1,\dots,8}$ to decompose the $(u,d,s)$ flavor sector into an isospin basis,
\begin{equation}
    j_\mu^m = \overline{\psi}\, T^m \gamma_\mu \psi\ ,\qquad 
    \overline{\psi} = (\overline{u},\overline{d},\overline{s})\ ,\qquad 
    T^m = \frac{1}{2}\lambda^m \oplus \mathbf{0}\ ,
    \label{eq:current_general}
\end{equation}
where $\mathbf{0}$ is the $2\times2$ zero matrix. For the heavy-quark sector $(c,b)$ we take $T^c = \mathrm{diag}(0,0,0,1,0)$ and $T^b = \mathrm{diag}(0,0,0,0,1)$. The corresponding flavor-separated correlator is then written as
\begin{equation}
    \delta_{kl}\, G^{(d,e)}(t) = - \int d^3x \, \langle j^d_k(t,\mathbf{x})\, j^e_l(0)\rangle\ .
\end{equation}
We associate $T^{m=\gamma}$ with the physical quark charge matrix, $T^{\gamma}=\mathrm{diag}(\frac{2}{3},-\frac{1}{3},-\frac{1}{3},\frac{2}{3},-\frac{1}{3})$. This leads to the flavor or isospin decomposition of the electromagnetic correlator.
\begin{equation}
    \begin{aligned}
            G \equiv G^{(\gamma,\gamma)} &= G^{(3,3)} + \frac{1}{3}G^{(8,8)} + \frac{4}{9}G^{(c,c)}_{\mathrm{conn}} + \frac{2}{3\sqrt{3}}G^{(c,8)}_{\mathrm{disc}} + \frac{4}{9}G^{(c,c)}_{\mathrm{disc}} + \frac{1}{9}G^{(b,b)} + \dots \\
            &= \frac{5}{9}G^{(l,l)} + \frac{1}{9}G^{(s,s)} + \frac{4}{9}G^{(c,c)}_{\mathrm{conn}} + \frac{1}{9}G^{(b,b)} + G_{\mathrm{disc}} + \dots \, .
    \end{aligned}
    \label{eq:decomp}
\end{equation}
Here, the subscript ``conn'' denotes quark-connected contributions, while ``disc'' denotes quark-disconnected contributions, and $(3,3)$ and $(8,8)$ denote the isovector and isoscalar components, respectively. Throughout this work, we adopt the isospin basis as our primary computational framework. The associated decomposition of the integrand in Eq.~\eqref{eq:general_def_hvp} for NLOb is shown in Fig.~\ref{fig:isospin_decomp}. Results obtained in this basis are then expressed in terms of the flavor decomposition. The convention used is defined as,
\begin{equation}
    \begin{aligned}
        & a_\mu^{d_1,e_1-\dots-d_{m_i},e_{m_i},\,(i)} \\
        = & {\left(\frac{\alpha}{\pi}\right)}^{2+N_i}\int dt_1\dots dt_{m_i}\, \tilde{f}^{(i)}(\hat{t}_1,\dots,\hat{t}_{m_i})\, 
        G^{(d_1,e_1)}(t_1)\times\dots\times G^{(d_{m_i},e_{m_i})}(t_{m_i})\, ,
    \end{aligned}
    \label{eq:general_def_comp}
\end{equation}
where the charge factors are factorized out of this definition.

\subsection{The electromagnetic current}
\label{sec:em_current}

The discretization of the electromagnetic current on the lattice is not unique, providing an opportunity to assess systematic uncertainties by comparing different implementations. In this work, we exploit this freedom by considering two discretizations of the current in Eq.~\eqref{eq:current_general}: the local (l) and point-split or conserved (c) variants,
\begin{equation}
    \begin{aligned}
            j_\mu^{(l),a}(x) & = \overline{\psi}(x)\gamma_\mu T^a\psi(x)\,, \\
            j_\mu^{(c),a}(x) & = \frac{1}{2}\left(\overline{\psi}(x+a\hat{\mu})(1+\gamma_\mu)U^\dagger_\mu(x)T^a\psi(x) - \overline{\psi}(x)(1-\gamma_\mu)U_\mu(x) T^a\psi(x+a\hat{\mu})\right)\,,
    \end{aligned}
\end{equation}
where $U_\mu(x)$ is the gauge link connecting sites $x$ and $x+a\hat{\mu}$. 

The $\mathrm{O}(a)$-improved versions of these currents are obtained by adding the derivative of the local tensor current $\Sigma_{\mu\nu}^a(x)=-\frac{1}{2}\overline{\psi}(x)[\gamma_\mu,\gamma_\nu]T^a\psi(x)$:
\begin{equation}
    j_\mu^{(\alpha),a,\,I}(x) = j_\mu^{(\alpha),a}(x) + a c_V^{(\alpha)}(g_0)\partial_\nu\Sigma_{\mu\nu}^a(x)\, ,\quad \alpha=l,c\, .
\end{equation}
The improvement coefficients $c_V^{(\alpha)}(g_0)$ can be determined non-perturbatively, this procedure introduces an ambiguity in the treatment of higher-order cutoff effects. We consider two different determinations, which we refer to as ``set 1'' and ``set 2''. Set 1 is obtained at finite quark masses directly on the CLS ensembles~\cite{Harris:2025xvk}, while set 2 is determined in the chiral limit using the Schr\"{o}dinger functional scheme~\cite{Heitger:2020zaq}.

For the correlator in Eq.~\eqref{eq:G_def}, we consider two discretization choices: both currents using the local definition (``ll'') or one current using the local and the other the conserved definition (``lc''). Combined with the two improvement sets, this yields four different data sets for each ensemble. By including all four variants in our chiral-continuum extrapolations and examining their spread through model averaging, we obtain a robust estimate of the systematic uncertainties associated with current discretization and improvement.

\subsection{NLO time-kernels}\label{sec:timekernels}

For a TMR calculation of the NLO contributions, one needs sufficiently precise knowledge of the time-kernels in~\eqref{eq:general_def_hvp}. These functions are in general not known in closed form, and they can only be related to the kernels in the time-like momentum representation through
\begin{equation}
        \tilde{f}^{(i)}(\hat{t}_1,\dots,\hat{t}_{m_i}) = \int_0^\infty d\hat{\omega}^2\, \hat{f}^{(i)}(\hat{\omega}^2)\prod_{j=1,\dots,m_i}\frac{4\pi^2}{m_\mu^2\hat{\omega}^2}\left[\hat{\omega}^2\hat{t}_j^2 - 4\sin^2\frac{\hat{\omega}\hat{t}_j}{2}\right]\, .
    \label{eq:timekernel_i}
\end{equation}
Here we introduced the dimensionless quantity $\hat{\omega}\equiv\omega/m_\mu$ and the dimensionless kernel function $\hat{f}^{(i)} \equiv m_\mu^2 f^{(i)}$, whose analytic structure is known up to NNLO. Eq.~\eqref{eq:timekernel_i} does not, in general, admit a closed-form solution. An analytic expression was obtained for the LO case in Ref.~\cite{DellaMorte:2017dyu}, while in this work we provide a solution for the NLOc contribution, presented in Appendix~\ref{app:timekern_NLOc}.

\begin{figure}[t]
    \centering
    \includegraphics[width=0.6\textwidth]{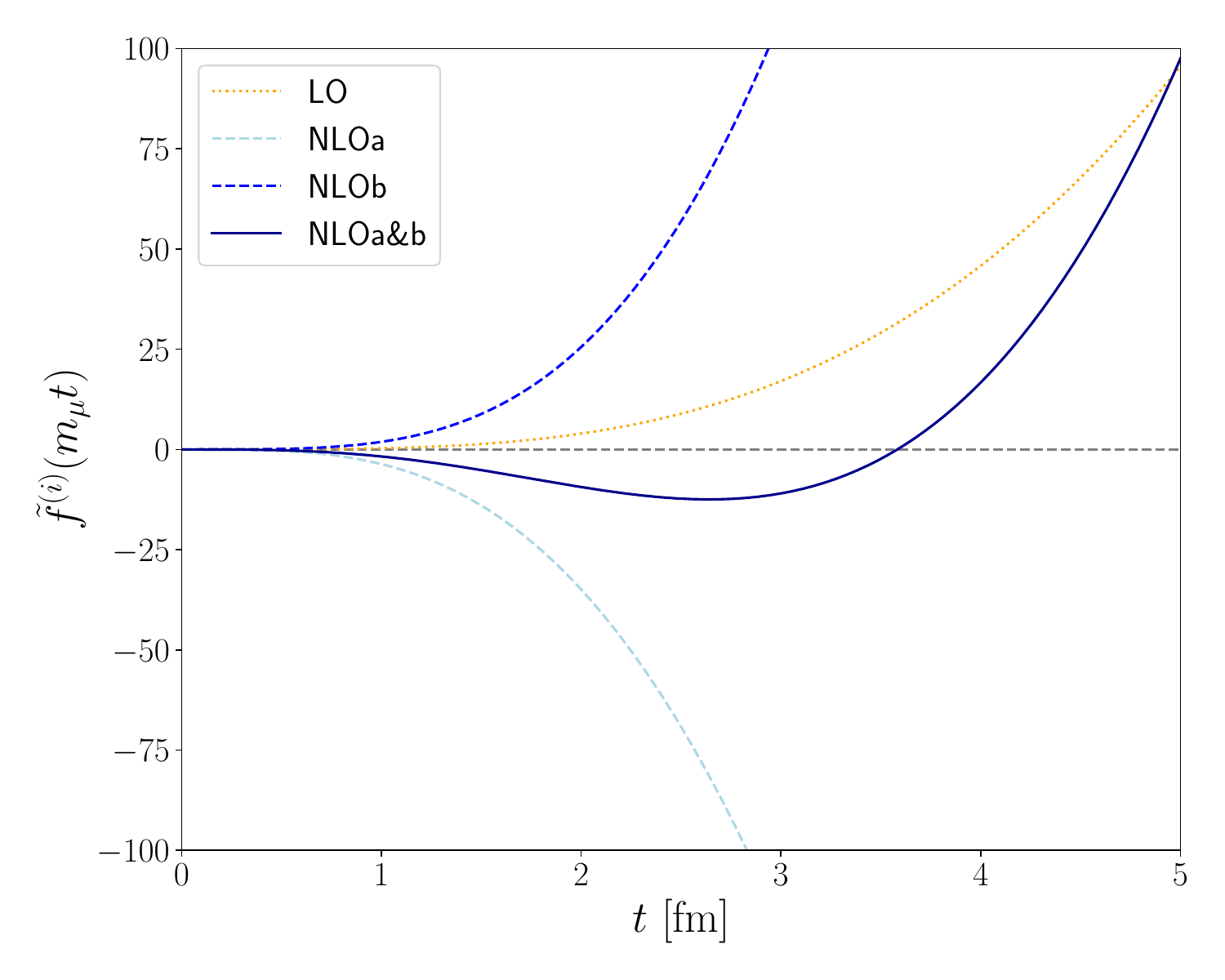}
    \caption{Comparison of the time-kernels for the LO and NLO diagrams, the addition NLOa\&b$\equiv$NLOa+NLOb is shown as a solid blue line.}
    \label{fig:kernel_comparision}
\end{figure}

It is, nevertheless, numerically convenient to work with polynomial representations of these results. Hence, expansions around $\hat{t}\ll 1$ are often used. These are fully adequate for lattice applications; such polynomial expansions typically reach an absolute precision better than $10^{-8}$ for $\hat{t}\lesssim 4$ (or $t\lesssim 7\ \mathrm{fm}$), which already covers the full range where lattice data for the correlation function $G(t)$ are usually available. We stress that the integrand beyond $7\,\mathrm{fm}$ can safely be neglected. In this region, the kernel function grows only quadratically, i.e. $\tilde{f}\sim t^2$ (see Fig.~\ref{fig:kernel_comparision}), such that the integrand is strongly suppressed by the exponentially decaying $G(t)$. We also remark that it is usually possible---although not needed here---to obtain an expansion around $\hat{t}\gg1$ and later interpolate between both regimes, thus covering the full Euclidean time.

We approximate Eq.~\eqref{eq:timekernel_i} for $(i)=(4a)$ and $(4b)$ by expanding the integral after separating it into a convenient form that admits such series representations. This technique was first introduced by E. Balzani, S. Laporta, and M. Passera in~\cite{Balzani:2024gmu} for diagram set NLOa, where a solution of the form
\begin{equation}
    \frac{m_\mu^2}{16\pi^2}\tilde{f}^{(4a)}(\hat{t}) = \sum_{n\geq4}\frac{\hat{t}^n}{n!}\Big(a_n + b_n\pi^2 + c_n{(\gamma_E+\ln\hat{t})} + d_n{(\gamma_E+\ln\hat{t})}^2\Big)\ ,
\end{equation}
was obtained, where $\gamma_E$ is the Euler-Mascheroni constant. We refer to~\cite{Balzani:2024gmu} for the analytic expressions of $a_n$, $b_n$, $c_n$, and $d_n$.

We follow a similar approach for NLOb. In this case, due to the presence of a lepton loop, a new energy scale enters the problem, requiring an additional splitting of the integral. Still, one can derive an analogous expansion. For the electron loop we find
\begin{equation}
    \frac{m_\mu^2}{16\pi^2}\tilde{f}_e^{(4b)}(\hat{t}) = 
    \sum_{n\geq4}\sum_{m\geq0}\frac{\hat{t}^n}{n!}M_e^m\Big(e_{nm}+f_{nm}\pi^2+g_{nm}{\left(\gamma_E+\ln\hat{t}\right)}+h_{nm}{\left(\gamma_E+\ln\hat{t}\right)}^2\Big)\ ,
    \label{eq:tildef_4b}
\end{equation}
where $M_e\equiv m_e/m_\mu$ is small enough to justify the double expansion. The values of the $e_{nm}$, $f_{nm}$, $g_{nm}$, and $h_{nm}$ matrices are shown in Tabs.~\ref{tab:NLOb_coefficients3},~\ref{tab:NLOb_coefficients1}, and~\ref{tab:NLOb_coefficients2}. Again, the expansion around $\hat{t}\ll1$ is sufficient for our purposes. It is nevertheless worth noting that because of the electron mass scale, the asymptotic regime is only reached when $M_e\hat{t}=m_e t\gg1$, which occurs only at very large Euclidean times. This makes an interpolation between the small- and large-$\hat{t}$ regions significantly more involved, and we do not attempt this here.

For the remainder of this work, we initially treat the tau-loop effect as negligible and use $\tilde{f}^{(4b)}=\tilde{f}^{(4b)}_e+\tilde{f}^{(4b)}_\tau\approx\tilde{f}^{(4b)}_e$ ($\mathrm{NLOb}\approx\mathrm{NLOb}(e)$). In Sec.~\ref{sec:NLOb_tau}, we briefly revisit this approximation and provide a rough estimate of the tau contribution $\mathrm{NLOb}(\tau)$.

In Fig.~\ref{fig:kernel_comparision} we show the time-kernels for NLOa and NLOb separately and compare them to their sum (solid line) and to the LO time-kernel. This confirms the strong numerical cancellation between NLOa and NLOb, which becomes exact around $t\approx3.6\,\mathrm{fm}$. An important consequence is that the long-distance contributions to the combination of NLOa and NLOb are more suppressed compared to the LO counterpart. This behavior is an important aspect of our analysis and will be discussed further in Sec.~\ref{sec:LD_control}. 

As mentioned above, the time-kernel for diagram NLOc admits an analytic solution of Eq.~\eqref{eq:timekernel_i}, which is shown in App.~\ref{app:timekern_NLOc}. Still, it is convenient to expand it in certain regions of the Euclidean-time plane. The expansion for $\hat{t},\hat{\tau}\ll1$ reads
\begin{equation}
    \begin{aligned}
        \frac{m_\mu^4}{32\pi^4}\tilde{f}^{(4c)}(\hat{t},\hat{\tau}) = &  
        \sum_{n\geq4}\frac{1}{n!}\bigg\{
            i_n\left(\hat{t}^n \hat{\tau}^2 + \hat{t}^2 \hat{\tau}^n\right)
            + j_n\left[\hat{t}^n \hat{\tau}^2(\gamma_E + \ln\hat{t}) + \hat{t}^2 \hat{\tau}^n(\gamma_E + \ln\hat{\tau})\right] \\
        & + k_n\left[{\left(\hat{t} + \hat{\tau}\right)}^n + {\left(\hat{t} - \hat{\tau}\right)}^n - 2\hat{t}^n - 2\hat{\tau}^n\right] \\
        & + l_n\Big[\left(\gamma_E + \ln(\hat{t} + \hat{\tau})\right){(\hat{t} + \hat{\tau})}^n 
                     + \left(\gamma_E + \ln|\hat{t} - \hat{\tau}|\right){(\hat{t} - \hat{\tau})}^n \\
        & - 2 (\gamma_E + \ln\hat{t}) \hat{t}^n 
                       - 2 (\gamma_E + \ln\hat{\tau}) \hat{\tau}^n\Big]
        \bigg\}\ .
    \end{aligned}
    \label{eq:tildef_4c}
\end{equation}
The analytic values for the coefficients $i_n$, $j_n$, $k_n$ and $l_n$ are listed in Tab.~\ref{tab:NLOc_small}.

We note that the convergence of this expansion is governed by $\hat{t} + \hat{\tau} \ll 1$ rather than $\sqrt{\hat{t}^2 + \hat{\tau}^2} \ll 1$. As a consequence, for ensembles with a large temporal extent, the expansion may not cover the entire numerically relevant region of the $(t,\tau)$ plane, particularly along the diagonal $t = \tau$ at large values. A complete treatment requires interpolation with a complementary high-Euclidean-time expansion to cover the full integration domain. Full details on the derivation of the time-kernels in Eqs.~\eqref{eq:tildef_4b} and~\eqref{eq:tildef_4c}, the closed-form solution in Eq.~\eqref{eq:diagram_c_exact}, and the possible interpolation procedure with large temporal extents will be provided in a forthcoming publication~\cite{beltran2026}.

\subsection{Window observables}
\label{sec:wind}

\begin{figure}[t]
    \centering
    \begin{subfigure}[b]{0.49\textwidth}
        \centering
        \includegraphics[width=\textwidth]{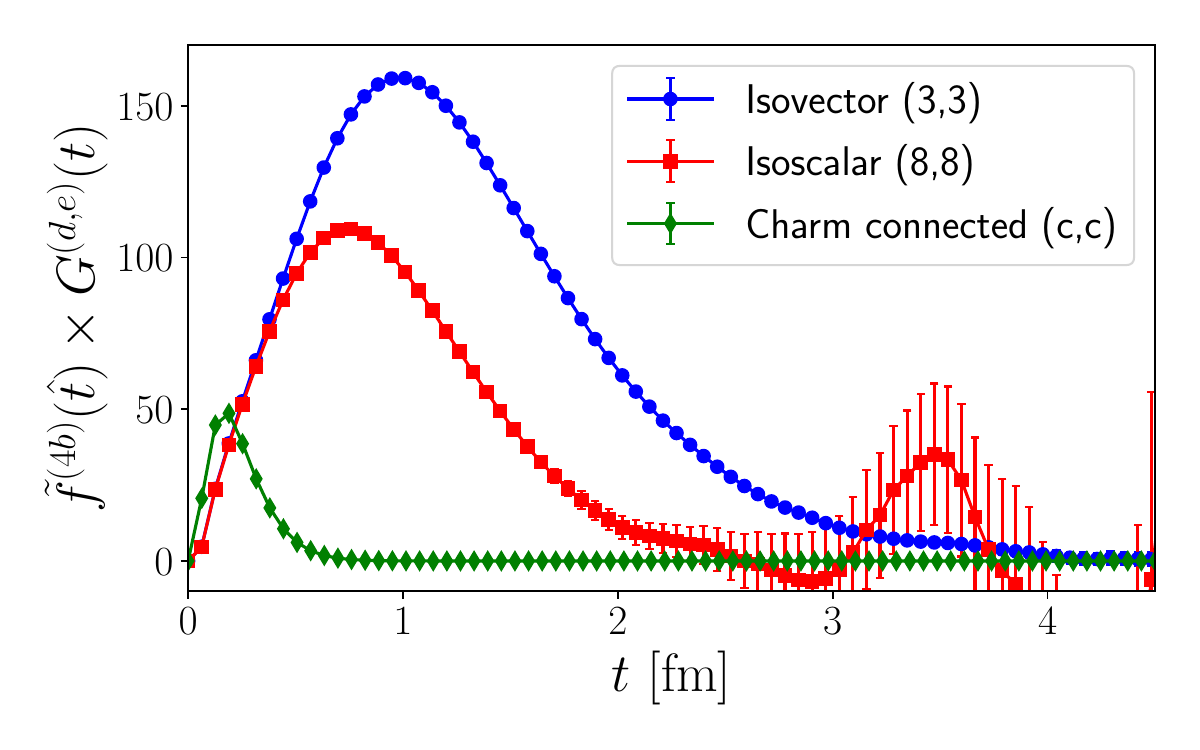}
        \caption{}
        \label{fig:isospin_decomp}
    \end{subfigure}
    \hfill
    \begin{subfigure}[b]{0.49\textwidth}
        \centering
        \includegraphics[width=\textwidth]{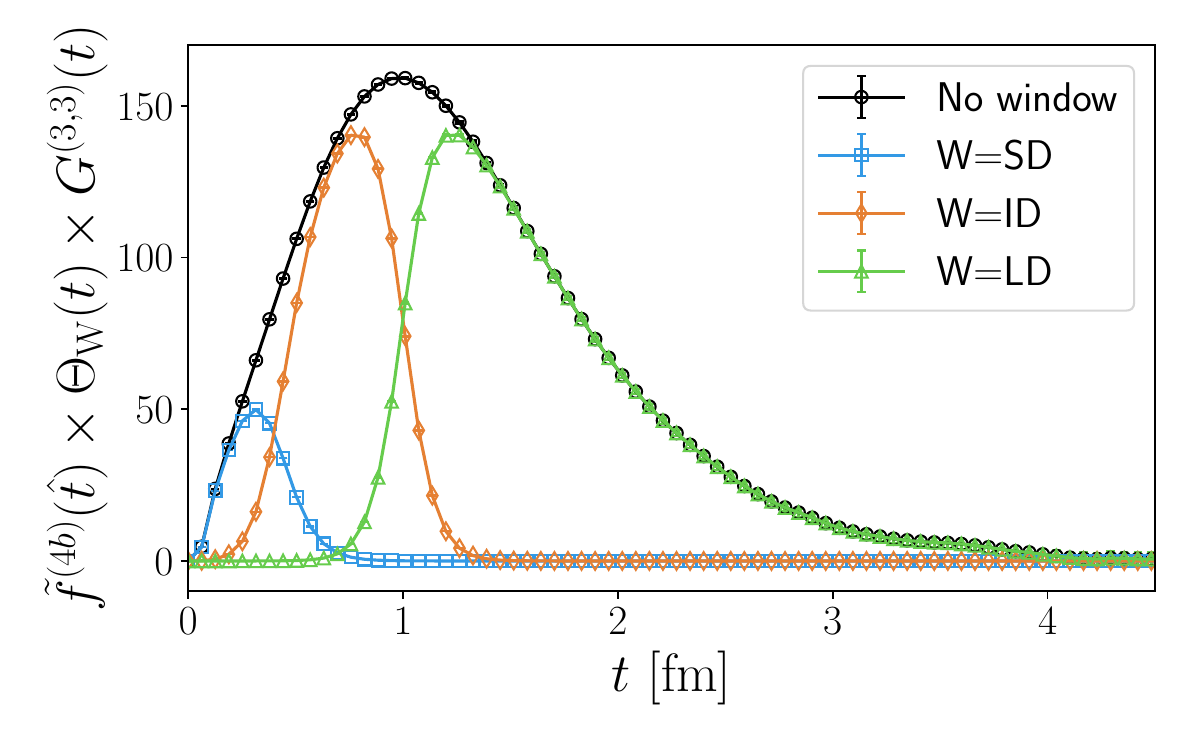}
        \caption{}
        \label{fig:wind_decomp}
    \end{subfigure}
    \caption{Decomposition of the integrand for NLOb in the isospin basis (left) and in the three characteristic windows for the isovector channel (right). The data used for these plots come from ensemble E250.}
  \label{fig:decomp}
\end{figure}

The use of window observables in determinations of the leading-order HVP is now standard in most lattice determinations based on the TMR. This practice, introduced in~\cite{RBC:2018dos}, allows for a more direct and transparent comparison with phenomenological estimates of the LO contribution. In addition, it has proven effective for isolating and controlling systematic uncertainties that naturally arise from the short- and long-distance regions of the integrand. Indeed, this systematic control is the primary motivation for adopting the window approach in this work. The integrand is then rewritten as a partition in which each component can be treated independently and subsequently combined,
\begin{equation}
    \tilde{f}^{(i)}(\hat{t}) = \tilde{f}^{(i)}(\hat{t})\left(\Theta_{\mathrm{SD}}(t) + \Theta_{\mathrm{ID}}(t) + \Theta_{\mathrm{LD}}(t)\right)\ .
\end{equation}
The locations and width of the smoothed step functions $\Theta_{W}(t)$ are defined in terms of the parameters $t_1$, $t_2$, and $\Delta$ according to~\eqref{eq:window_def}, with standard values $t_1=0.4\, \mathrm{fm}$, $t_2=1.0\, \mathrm{fm}$, and $\Delta = 0.15\, \mathrm{fm}$~\cite{RBC:2018dos}.
\begin{equation}
    \begin{aligned}
        \Theta_{\mathrm{SD}}(t) &= 1 - \frac{1}{2}\left(1 + \tanh\frac{t-t_1}{\Delta}\right)\, ,\\
        \Theta_{\mathrm{ID}}(t) &= \frac{1}{2}\left(1 + \tanh\frac{t-t_1}{\Delta}\right) - \frac{1}{2}\left(1 + \tanh\frac{t-t_2}{\Delta}\right)\, ,\\
        \Theta_{\mathrm{LD}}(t) &= \frac{1}{2}\left(1 + \tanh\frac{t-t_2}{\Delta}\right)\, .\\
    \end{aligned}
    \label{eq:window_def}
\end{equation}
Figure~\ref{fig:wind_decomp} shows the window decomposition of the integrand.

In this work, we employ this technique to better control systematic effects. For the most relevant diagram sets, NLOa and NLOb, we separate the integrand into the usual three windows and perform a dedicated analysis for each of them,
\begin{equation}
    a^{\mathrm{hvp},\,(i)}_\mu = {\left(a^{\mathrm{hvp},\,(i)}_\mu\right)}^{\mathrm{SD}} + {\left(a^{\mathrm{hvp},\,(i)}_\mu\right)}^{\mathrm{ID}} + {\left(a^{\mathrm{hvp},\,(i)}_\mu\right)}^{\mathrm{LD}}\ ,
\end{equation}
for $(i)=(4a)$ and $(4b)$. Diagram NLOc is both subleading and computationally the most demanding of the three; given its much smaller numerical impact relative to NLOa and NLOb, we do not apply a window decomposition to it. A more direct and less refined estimate is sufficient for our purposes.

Each window presents its own challenges. The short-distance contribution is most susceptible to lattice artifacts, making its continuum extrapolation particularly delicate. As will be discussed in more detail in Sec.~\ref{sec:SD_sub}, we apply a multiplicative tree-level improvement and allow for quartic lattice-spacing terms in our fits. Logarithmically enhanced lattice-spacing effects are expected to appear when $t\rightarrow 0$. These terms cannot be constrained with sufficient precision and are cancelled through the introduction of subtracted kernels, as considered in~\cite{Kuberski:2024bcj}. The short-distance window is dominated by high-energy contributions to the spectral function. At these scales, $\mathrm{SU}{(3)}_{\mathrm{f}}$ symmetry is approximately restored, which we exploit to improve our determinations.

The long-distance window is most strongly affected by the scale-setting observable. As described in more detail in Sec.~\ref{sec:extrap}, we set the scale using the gradient-flow observable $t_0$~\cite{t0madrid}. This part of the integrand also suffers from the well-known noise-to-signal problem, which we discuss in Sec.~\ref{sec:LD_control}. This window typically dominates the uncertainties and must be handled with particular care. The significant cancellation between the NLOa and NLOb contributions implies a reduced sensitivity to scale setting, statistical noise, and finite-volume corrections in the long-distance region. As discussed in more detail in Sec.~\ref{sec:LD_control}, this feature can be exploited to improve the precision in our final result.

To protect this work from unconscious bias, we apply a multiplicative blinding factor to the long-distance window, which is reversed only after the analysis is complete and frozen:
\begin{equation}
    \Theta_{\mathrm{LD}}(t) \longrightarrow \mathrm{BLIND}\times\Theta_{\mathrm{LD}}(t)\, .
    \label{eq:blinding}
\end{equation}

\begin{table}[!htbp]
\vskip 0.1in
\begin{tabular}{l@{\hskip -.15em}c@{\hskip -.1em}c@{\hskip 01em}c@{\hskip 01em}c@{\hskip 01em}c@{\hskip 01em}c@{\hskip 01em}c@{\hskip 01em}c@{\hskip 01em}c@{\hskip 01em}
	}
\hline
Id   & $\quad\beta\phantom{\Big|}\quad$   & bc & $\textstyle{\big(\frac{L}{a}\big)}^3\times\frac{T}{a}$   & $a\,[{\rm fm}]$   & $m_\pi\,[{\rm MeV}]$   & $m_K\,[{\rm MeV}]$   &   $m_\pi L$ &   $L\,[{\rm fm}]$ 
& MDU
\\
\hline
A653 & 3.34 & p  & $24^3 \times 48$   & 0.097 & 430 & 430 & 5.1 & 2.3 & 20200 \\
A654$^{*}$ &      & p  & $24^3 \times 48$   &       & 338 & 462 & 4.0 & 2.3 & 16000 \\
H650 &      & p  & $32^3 \times 96$   &       & 258 & 470 & 4.1 & 3.1 & 6152 \\
\hline
H101 & 3.4  & o  & $32^3 \times 96$   & 0.085 & 424 & 424 & 5.8 & 2.7 &  8064 \\
H102 &      & o  & $32^3 \times 96$   &       & 358 & 445 & 4.9 & 2.7 &  7832 \\
N101 &      & o  & $48^3 \times 128$  &       & 282 & 468 & 5.8 & 4.1 &  6376 \\
C101 &      & o  & $48^3 \times 96$   &       & 222 & 478 & 4.6 & 4.1 &  8000 \\
C102$^{\dag}$ &      & o  & $48^3 \times 96$   &       & 224 & 506 & 4.6 & 4.1 &  6000 \\
D150$^{\dag}$ &      & p  & $64^3 \times 128$  &       & 131 & 484 & 3.6 & 5.4 &  1616 \\
\hline
B450 & 3.46 & p  & $32^3 \times 64$   & 0.075 & 422 & 422 & 5.1 & 2.4 &  6448 \\
N452 &      & p  & $48^3 \times 128$  &       & 356 & 447 & 6.5 & 3.6 &  4000 \\
N451 &      & p  & $48^3 \times 128$  &       & 291 & 468 & 5.3 & 3.6 &  4044 \\
D450 &      & p  & $64^3 \times 128$  &       & 219 & 483 & 5.3 & 4.8 &  2000 \\
D451$^{\dag}$ &      & p  & $64^3 \times 128$  &       & 219 & 509 & 5.3 & 4.8 &  2056 \\
D452 &      & p  & $64^3 \times 128$  &       & 156 & 490 & 3.8 & 4.8 &  4000 \\
\hline
N202 & 3.55 & o  & $48^3 \times 128$  & 0.064 & 417 & 417 & 6.4 & 3.0 &  7608 \\
N203 &      & o  & $48^3 \times 128$  &       & 349 & 447 & 5.4 & 3.0 &  6172 \\
N200 &      & o  & $48^3 \times 128$  &       & 286 & 468 & 4.4 & 3.0 &  6848 \\
D251 &      & p  & $64^3 \times 128$  &       & 286 & 467 & 5.9 & 4.1 &  5968 \\
D200 &      & o  & $64^3 \times 128$  &       & 202 & 486 & 4.2 & 4.1 &  8004 \\
D201$^{\dag}$ &      & o  & $64^3 \times 128$  &       & 202 & 507 & 4.2 & 4.1 &  4312 \\
E250$^{\dag}$ &      & p  & $96^3 \times 192$  &       & 131 & 495 & 4.1 & 6.1 &  4496 \\
\hline
N300$^{*}$ & 3.7  & o  & $48^3 \times 128$  & 0.049 & 425 & 425 & 5.1 & 2.4 &  8188 \\
J307 &      & o  & $64^3 \times 192$  &       & 424 & 424 & 6.7 & 3.1 &  15896 \\
N302$^{*}$ &      & o  & $48^3 \times 128$  &       & 350 & 456 & 4.2 & 2.4 &  8804 \\
J306 &      & o  & $64^3 \times 192$  &       & 349 & 455 & 5.6 & 3.1 &  16000 \\
J303 &      & o  & $64^3 \times 192$  &       & 260 & 480 & 4.1 & 3.1 &  8584 \\
J304$^{\dag}$ &      & o  & $64^3 \times 192$  &       & 263 & 530 & 4.2 & 3.1 &  6508 \\
E300 &      & o  & $96^3 \times 192$  &       & 177 & 498 & 4.2 & 4.7 &  7180 \\
F300$^{\dag}$ &      & o  & $128^3 \times 256$ &       & 136 & 496 & 4.3 & 6.3 &  2400 \\
\hline
J500 & 3.85 & o  & $64^3 \times 192$  & 0.039 & 417 & 417 & 5.2 & 2.5 & 15000 \\
J501 &      & o  & $64^3 \times 192$  &       & 337 & 450 & 4.2 & 2.5 & 15680 \\
\hline
\end{tabular} 
\caption{Overview of ensembles used in this work, characterized by the bare coupling $\beta =
	6/g_0^2$, the temporal boundary conditions, open (o) or
	anti-periodic (p), the lattice dimensions, the lattice spacing $a$
	in physical units based on~\cite{scalesettingcls_straßberger2021,RQCD:2022xux}, the approximate pion and kaon masses, the physical size of the
	lattice and the length of the Monte Carlo chain in Molecular
	Dynamics Units (MDU). Ensembles with an asterisk are excluded from the final analysis of the light quark (isovector and isoscalar) channels but are included in the determination of charm quark contributions. Ensembles marked by a dagger lie on a second chiral trajectory where $m_{\rm s} \approx m_{\rm s}^{\rm phys}$.
}
\label{tab:ensembles}
\end{table}

\subsection{Lattice setup and extrapolation}
\label{sec:extrap}
Our calculations are performed on a subset of $2+1$-flavor ensembles generated by the CLS effort~\cite{Bruno:2014jqa,Bali:2016umi} using a tree-level Symanzik-improved L\"{u}scher-Weisz gauge action and a non-perturbatively $\mathrm{O}(a)$-improved Wilson fermion action~\cite{Bulava:2013cta}. The light quark sector is stabilized through a mass twist in the Dirac operator and is subsequently reweighted back to the target action using appropriate reweighting factors~\cite{Clark:2006fx,RW2_kuberski2024}. The strange quark sector is simulated using the RHMC algorithm.

We make use of two chiral trajectories. Most ensembles lie on a trajectory along which the sum of the bare sea-quark masses is held constant, corresponding approximately to
$m_K^2 + \frac{1}{2}m_\pi^2 \approx \mathrm{cst}$. For each lattice spacing, we evolve the pion and kaon masses from the $\mathrm{SU}{(3)}_{\mathrm{f}}$-symmetric point toward their physical values while maintaining this condition. To account for possible mistunings and to improve control over the strange quark mass dependence, we also include a second trajectory along which the strange quark mass
is kept close to its physical value.

The ensemble selection and statistics are similar to those used in our recent works~\cite{Djukanovic:2024cmq,Conigli:2025qvh}. Relative to those works, we include a new ensemble at the coarsest lattice spacing (H650), which improves our sensitivity to higher-order cutoff effects and reduces our reliance on the A-box ensembles, whose spatial and temporal extents are relatively small. In addition, we have increased the statistics for  the J306 and J307 ensembles, as well as for F300, a near-physical-point ensemble with fine lattice spacing. For E300, we have significantly improved the statistics of the disconnected contributions. The complete set of ensembles used in this work is listed in Tab.~\ref{tab:ensembles}.

As in many of our previous works, this ensemble set is used to constrain a global chiral-continuum extrapolation to the physical point. We set the scale using the gradient-flow observable $t_0/a^2$~\cite{Luscher:2010iy}, which can be determined with high precision on each ensemble. For its physical value, we take $\sqrt{t_0^{\mathrm{ph}}} = 0.1440(7)\,\mathrm{fm}$ from~\cite{t0madrid}, which is based on a combination of the pion and kaon decay constants $f_{\pi K}$.

We define the physical point in the isospin-symmetric QCD world by
\begin{equation*}
    m_\pi = {(m_\pi^0)}^{\mathrm{ph}}\, ,\qquad 
    2m_K^2 - m_\pi^2 = {\left(m_{K^+}^2 + m_{K^0}^2 - m_{\pi^+}^2\right)}^{\mathrm{ph}}\, ,
\end{equation*}
which is equivalent to~\cite{ParticleDataGroup:2024cfk}
\begin{equation*}
    m_\pi = 134.9768(5)\, \mathrm{MeV}\, ,\qquad 
    m_K = 495.011(10)\, \mathrm{MeV}\, .
\end{equation*}
We introduce the dimensionless variables
\begin{equation}
    X_a = \sqrt{\frac{a^2}{8t_0}}\, ,\quad 
    \Phi_2 = 8t_0 m_\pi^2\, ,\quad 
    \Phi_4 = 8t_0\left(m_K^2 + \frac{1}{2}m_\pi^2\right)\, ,
\end{equation}
to parametrize the dependence of the observables on the lattice spacing, the light quark mass, and the strange quark mass, respectively.

Our baseline ansatz for the lattice-spacing and chiral dependence of an observable
$\mathcal{O} = \mathcal{O}(X_a,\Phi_2,\Phi_4)$ around the physical point
$\mathcal{O}^{\mathrm{ph}} = \mathcal{O}(0,\Phi_2^{\mathrm{ph}},\Phi_4^{\mathrm{ph}})$
is given by
\begin{equation}
    \mathcal{O} = \mathcal{O}^{\mathrm{ph}} 
    + \alpha_0 X_a^2 
    + \beta_0\left(\Phi_2 - \Phi_2^{\mathrm{ph}}\right) 
    + \gamma_0\left(\Phi_4 - \Phi_4^{\mathrm{ph}}\right) \, .
    \label{eq:ansatz_base}
\end{equation}
This baseline form assumes a quadratic dependence on the lattice spacing and quark masses. To account for possible higher-order effects, we consider extensions of Eq.~\eqref{eq:ansatz_base} that include additional terms:
\begin{equation}
    \cdots + \alpha_1 X_a^3 
    + \alpha_2 X_a^2\left(\Phi_2 - \Phi_2^{\mathrm{ph}}\right) 
    + \alpha_3 X_a^2\left(\Phi_4 - \Phi_4^{\mathrm{ph}}\right) 
    + \beta_1\left(f_{\mathrm{ch}}(\Phi_2) - f_{\mathrm{ch}}(\Phi_2^{\mathrm{ph}})\right)\, ,
\end{equation}
where $f_{\mathrm{ch}}(\Phi_2) \in \{\Phi_2^2,\; \Phi_2\ln\Phi_2\}$ represents different functional forms for the chiral dependence. We explore various combinations of these terms and weight them through a model-averaging procedure, as discussed below. This approach allows the data to determine which higher-order corrections are statistically significant.

For observables in the SD window, where stronger cutoff effects are expected, we also allow for a quartic lattice-spacing term $\alpha_4 X_a^4$. In the isovector channel of the ID and LD windows, larger chiral effects are anticipated. In this case, we allow for a second higher-order chiral term, $\beta_2\big(f_{\mathrm{ch}}(\Phi_2) - f_{\mathrm{ch}}(\Phi_2^{\mathrm{ph}})\big)$, and we further include asymptotic chiral behaviors $f_{\mathrm{ch}}(\Phi_2) \in \{\Phi_2^2,\; \Phi_2\ln\Phi_2,\; 1/\Phi_2,\; \ln\Phi_2\}$.

We also consider the possibility of cutoff effects with a non-zero anomalous dimension $\hat{\Gamma}$~\cite{Husung:2019ytz,Husung:2021mfl,anomalousdim_Husung_2024}. For each fit, we then consider two variants of the leading cutoff term, $\alpha_0 X_a^2 \rightarrow \alpha_0 {[\alpha_s(1/a)]}^{\hat{\Gamma}} X_a^2$, with $\hat{\Gamma} \in \{0;\,0.395\}$.

To obtain a final estimate at the physical point and to quantify the systematic uncertainty associated with the extrapolation, we perform a model average over all fit ans\"{a}tze considered. To this end, we employ the Akaike Information Criterion (AIC)~\cite{Akaike:1998zah,Jay:2020jkz,Neil:2022joj} and assign a weight to each fit, based on the observed $\chi^2$ and the expected $\chi^2_{\mathrm{exp}}$~\cite{CorrFits_MattiaBruno}. The central value and its statistical uncertainty are determined from a weighted average of the continuum results from all data sets, while the systematic uncertainty is estimated from the weighted variance across the model set following Ref.~\cite{Jay:2020jkz}, reflecting the distribution of the models included in the model average.

In addition, we consider three different cuts on the ensemble set within the model average. Cuts are implemented by (1) excluding ensembles at $\beta = 3.34$ (the coarsest lattice spacing), (2) excluding ensembles at the $\mathrm{SU}{(3)}_{\mathrm{f}}$-symmetric point, and (3) by combining both criteria. For some short-distance observables that show more pronounced cut-off effects, we also explore a further cut in the lattice spacing, excluding $\beta\leq3.4$. For the determination and propagation of uncertainties throughout this work, we use the \texttt{ADerrors} package~\cite{Ramos:2018vgu}, which implements the $\Gamma$-method~\cite{GammaMethod_UlliWolff_2003} to account for autocorrelations.

\section{Strategy}
\label{sec:strategy}

\subsection{NLOa and NLOb diagram sets}
\label{sec:NLOab}

In this section, we describe the strategy adopted to reduce systematic effects in the determination of the leading NLOa and NLOb sets.

\subsubsection{Short-distance window}
\label{sec:SD_sub}

The short-distance window exhibits a strong dependence on the lattice spacing and presents several additional challenges. Here, we present the approach followed for this window. Most of the techniques described below were introduced previously in Ref.~\cite{Kuberski:2024bcj}. However, the presence of photon lines in diagram set NLOa further complicates the extrapolation, making a straightforward application of the methods of Ref.~\cite{Kuberski:2024bcj} impossible.

As briefly discussed in Sec.~\ref{sec:wind}, we expect that the continuum extrapolation of the small-Euclidean-time region of the integrand is affected by logarithmically enhanced cutoff effects. In practice, such terms cannot be reliably constrained through fits to a global continuum limit. Due to the photon lines appearing in diagram set NLOa, we expect these logarithmically enhanced effects to be even more problematic for this observable. In particular, in addition to the more common $\mathrm{O}(a^2\ln a)$ terms, we also expect terms like $\mathrm{O}(a^2\ln^2 a)$ to arise, this is discussed and motivated in App.~\ref{app:SDsub_NLOa}.

\begin{figure}[t]
    \centering
    \includegraphics[width=0.625\textwidth]{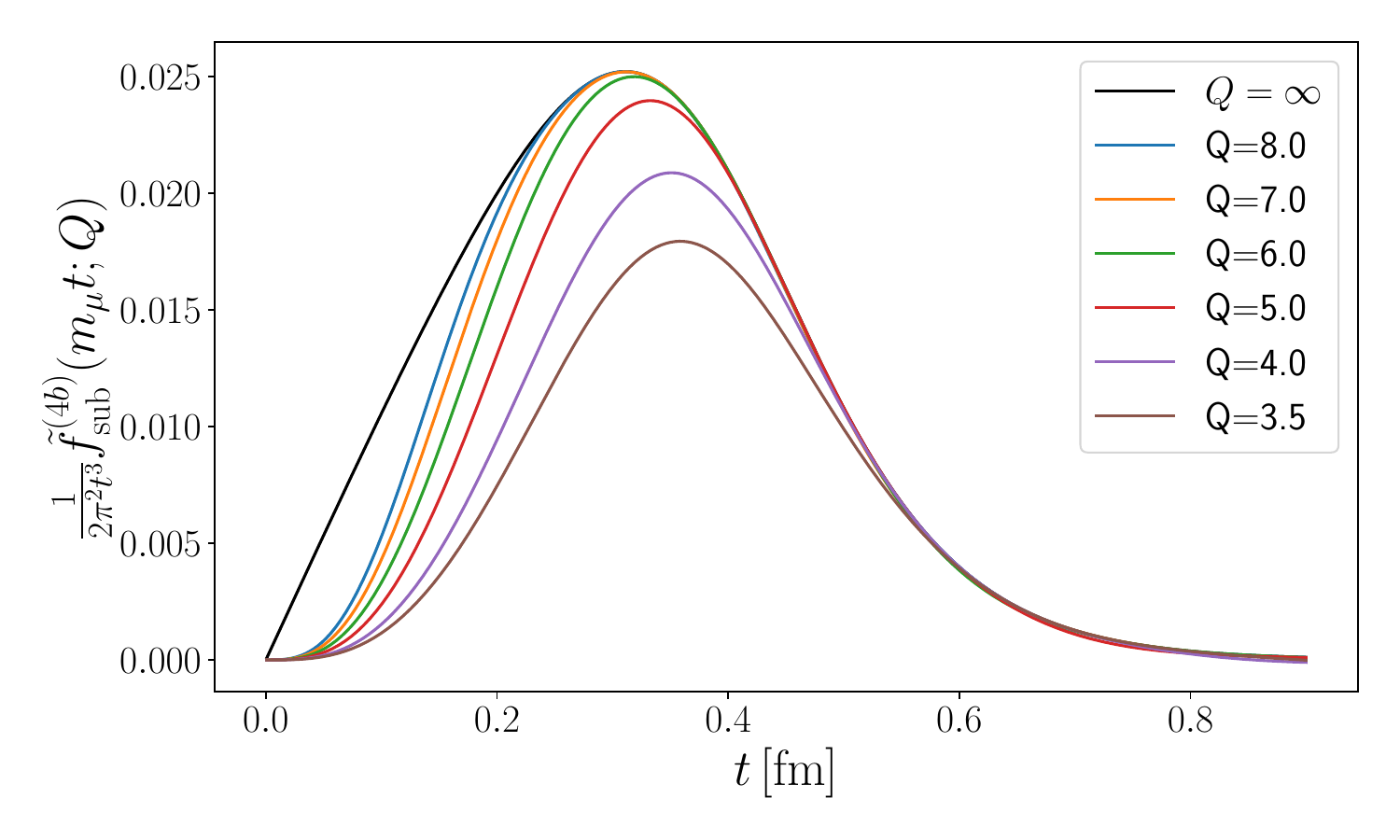}
    \caption{Modification of the short-distance window applied to the time-kernel for diagram set NLOb following the subtraction in Eq.~\eqref{eq:subtraction_def} for all explored auxiliary virtualities in this work.}
  \label{fig:SDsubtraction}
\end{figure}

Following Ref.~\cite{Kuberski:2024bcj}, we remove the leading short-distance behavior through a redefinition of the kernel, which cancels these unwanted contributions (see Fig.~\ref{fig:SDsubtraction}):
\begin{equation}
    \Theta_{\mathrm{SD}}(t)\tilde{f}^{(i)}(\hat{t})
    \;\longrightarrow\;
    \tilde{f}^{(i)}_{\mathrm{sub}}(\hat{t};Q)
    =
    \Theta_{\mathrm{SD}}(t)\tilde{f}^{(i)}(\hat{t})
    -
    \Theta_{\mathrm{SD}}(0)
    \left(\frac{16\pi m_\mu}{Q^2}\right)^{2}
    C_4^{(i)} \sin^4\!\left(\frac{Qt}{4}\right)\, ,
    \label{eq:subtraction_def}
\end{equation}
where $C_4^{(i)}$ is the coefficient of the leading term in the small-$\hat{t}$ expansion,
\begin{equation}
    \tilde{f}^{(i)}(\hat{t}) \;=\; \frac{\pi^2}{m_\mu^2}\, C_4^{(i)}\, \hat{t}^4 + \mathrm{O}(\hat{t}^6)\, .
\end{equation}

For sufficiently small $Q^2$, this subtraction removes the $\mathrm{O}(a^2 \ln a)$ and $\mathrm{O}(a^2 \ln^2 a)$ behavior in the continuum extrapolation. We then estimate the subtracted piece perturbatively and generally write the combined contribution as
\begin{equation}
    \big(a^{d,e,\,(i)}_\mu\big)^{\mathrm{SD}}
    =
    \big(a^{d,e,\,(i)}_\mu\big)^{\mathrm{SD}}_{\mathrm{sub}}(Q^2)
    +
    \Theta_{\mathrm{SD}}(0)
    \left(\frac{\alpha}{\pi}\right)^{N_i}
    \tilde{b}^{(d,e),\,(i)}(Q^2)\, ,
    \label{eq:subtraction_comb}
\end{equation}
where the dependence on the auxiliary virtuality $Q$ must cancel on the right-hand side. The function $\tilde{b}$ is given by
\begin{equation}
    \tilde{b}^{(d,e),\,(i)}(Q^2) = \left(\frac{16 m_\mu \alpha}{Q^2}\right)^{2}
    \int_0^\infty dt\, C_4^{(i)}\, \sin^4\!\left(\frac{Qt}{4}\right)\, G(t)\, .
    \label{eq:b_tilde}
\end{equation}
For the LO and NLOb diagrams, i.e. for $(i)=(2)$ and $(4b)$, the time-kernel expansion at leading order in $\hat{t}$ is analytic with expansion coefficients
\begin{equation*}
    \begin{aligned}
        C_4^{(2)} =& \frac{1}{9}\, ,\\[4pt]
        C_4^{(4b)} =& - \frac{1}{27} (1 + 4 \ln M_e)
        + \frac{2}{3}M_e^2
        - \frac{4\pi^2}{9} M_e^3
        + \frac{2}{3}\!\left(1 + \frac{\pi^2}{6} + \ln^2 M_e\right) M_e^4 \\
        &
        - \frac{4}{27}
        \left(\frac{23}{6} + \pi^2 - 7\ln M_e + 6\ln^2 M_e\right) M_e^6
        + \mathrm{O}(M_e^8)\, .
    \end{aligned}
\end{equation*}
Thus, the coefficient $C_4^{(i)}$ can be factored out of the integral in Eq.~\eqref{eq:b_tilde}, and hence for $(i) = (2)$ and $(4b)$, the computation reduces to $\tilde{b}^{(d,e),\,(i)}(Q^2) = C_4^{(i)} b^{(d,e)}(Q^2)$, where
\begin{equation}
    b^{(d,e)}(Q^2)
    =
    \left(\frac{4m_\mu \alpha}{Q}\right)^{2}
    \left[
        \Pi^{(d,e)}(Q^2) - \Pi^{(d,e)}(Q^2/4)
    \right]
    \label{eq:b_momentum}
\end{equation}
can be computed perturbatively to high precision for sufficiently large $Q^2$.

For diagram set NLOa, the presence of photon lines modifies the short-distance behavior. In this case, the first term in the time-kernel expansion depends logarithmically on the Euclidean time,
\begin{equation}
    C_4^{(4a)} = 
    C_4^{(4a)}(\hat{t})
    =
    \frac{317}{324}
    - \frac{2\pi^2}{9}
    + \frac{23}{27}\bigl(\gamma_E + \ln \hat{t}\bigr)\, .
    \label{eq:C4_NLOa_def}
\end{equation}
This means that the coefficient  $C_4^{(4a)}$ cannot be factored out. In addition, it produces an even more problematic $\mathrm{O}(a^2 \ln^2 a)$ contribution. For $(i)=(4a)$, we then have
\begin{equation}
    \begin{aligned}
        \tilde{b}^{(d,e),\,\mathrm{nlo(a)}}(Q^2) = &
        \left[
            \frac{317}{324}
            - \frac{2\pi^2}{9}
            + \frac{23}{27}\!\left(\gamma_E + \ln \frac{m_\mu}{\Lambda}\right)
        \right] b^{(d,e)}(Q^2)
        \\
        &
        +\;
        \frac{23}{27}
        \left(\frac{16m_\mu\alpha}{Q^2}\right)^{2}
        \int_0^\infty dt\,
        \ln(\Lambda t)\;
        G^{(d,e)}(t)\,
        \sin^4\!\left(\frac{Qt}{4}\right)\, .
    \end{aligned}
    \label{eq:b_tilde_NLOa}
\end{equation}
In order to isolate and evaluate the effect of the logarithmic contribution to Eq.~\eqref{eq:C4_NLOa_def}, we introduce a scale $\Lambda$ via $\hat{t}\equiv m_\mu t=(m_\mu/\Lambda)(\Lambda t)$. Equation~\eqref{eq:b_tilde_NLOa} should be independent of $\Lambda$ for all virtualities. Hence, for a suitable choice of $\Lambda$, the second term in Eq.~\eqref{eq:b_tilde_NLOa} is subleading, and the variation of the result under changes of $\Lambda$ can be used to obtain a conservative estimate of the associated uncertainty. 

To evaluate the integral, we employ the perturbative expansion of the position-space vector correlator from Ref.~\cite{Chetyrkin_pQCD}, which is known up to $\mathrm{O}(\alpha_s^4)$. Further details are provided in App.~\ref{app:SDsub_NLOa}. In Tab.~\ref{tab:sub_piece}, we present the perturbative results for $\tilde{b}^{(3,3)}(Q^2)$ at several values of the virtuality.
\begin{table}[t]
    \centering
    \small
    \renewcommand{\arraystretch}{1.1}
    \setlength{\tabcolsep}{4.9pt}    
    \begin{tabular}{c | r@{.}l r@{.}l r@{.}l r@{.}l r@{.}l r@{.}l}
        diag & \multicolumn{2}{c}{$Q=3.5$} & \multicolumn{2}{c}{$Q=4.0$} & \multicolumn{2}{c}{$Q=5.0$} & \multicolumn{2}{c}{$Q=6.0$} & \multicolumn{2}{c}{$Q=7.0$} & \multicolumn{2}{c}{$Q=8.0$} \\
        \noalign{\smallskip}\hline\noalign{\smallskip}
        NLOa    & $-$9&162(77) & $-$7&264(50) &  $-$4&922(24) & $-$3&576(14) & $-$2&7261(91) & $-$2&1528(68) \\
        NLOb    &  2&635(12) &  2&0031(74) &  1&2698(35) & 0&8762(19) &  0&6408(11) &  0&48891(79) \\
        NLOa\&b & $-$6&524(69) & $-$5&259(45) &  $-$3&653(22) & $-$2&701(13) &  $-$2&0858(83) & $-$1&6642(62) \\
    \end{tabular}
    \caption{Values for the subtracted piece $(\alpha/\pi)\,\tilde{b}^{(3,3)}(Q^2)$, for the leading diagrams and the virtualities that we explore in this work (all virtualities are given in GeV). The central values and uncertainties are estimated following the discussion in App.~\ref{app:SDsub_NLOa}.}
    \label{tab:sub_piece}
\end{table}

By construction, the SD window is dominated by the high-energy regime, where $\mathrm{SU}(3)_{\mathrm{f}}$ is approximately restored, i.e. we expect the quark-disconnected contributions to be small and $G^{(3,3)} \approx G^{(8,8)} \approx G_{\mathrm{conn}}^{(8,8)} \approx G^{(l,l)}/2 \approx G^{(s,s)}/2$. To exploit this symmetry, we determine the isoscalar $(8,8)$ and strange $(s,s)$ channels from the isovector $(3,3)$ contribution. We thus introduce the suppressed quantities
\begin{equation}
    \begin{aligned}
            \Delta_{ls}\left(a_\mu^{(i)}\right)^{\mathrm{SD}} &= \left(\frac{\alpha}{\pi}\right)^{2+N_i}\int_0^\infty dt\, \tilde{f}^{(i)}(\hat{t})\Theta_{\mathrm{SD}}(t)\left(G^{(8,8)}(t)-G^{(3,3)}(t)\right)\, , \\
            \Delta^{\mathrm{conn}}_{ls}\left(a_\mu^{(i)}\right)^{\mathrm{SD}} &= \left(\frac{\alpha}{\pi}\right)^{2+N_i}\int_0^\infty dt\, \tilde{f}^{(i)}(\hat{t})\Theta_{\mathrm{SD}}(t)\left(G_{\mathrm{conn}}^{(8,8)}(t)-G^{(3,3)}(t)\right)\, ,
    \end{aligned}
\end{equation}
and extract the isoscalar and strange contributions from
\begin{equation}
    \begin{aligned}
        \left(a_\mu^{8,8,\,(i)}\right)^{\mathrm{SD}} &= \left(a_\mu^{3,3,\,(i)}\right)^{\mathrm{SD}} + \Delta_{ls}\left(a_\mu^{(i)}\right)^{\mathrm{SD}}\, ,\\
        \left(a_\mu^{s,s,\,(i)}\right)^{\mathrm{SD}} &= 2\left(a_\mu^{3,3,\,(i)}\right)^{\mathrm{SD}} + 3\Delta_{ls}^{\mathrm{conn}}\left(a_\mu^{(i)}\right)^{\mathrm{SD}}\, .
    \end{aligned}
    \label{eq:SDsub_88_ss}
\end{equation}
To evaluate $\tilde{b}^{(c,c)}$, we relate to the massless perturbative determination of $\tilde{b}^{(3,3)}$. To this end, we proceed with a non-perturbative evaluation of the suppressed quantity
\begin{equation}
    \Delta_{lc}\left(\tilde{b}^{(i)}\right)(Q^2) = \left(\frac{16 m_\mu \alpha}{Q^2}\right)^{2}
    \int_0^\infty dt\, C_4^{(i)}(\hat{t})\, \sin^4\!\left(\frac{Qt}{4}\right)\, \left(G^{(c,c)}(t)-2G^{(3,3)}(t)\right)\, ,
\end{equation}
where the effect from log-enhanced terms is again suppressed. The charm quark contribution is then obtained by combining all individual pieces,
\begin{equation}
    \left(a_\mu^{c,c,\,(i)}\right)^{\mathrm{SD}} = \left(a_\mu^{c,c,\,(i)}\right)^{\mathrm{SD}}_{\mathrm{sub}}(Q^2) + \Theta_{\mathrm{SD}}(0)\left(\frac{\alpha}{\pi}\right)^{N_i}\tilde{b}^{(c,c),\,(i)}(Q^2)\, ,
    \label{eq:SDsub_cc}
\end{equation}
where
\begin{equation}
    \tilde{b}^{(c,c),\,(i)}(Q^2) = 2\tilde{b}^{(3,3),\,(i)}(Q^2) + \Delta_{lc}\left(\tilde{b}^{(i)}\right)(Q^2)\, .
    \label{eq:SDsub_cc_deltalc}
\end{equation}
In this way, most of the contributions to the SD window are determined from the isovector channel and its constituents. 

Finally, to improve the non-perturbative determination of $\big(a_\mu^{3,3,\,(i)}\big)^{\mathrm{SD}}_{\mathrm{sub}}$, we apply a multiplicative tree-level improvement to the correlator following Ref.~\cite{Kuberski:2024bcj}. Denoting by $G^{(3,3)}_{tl}$ the tree-level estimate of $G^{(3,3)}$, we perform
\begin{equation}
    G^{(3,3)}(t;a)\, \rightarrow\, G^{(3,3)}(t;a)\, \frac{G_{tl}^{(3,3)}(t;0)}{G_{tl}^{(3,3)}(t;a)}\, .
\end{equation}
\subsubsection{Long-distance window}
\label{sec:LD_control}
As is typical for HVP observables, the uncertainty of our final result is expected to be dominated by long-distance contributions. These arise from several sources, including the signal-to-noise problem, finite-volume (FV) effects, isospin-breaking (IB) corrections and the scale setting. FV effects are discussed in Sec.~\ref{sec:fvc}, while IB corrections are addressed in Sec.~\ref{sec:isobreak}. The present section focuses on the signal-to-noise problem and the strategies we employ to mitigate it. Following Ref.~\cite{Djukanovic:2024cmq}, our approach combines low-mode averaging (LMA) estimators, which are based on an exact treatment of the low modes of
the Dirac operator~\cite{Giusti:2004yp, DeGrand:2004qw}, with a dedicated spectral reconstruction of the isovector channel~\cite{Gerardin:2019rua,Andersen:2018mau,Paul:2021pjz,Paul:2023ksa} for two ensembles in the vicinity of the physical point, supplemented by the standard bounding method~\cite{RBC:2018dos}. LMA is applied exclusively to ensembles with near-physical pion masses ($m_\pi\lesssim280\,\mathrm{MeV}$), where it provides a significantly improved estimator for the light-connected correlation function. In addition, we show how cancellations between NLOa and NLOb contributions at large distances can be exploited to suppress statistical noise.

Both the bounding method  and the spectral reconstruction arise from the spectral decomposition of the correlators;
\begin{equation}
    G^{(d,e)}(t) = \sum_{n\geq0}\frac{Z_n^2}{2E_n}e^{-E_n t}\ ,
    \label{eq:spec_rec}
\end{equation}
where $E_n$ is the energy of state $n$ and $Z_n$ is the matrix element of the electromagnetic current between the vacuum and such state. Equation~\eqref{eq:spec_rec} can be used to impose bounds on both the isovector and
isoscalar correlators in the large Euclidean-time region~\cite{RBC:2018dos,Gerardin:2019rua,Borsanyi:2016lpl}. The lower and upper bounds read as follows:
\begin{equation*}
    0 \leq G^{(d,e)}(t_c)\, e^{-E_{\mathrm{eff}}^*(t-t_c)} \leq G^{(d,e)}(t) \leq G^{(d,e)}(t_c)\, e^{-E_0(t-t_c)}\, ,\qquad t \geq t_c
\end{equation*}
where $E_0 = \mathrm{min}(E_{2\pi},m_\rho)$ for $(d,e)=(3,3)$ and $E_0 = m_\rho$ for $(8,8)$. Here $E_{2\pi}$ refers to the ground state energy of two non-interacting pions on the lattice, and $m_\rho$ is the $\rho$ meson mass. The latter is a sufficiently conservative choice for the isoscalar channel, based on the fact that $m_\rho\lesssim m_\omega$. The energy $E_{\mathrm{eff}}^*$ is determined from the logarithmic derivative of the correlator function at some fixed $t_{\mathrm{eff}}\leq t_c$, such that the effective mass $E_{\mathrm{eff}}^*$ at $t_{\mathrm{eff}}$ is clearly larger than in the region where the bounding method is applied.

Lastly, for two of our ensembles nearest to the physical point, E250 and D200, we make use of the generalized eigenvalue problem (GEVP) to perform a dedicated spectroscopy study of the lowest energy levels (see App.~D of Ref.~\cite{Djukanovic:2024cmq}). This spectroscopic information is then used to reconstruct the tail of the isovector channel by replacing $G^{(d,e)}(t)$ for the first few terms in Eq.~\eqref{eq:spec_rec} in the region where only the low-lying states dominate. This produces a substantial increase in precision on the physical-pion-mass ensemble E250, which has a significant effect on constraining the strong chiral dependence of the long-distance window.

In this work, we propose to combine the diagram sets NLOa and NLOb into a single contribution denoted NLOa\&b, where all lepton loops and photon lines are included. This takes advantage of the cancellation in the long-distance regime of both diagram sets, as discussed in Sec.~\ref{sec:timekernels} and shown in Fig.~\ref{fig:kernel_comparision}. This enhances the relevance of the shorter windows and thus naturally suppresses the effect of the signal-to-noise problem on the full NLO (or NLOa\&b) contribution.
\begin{equation}
    a^{\mathrm{hvp,\,nlo(a\&b)}}_\mu = \left(\frac{\alpha}{\pi}\right)^3\int_0^\infty dt\left(\tilde{f}^{(4a)}(\hat{t}) + \tilde{f}^{(4b)}(\hat{t})\right) G(t)\ .
\end{equation}
\subsubsection{Finite-volume corrections}
\label{sec:fvc}

Finite-volume effects must be accounted for, as they are non-negligible and primarily impact the long-distance window. They originate from the distortion of the low-energy hadron spectrum due to the finite spatial volume used in lattice simulations. In this work, we follow the strategy developed in Ref.~\cite{Djukanovic:2024cmq}, extending it to the NLO contributions considered here.

For the dominant diagram sets NLOa and NLOb, we adopt the same procedure as in Refs.~\cite{DellaMorte:2017dyu,Ce:2022eix,Ce:2022kxy,Djukanovic:2024cmq}. In the isovector channel, the FV effects are governed by two-pion states and are therefore sizeable. To control them, we rely on a combination of the Hansen-Patella (HP) formalism~\cite{HP1,HP2} and the Meyer-Lellouch-L\"{u}scher (MLL) method~\cite{MLL}.

The HP method expresses the finite-volume distortion of the correlator in terms of the pion electromagnetic form factor. It is particularly reliable at short and intermediate Euclidean times, where the relevant pion momenta remain in a regime where the form-factor description is accurate. At larger Euclidean times the MLL method becomes more robust, as it incorporates the time-like pion form factor and the correct two-particle finite-volume quantization condition. For this reason, we employ the HP framework for small~$t$ and switch to the MLL description at
\[
    t^* = m_\pi \left(\frac{L}{4}\right)^2\,,
\]
following Refs.~\cite{DellaMorte:2017dyu,Ce:2022eix,Ce:2022kxy,Djukanovic:2024cmq}. This scale ensures a smooth transition between the regions where each method is theoretically well justified. 

To minimize the finite-volume shifts applied to individual ensembles, we first correct each ensemble to a common reference volume $L_{\mathrm{ref}}$, as first introduced in Ref.~\cite{Borsanyi:2020mff}. The residual finite-volume correction (FVC) needed to shift the result from this reference volume to infinite volume is then computed in the continuum limit using chiral perturbation theory (ChPT) at NNLO~\cite{hvp_milc_ChPT}. This two-step strategy (HP\&MLL corrections to $L_{\mathrm{ref}}$ on each ensemble, followed by NNLO ChPT correction to infinite volume in the continuum) ensures that the ensembles closest to the physical point---which carry the most weight in the extrapolation---receive only small shifts, thereby avoiding distortions to the fit from large correlated uncertainties. The residual FV shifts from $L_{\mathrm{ref}}$ to infinite volume for NLOa, NLOb, and their sum are listed in Tab.~\ref{tab:FVC_cont}, to which we assign a 10\% uncertainty. This uncertainty estimate is motivated by the agreement between different FVC approaches observed for the LO HVP determination in Ref.~\cite{Djukanovic:2024cmq}, where variations among methods were found to be at the 10\% level.

The reference point is chosen close to the spatial extent of our finest physical-mass ensemble (F300) and matches the one used in~\cite{Borsanyi:2020mff},
\begin{equation}
    \left(m_\pi L\right)^{\mathrm{ref}}
    =
    \left(m_{\pi^0}\right)_{\mathrm{phys}}
    \times 6.272\ \mathrm{fm}
    \approx 4.290 \, .
\end{equation}
In addition to the dominant pion effects, we also account for subleading FVC from kaon loops. These contributions are typically very small; however, on ensembles along the $\mathrm{Tr}\,M=\mathrm{const}$ trajectory and near the $\mathrm{SU}(3)_{\mathrm{f}}$-symmetric point they can become non-negligible, and are therefore included.

\begin{table}[t]
    \centering
    \small
    \renewcommand{\arraystretch}{1.1}
    \begin{tabular}{c | r@{.}l r@{.}l r@{.}l | r@{.}l}
        diag & \multicolumn{2}{c}{SD} & \multicolumn{2}{c}{ID} & \multicolumn{2}{c|}{LD} & \multicolumn{2}{c}{Total} \\
        \noalign{\smallskip}\hline\noalign{\smallskip}
        NLOa    & $-$0&00990 & $-$0&131 &  $-$2&93 &  $-$3&07 \\
        NLOb    &  0&00340 &  0&0633 &  2&45 &   2&52 \\
        NLOa\&b & $-$0&00650 & $-$0&0680 & $-$0&479 & $-$0&553 \\
    \end{tabular}
    \caption{Finite-volume corrections $(a^{3,3}_\mu - a^{3,3}_\mu(L_{\mathrm{ref}}))\times10^{11}$ relating $L_{\mathrm{ref}}$ to infinite volume in the isovector channel, computed in the continuum limit using ChPT at NNLO. We decompose this correction by window and diagram set.}
    \label{tab:FVC_cont}
\end{table}

\subsection{NLOc diagram}
\label{sec:NLOc}

The NLOc diagram presents additional challenges compared to the NLOa and NLOb contributions. Its defining feature is the presence of two QCD insertions, which in the TMR formalism results in a double integral over Euclidean time. The overall magnitude of this contribution is expected to be significantly smaller than that of the other NLO diagrams. Consequently, achieving the target precision does not require the same level of control over the short- and long-distance regions, and therefore we do not employ a window decomposition for the NLOc integrand. As for the other diagrams, finite-volume corrections are implemented using the HP\&MLL method. However, due to the smallness of the contribution, we drop the matching step involving the intermediate reference volume $L_{\mathrm{ref}}$ and correct directly to infinite volume.

Exploiting the symmetry $\tilde{f}^{(4c)}(\hat{t},\hat{\tau})=\tilde{f}^{(4c)}(\hat{\tau},\hat{t})$ and accounting for all crossed terms, the full observable can be written as a sum over isospin and flavor components
\begin{equation}
    a_\mu^{\mathrm{hvp,\,nlo(c)}} = 
    a_\mu^{3,3-3,3} + 
    \frac{2}{3}a_\mu^{3,3-8,8} + 
    \frac{8}{9}a_\mu^{3,3-c,c} + 
    \frac{1}{9}a_\mu^{8,8-8,8} + 
    \frac{8}{27}a_\mu^{8,8-c,c} + 
    \frac{16}{81}a_\mu^{c,c-c,c} +
    \dots\ ,
\end{equation}
where the ellipsis denotes additional contributions (including bottom quark terms discussed in Sec.~\ref{sec:bottom_quark}) that are numerically subleading. All individual contributions are defined as
\begin{equation}
    a_\mu^{a-b} \equiv a_\mu^{a-b,\,\mathrm{nlo(c)}}
    = \left(\frac{\alpha}{\pi}\right)^3
      \int_0^\infty \!\! dt \int_0^\infty \!\! d\tau\ 
      \tilde{f}^{(4c)}(\hat{t},\hat{\tau})\,
      G^{a}(t)\,G^{b}(\tau)\, .
\end{equation}
Since the NLOc integral is bilinear in two correlators, the finite-volume corrections also contain crossed terms. Writing the FV shift of each flavor combination as
\begin{equation}
    \Delta^{\mathrm{fvc}} a_\mu^{a-b} = 
    \left(\frac{\alpha}{\pi}\right)^3
    \int_0^\infty \!\! dt \int_0^\infty \!\! d\tau\ 
    \tilde{f}^{(4c)}(\hat{t},\hat{\tau})\,
    \Delta^{\mathrm{fvc}}\!\left(G^{a}(t)\,G^{b}(\tau)\right),
\end{equation}
the finite-volume correction to the product of correlators decomposes as
\begin{equation}
    \Delta^{\mathrm{fvc}}\big(G^{a}G^{b}\big)
    =
    \Delta^{\mathrm{fvc}}\!\big(G^{a}\big)G^{b}
    + 
    G^{a}\,\Delta^{\mathrm{fvc}}\!\big(G^{b}\big)
    + 
    \Delta^{\mathrm{fvc}}\!\big(G^{a}\big)\,
      \Delta^{\mathrm{fvc}}\!\big(G^{b}\big)\, .
\end{equation}
This structure is characteristic of bilinear observables in the TMR and must be handled consistently when applying HP\&MLL corrections. In practice, the last term is numerically small, but we retain it for completeness to ensure a controlled systematic treatment of the full NLOc contribution.

\section{Results}
\label{sec:results}

In this section, we present the results of our analysis. In Sec.~\ref{sec:isoQCD_result}, we begin with the determination of the isospin-symmetric QCD (isoQCD) contributions for diagram sets NLOa and NLOb. We first discuss the estimates obtained for each window and then provide the final numerical results for $a_\mu^{\mathrm{hvp},\,(i)}$. The isoQCD results for diagram NLOc are presented in Sec.~\ref{sec:isoQCD_results_NLOc}. In Sec.~\ref{sec:isobreak}, we explain how the electromagnetic and strong isospin-breaking corrections are determined for each diagram and how they are incorporated into the isoQCD results of Secs.~\ref{sec:isoQCD_result} and~\ref{sec:isoQCD_results_NLOc}. In Sec.~\ref{sec:smallcontr}, we address additional subleading contributions to the HVP. Finally, in Sec.~\ref{sec:finalresult}, we present the complete NLO HVP result.

\subsection{NLOa and NLOb results in isoQCD}
\label{sec:isoQCD_result}

In this section, we present the results obtained within the isoQCD framework for NLOa and NLOb, as well as for their combination NLOa\&b. In Secs.~\ref{sec:SD_result},~\ref{sec:ID_result}, and~\ref{sec:LD_result}, we show the partial results for the SD, ID, and LD windows respectively, and we then combine them in Sec.~\ref{sec:NLOa_NLOb_result}.

\subsubsection{Short-distance window}
\label{sec:SD_result}

The main challenge in the SD window arises from the continuum extrapolation, since most higher-order and logarithmically enhanced discretization effects are concentrated in the small Euclidean-time region. As described above, we perform a tree-level improvement of the isovector data together with a subtraction of the leading time-kernel behavior at small Euclidean times.

\begin{figure}[t]
    \centering
    \begin{subfigure}[b]{0.49\textwidth}
        \centering
        \includegraphics[width=\textwidth]{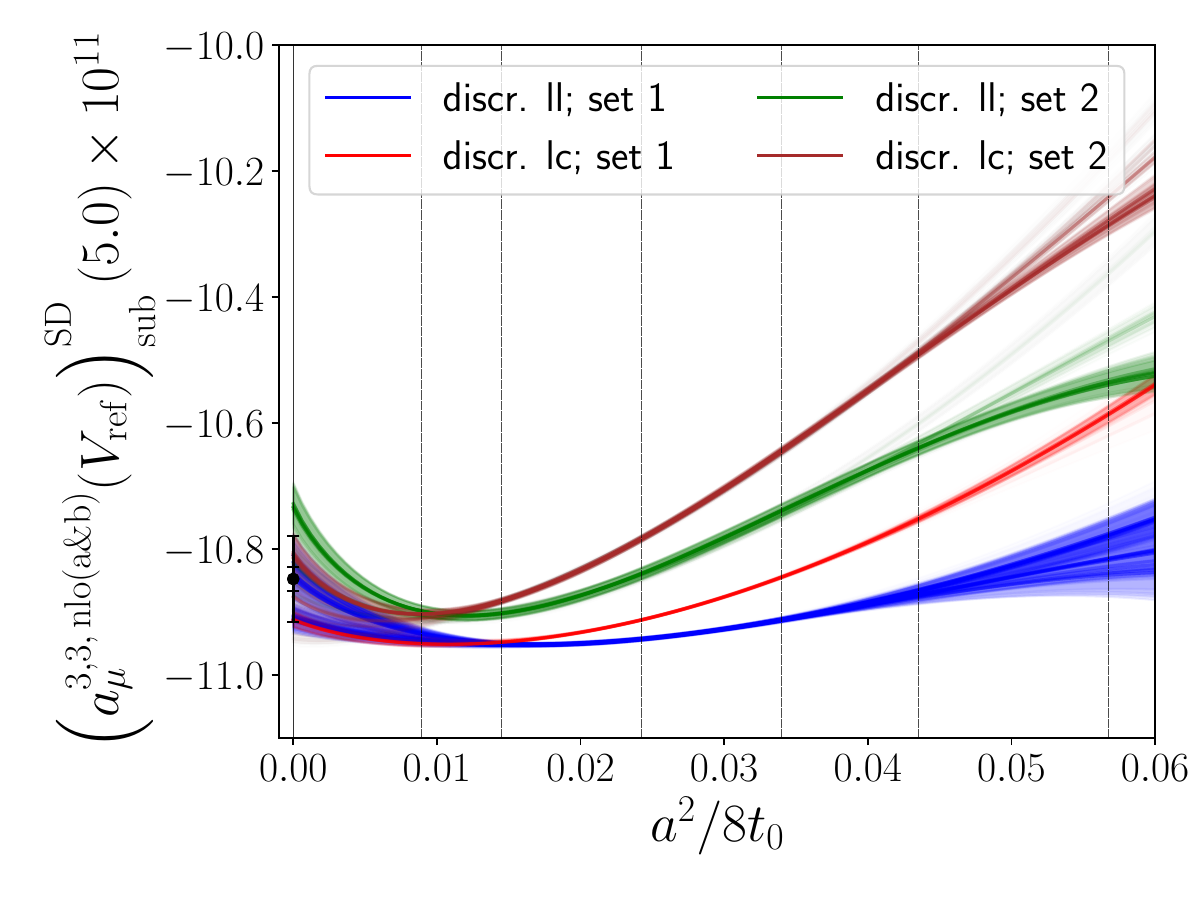}
        \caption{}
        \label{fig:SDsub_33_extr}
    \end{subfigure}
    \hfill
    \begin{subfigure}[b]{0.49\textwidth}
        \centering
        \includegraphics[width=\textwidth]{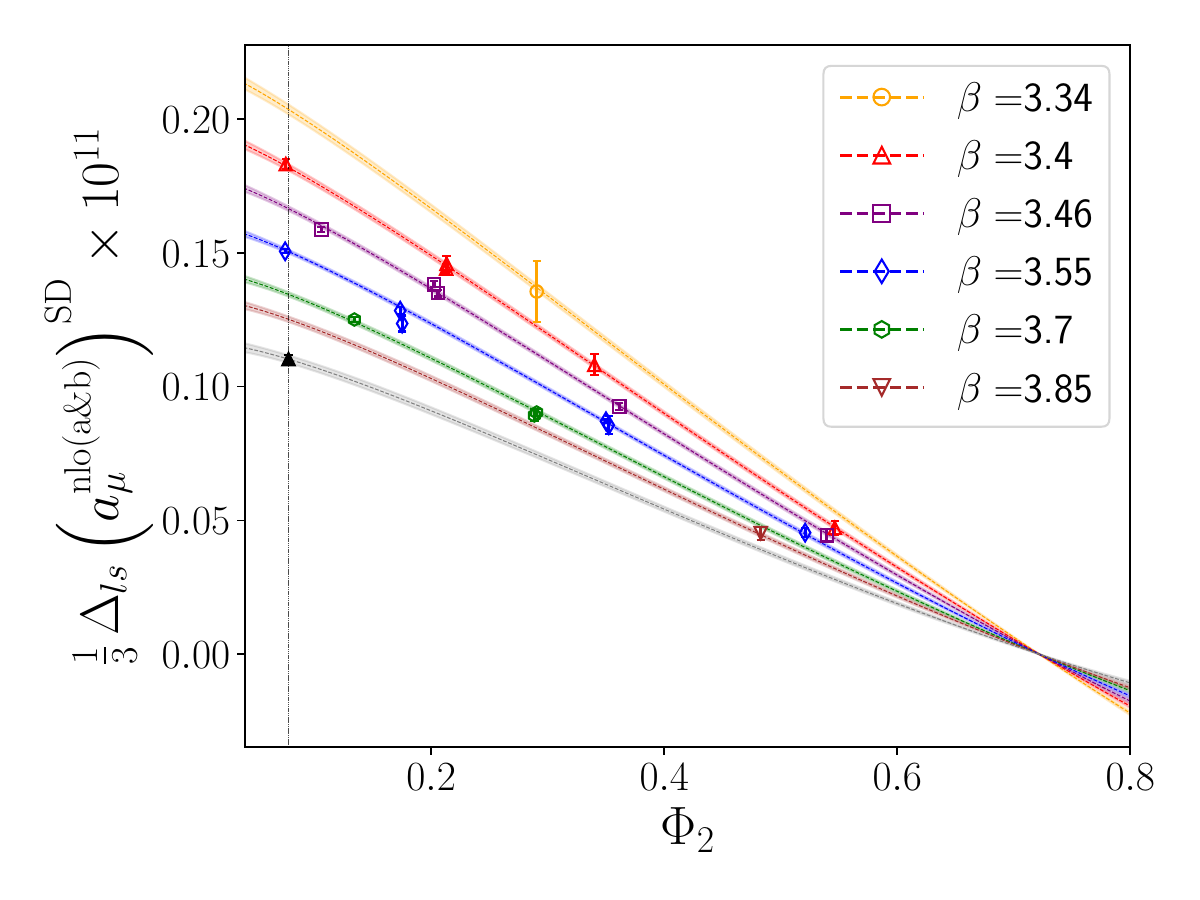}
        \caption{}
        \label{fig:SDsub_dells_extr}
    \end{subfigure}
    \caption{On the left-hand panel we show the continuum extrapolation of the SD window for the isovector channel of the combined NLOa\&b contribution. Each line and band corresponds to a fit, with its opacity proportional to the AIC weight. The leftmost dark vertical line indicates the continuum limit, while the six vertical dotted lines correspond to the lattice spacings available in the CLS ensembles. The black point denotes the final estimate. On the right-hand panel we show a representative chiral-continuum extrapolation of $\Delta_{ls}(a_\mu)$ also for NLOa\&b, shown here for the local-conserved (lc) discretization and improvement set~1. The solid vertical line indicates the physical pion mass, and the black point corresponds to the physical-point estimate for this fit.}
    \label{fig:SDsub_extr}
\end{figure}

Figure~\ref{fig:SDsub_33_extr} shows the continuum extrapolation of the isovector $(3,3)$ channel with the subtracted kernel and $Q=5\,\mathrm{GeV}$ for diagram set NLOa\&b. The model average assigns significant weight to extrapolations containing $a^4$ terms.  These terms are typically hard to constrain, which translates into an increase of the spread of the weighted average. This is accounted for in our estimate for the systematic uncertainty. We observe a mild dependence on the pion mass for this quantity. Nevertheless, fit ans\"{a}tze including $a^2\Phi_2$ and $\Phi_2^2$ receive a non-negligible weight by the AIC. We note that the continuum extrapolation of this quantity exhibits significant curvature in the region below the smallest lattice spacing. Although this feature appears in the extrapolation region devoid of lattice data, it reflects the genuine higher-order cutoff effects that the data requires. We have verified that suppressing this curvature by imposing a suppression on higher-order cutoff corrections and thus imposing a hierarchy on the relative relevance of the cutoff corrections yields results within $1\sigma$ of our final estimate, while also producing increased values for the reduced $\chi^2$.

We exploit the residual $\mathrm{SU}{(3)}_{\mathrm{f}}$ symmetry by computing the suppressed difference $\Delta_{ls}(a_\mu)\equiv a_\mu^{8,8}-a_\mu^{3,3}$, which allows for a straightforward determination of the isoscalar contribution from the isovector one. In an analogous manner~\eqref{eq:SDsub_88_ss}, the strange quark channel is extracted from $\Delta^{\mathrm{conn}}_{ls}(a_\mu)$.

A representative global extrapolation of $\Delta_{ls}(a_\mu)$ is shown in Fig.~\ref{fig:SDsub_dells_extr}. In this example, the extrapolation employs the local--conserved discretization. As expected, this quantity vanishes at the $\mathrm{SU}{(3)}_{\mathrm{f}}$-symmetric point. To enforce this behavior, all fit ans\"{a}tze for this observable are multiplied by $\Phi_4-\frac{3}{2}\Phi_2$, which vanishes identically at the $\mathrm{SU}{(3)}_{\mathrm{f}}$-symmetric point.
\begin{figure}[t]
    \centering
    \begin{subfigure}[b]{0.49\textwidth}
        \centering
        \includegraphics[width=\textwidth]{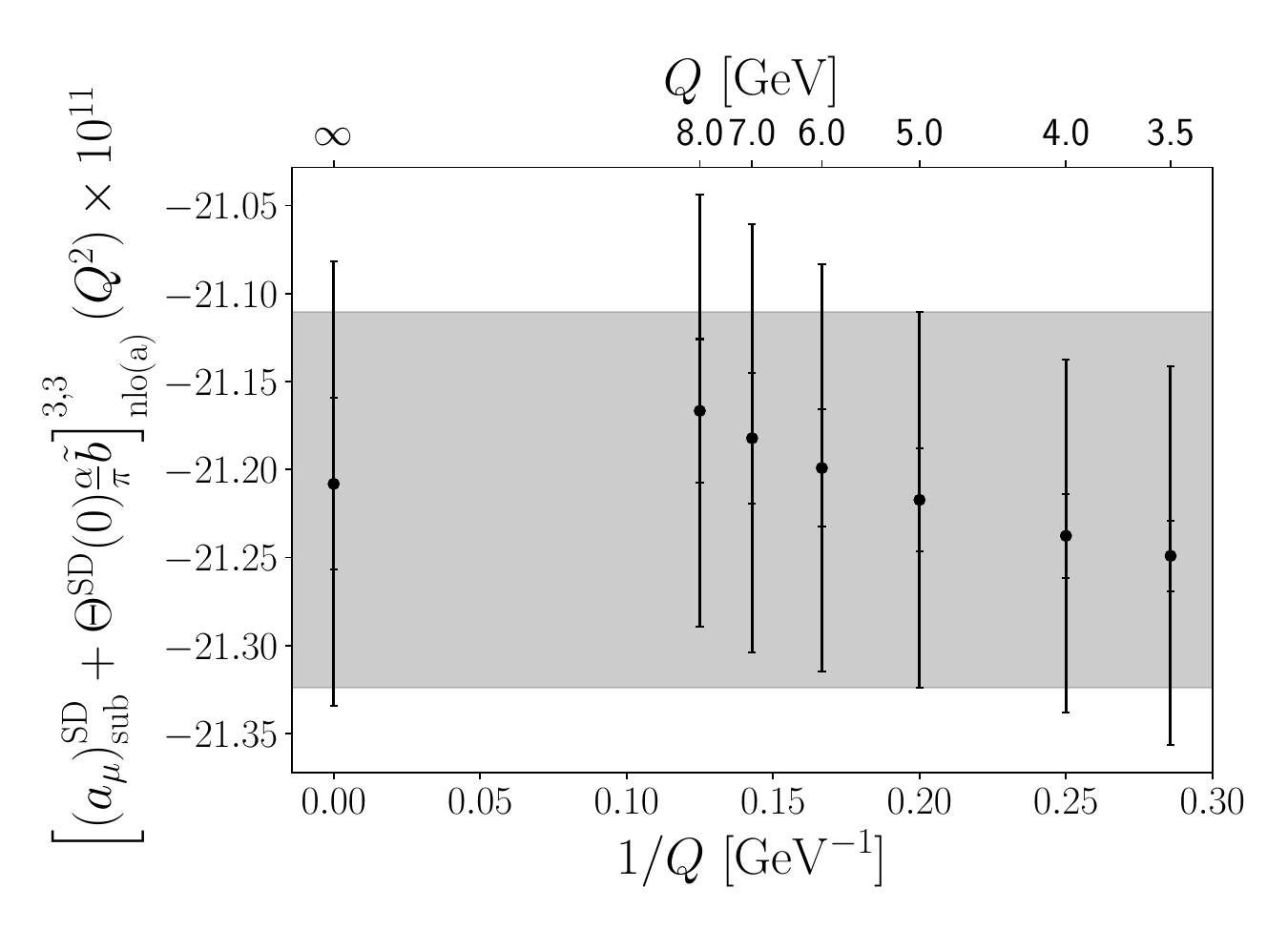}
    \end{subfigure}
    \hfill
    \begin{subfigure}[b]{0.49\textwidth}
        \centering
        \includegraphics[width=\textwidth]{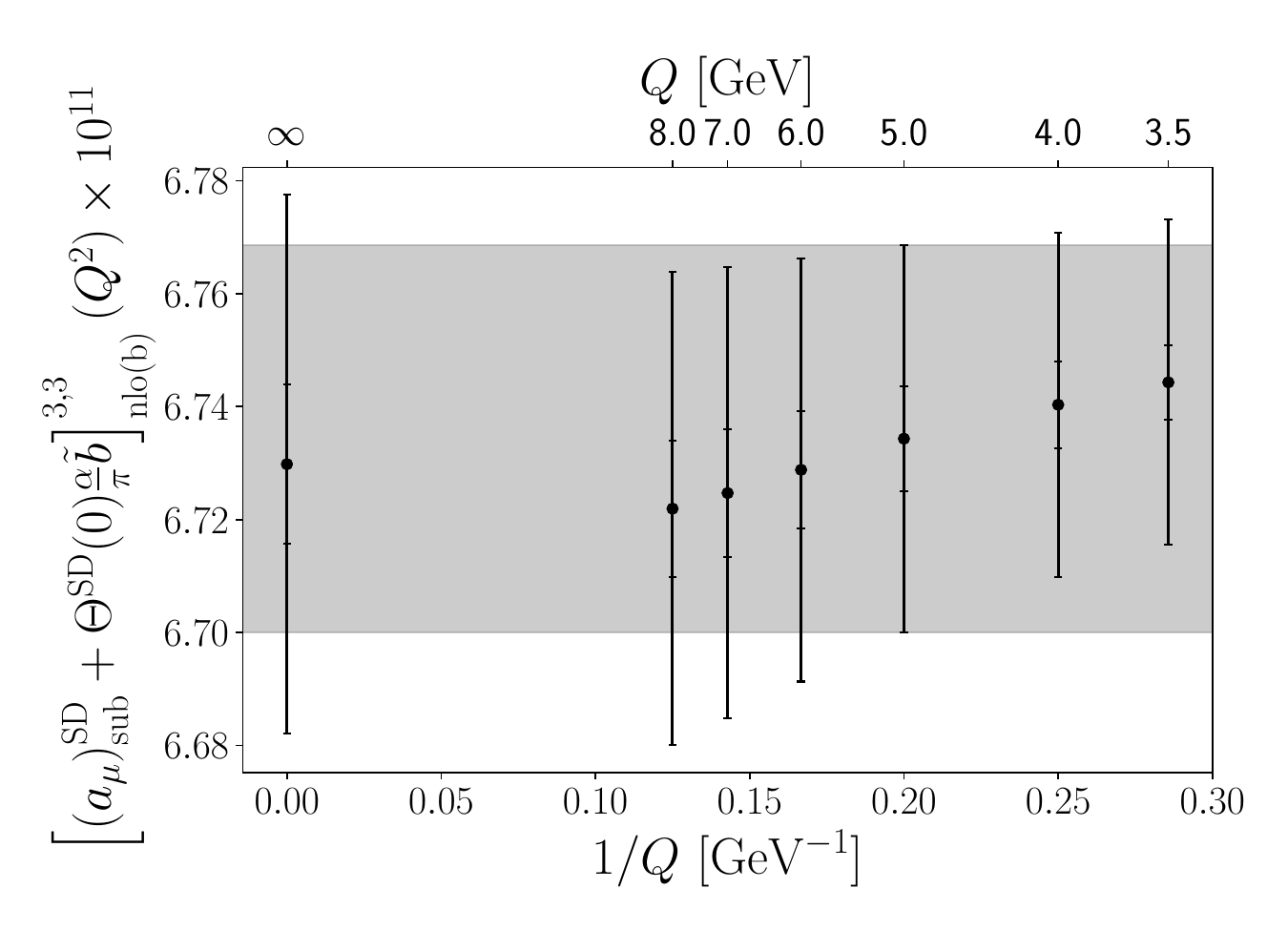}
    \end{subfigure}
        \begin{subfigure}[b]{0.49\textwidth}
        \centering
        \includegraphics[width=\textwidth]{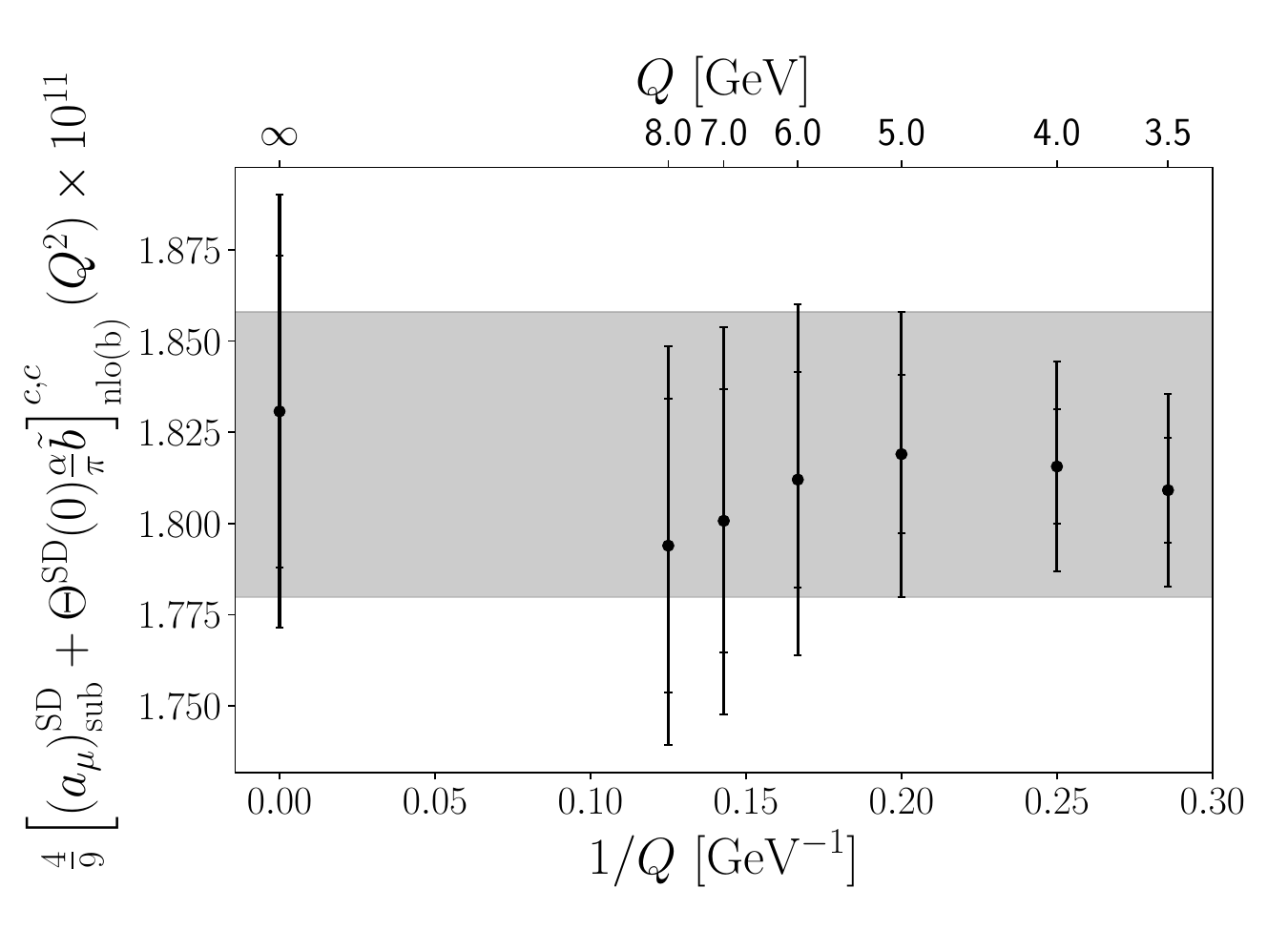}
    \end{subfigure}
    \hfill
    \begin{subfigure}[b]{0.49\textwidth}
        \centering
        \includegraphics[width=\textwidth]{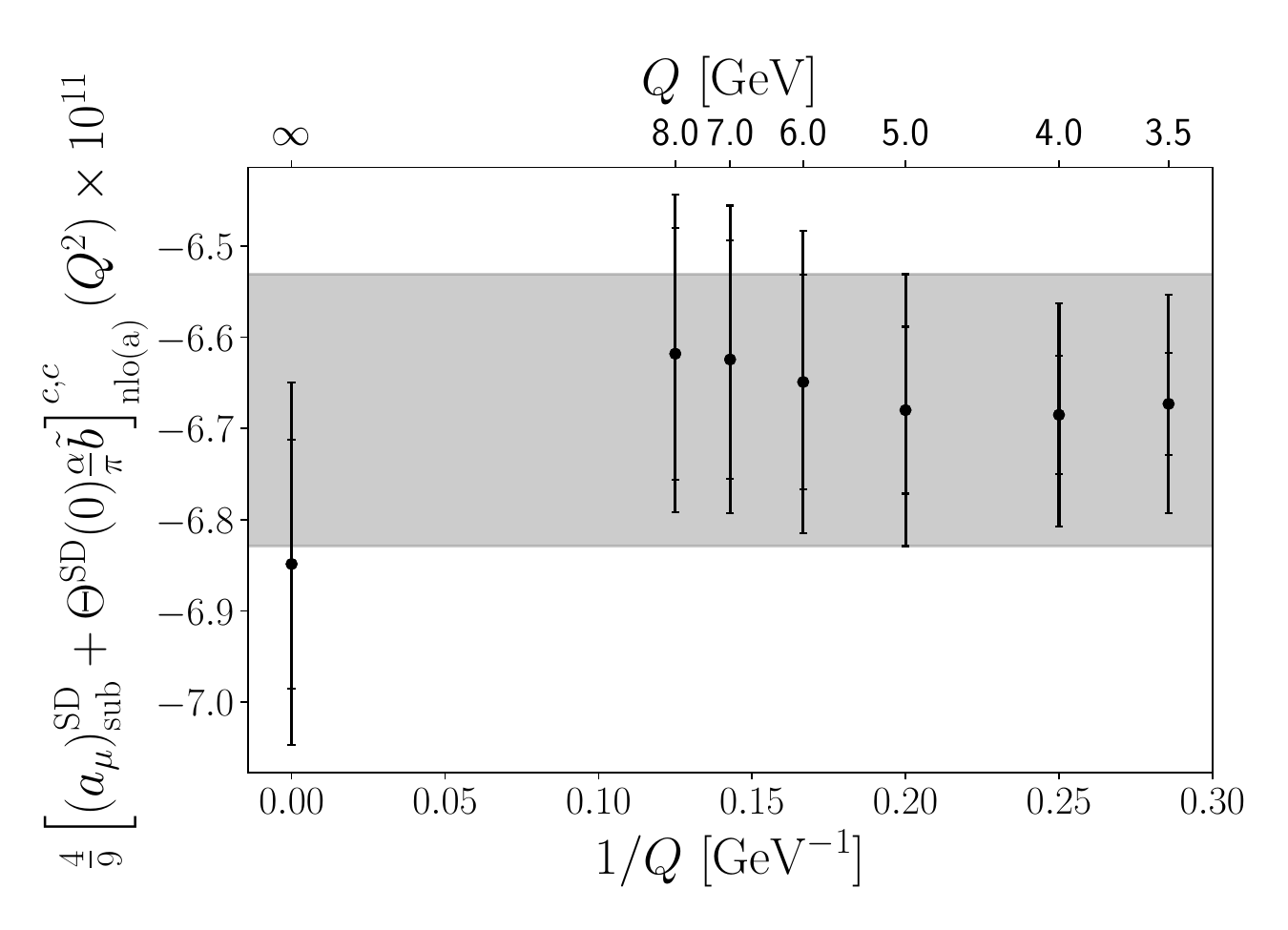}
    \end{subfigure}
    \caption{Stability plot of ${\left(a_\mu\right)}^{\mathrm{SD}}_{\mathrm{sub}}(Q^2)+\Theta_{\mathrm{SD}}(0)\frac{\alpha}{\pi}\tilde{b}(Q^2)$ for all virtualities explored in this work. Results for NLOa and NLOb are shown on the left and right panels, respectively, while the isovector and charm quark channels are displayed in the top and bottom panels. The gray horizontal bands correspond to the final estimates adopted in this work.}
    \label{fig:SDsub_stability}
\end{figure}

In the top panel of Fig.~\ref{fig:SDsub_stability}, we show the stability of ${(a^{3,3}_\mu)}^{\mathrm{SD}}_{\mathrm{sub}}(Q^2)+\Theta_{\mathrm{SD}}(0)\frac{\alpha}{\pi}\tilde{b}^{3,3}(Q^2)$ as a function of the virtuality $Q$. As $1/Q$ increases, the subtracted piece becomes increasingly relevant, until the logarithmically enhanced discretization effects are fully suppressed. For sufficiently large $1/Q$, however, perturbative QCD ceases to be reliable. We therefore choose $Q=5\,\mathrm{GeV}$ as a compromise, where logarithmically enhanced effects are negligible while perturbation theory remains under control.

For the charm contribution, we compute the subtracted piece and determine $\Delta_{lc}(b)\equiv b^{(c,c)}_{\mathrm{conn.}}(Q^2)-2b^{(3,3)}(Q^2)$ non-perturbatively. For all charm-related quantities, we again choose $Q=5\,\mathrm{GeV}$. The corresponding partial results are summarized in Tab.~\ref{tab:isoQCD_SDpre}.
\begin{table}[H]
    \centering
    \small
    \renewcommand{\arraystretch}{1.1}
    \setlength{\tabcolsep}{5pt}
    \begin{tabular}{c | r@{.}l r@{.}l}
        $(i)$ & \multicolumn{2}{c}{${(a_\mu^{3,3})}^{\mathrm{SD}}_{\mathrm{sub}}(Q)$} & \multicolumn{2}{c}{$\frac{4}{9}{(a_\mu^{c,c})}^{\mathrm{SD}}_{\mathrm{sub}}(Q)$} \\
        \noalign{\smallskip}\hline\noalign{\smallskip}
        (4a)    & $-$16&327(28)(99) & $-$3&567(83)(107) \\
        (4b)    & 5&474(8)(32) & 1&068(19)(30) \\
        (4a\&b) & $-$10&853(19)(66) & $-$2&497(63)(72) \\
    \end{tabular}
    \break\break\break
    \begin{tabular}{c | r@{.}l r@{.}l r@{.}l}
        $(i)$ & \multicolumn{2}{c}{$\frac{1}{3}\Delta_{ls}{(a_\mu)}^{\mathrm{SD}}$} & \multicolumn{2}{c}{$\frac{1}{3}\Delta_{ls}^{\mathrm{conn}}{(a_\mu)}^{\mathrm{SD}}$} & \multicolumn{2}{c}{$\frac{4}{9}\Delta_{lc}(b)(Q)$}  \\
        \noalign{\smallskip}\hline\noalign{\smallskip}
        (4a)    & 0&1906(40)(72) & 0&1911(30)(58) & 1&261(15)(45) \\
        (4b)    & $-$0&0756(14)(24) & $-$0&0755(11)(19) & $-$0&378(3)(11) \\
        (4a\&b) & 0&1149(26)(45) & 0&1156(19)(39) & 0&883(12)(34) \\
    \end{tabular}
    \caption{Partial results for the SD subtracted quantities. All results are given in units of $10^{-11}$, and $Q=5.0\, \mathrm{GeV}$.}
    \label{tab:isoQCD_SDpre}
\end{table}
\noindent Finally, all results presented in Tab.~\ref{tab:isoQCD_SDpre} are combined according to Eqs.~\eqref{eq:subtraction_comb},~\eqref{eq:SDsub_88_ss},~\eqref{eq:SDsub_cc}, and~\eqref{eq:SDsub_cc_deltalc}, together with the values listed in Tab.~\ref{tab:sub_piece}, to obtain the final SD-window contributions. The resulting isospin and flavor decomposition\footnote{Note that the results for  ${(a_\mu^{\mathrm{disc}})}^{\mathrm{SD}}$ collected in Tab.~\ref{tab:isoQCD_SD} do not include charm quark-disconnected contributions. These are studied later in Sec.~\ref{sec:disccharm}.} is shown in Tab.~\ref{tab:isoQCD_SD}.
\begin{table}[H]
    \centering
    \small
    \renewcommand{\arraystretch}{1.1}
    \setlength{\tabcolsep}{5pt}
    \begin{tabular}{c | r@{.}l r@{.}l r@{.}l | r@{.}l r@{.}l}
        $(i)$ & \multicolumn{2}{c}{${(a_\mu^{3,3})}^{\mathrm{SD}}$} & \multicolumn{2}{c}{$\frac{1}{3}{(a_\mu^{8,8})}^{\mathrm{SD}}$} & \multicolumn{2}{c|}{$\frac{4}{9}{(a_\mu^{c,c})}^{\mathrm{SD}}$} & \multicolumn{2}{c}{$\frac{1}{9}{(a_\mu^{s,s})}^{\mathrm{SD}}$} & \multicolumn{2}{c}{${(a_\mu^{\mathrm{disc}})}^{\mathrm{SD}}$} \\
        \noalign{\smallskip}\hline\noalign{\smallskip}
        (4a)    & $-$21&225(28)(102) & $-$6&885(8)(35) & $-$6&667(91)(118)  & $-$4&526(5)(23)  & $-$0&001(2)(43) \\
        (4b)    & 6&737(8)(32) & 2&170(2)(11) & 1&815(22)(32)  & 1&422(2)(7)  & $-$0&000(0)(14) \\
        (4a\&b) & $-$14&488(19)(70) & $-$4&715(5)(24) & $-$4&850(69)(82)  & $-$3&104(3)(16)  & $-$0&001(1)(30) \\
    \end{tabular}
    \caption{Isospin and flavor decomposition of the SD window for diagram sets NLOa, NLOb and their combination. The light contribution can be obtained from the isovector channel via $(5/9)\, a_\mu^{l,l} = 10/9\, a_\mu^{3,3}$.  All results are given in units of $10^{-11}$.}
    \label{tab:isoQCD_SD}
\end{table}

\subsubsection{Intermediate-distance window}
\label{sec:ID_result}

\begin{figure}[t]
    \centering
    \begin{subfigure}[b]{0.49\textwidth}
        \centering
        \includegraphics[width=\textwidth]{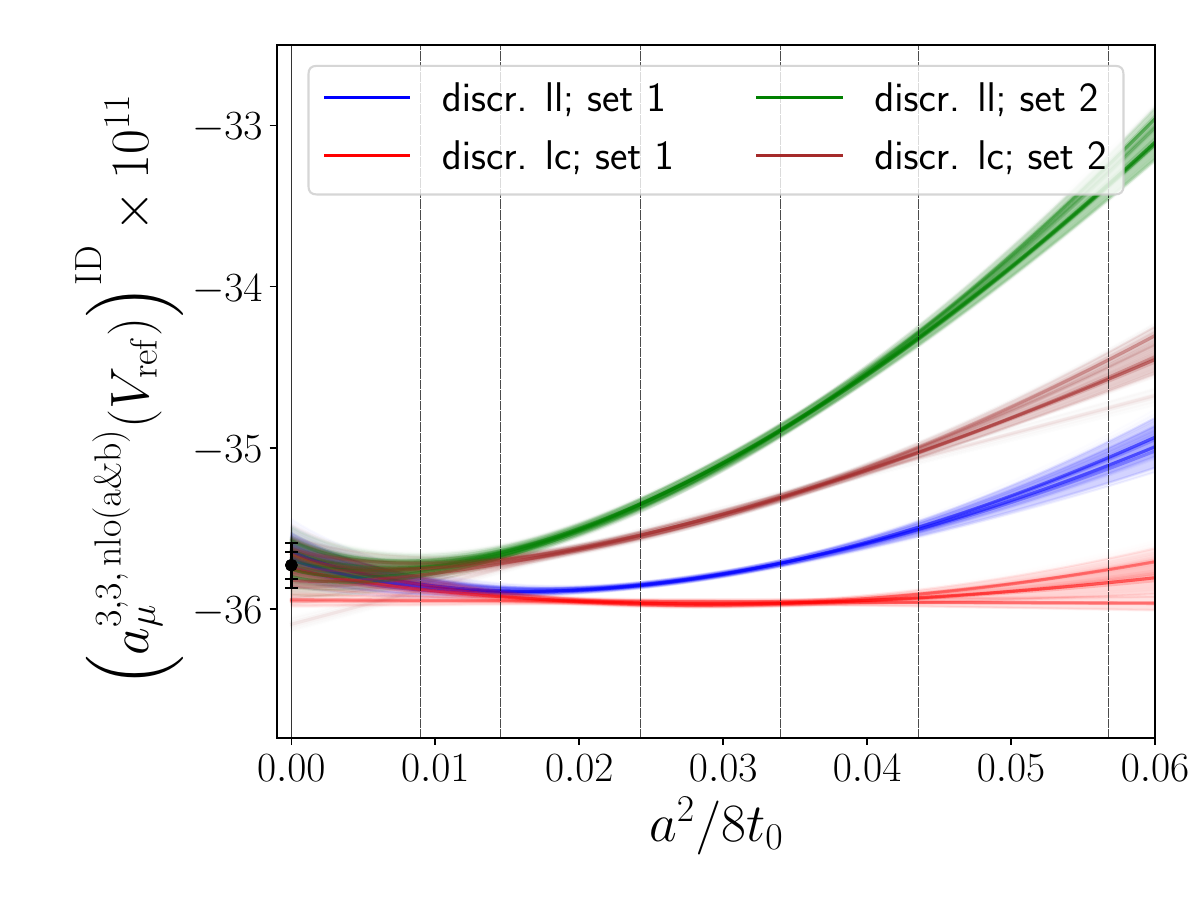}
        \caption{}
        \label{fig:ID_cont}
    \end{subfigure}
    \hfill
    \begin{subfigure}[b]{0.49\textwidth}
        \centering
        \includegraphics[width=\textwidth]{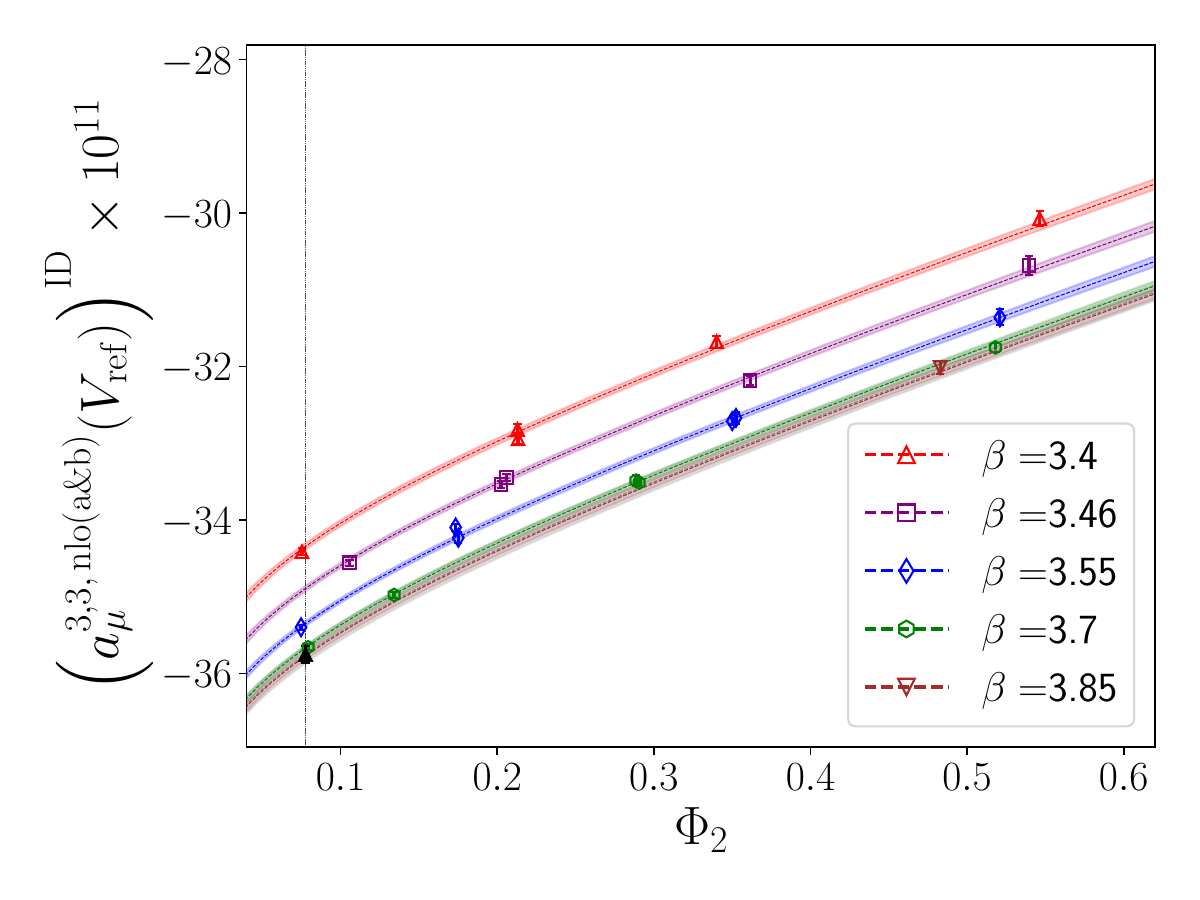}
        \caption{}
        \label{fig:ID_chircont}
    \end{subfigure}
    \caption{Similar to Fig.~\ref{fig:SDsub_extr} but for the ID window. In the left panel, we show the continuum extrapolation of the isovector contribution for the combined NLOa\&b. In the right panel, a representative chiral--continuum extrapolation for the same component is shown. The extrapolation corresponds to the best fit according to the model average for the local-local (ll) discretization and improvement set~2, for which $\chi^2/\chi^2_{\mathrm{exp}}\approx1.3$.}
  \label{fig:ID_extrap}
\end{figure}

The extrapolation for the ID window remains sensitive to higher-order cutoff effects. Fits including $a^3$ terms are favored by the model average, although, in some cases, fits with only leading $a^2$ discretization effects also receive non-negligible weight. This translates into a slight increase in the spread of the weighted average which is reflected by an increase of the systematic uncertainty (See Fig.~\ref{fig:ID_cont}). With respect to the chiral behavior of the isovector channel, we observe an increased sensitivity to higher-order effects, as can be seen in Fig.~\ref{fig:ID_chircont}. The final decomposition into isospin and flavor channels for this window is listed in Tab.~\ref{tab:isoQCD_ID}.
\begin{table}[H]
    \centering
    \small
    \renewcommand{\arraystretch}{1.1}
    \setlength{\tabcolsep}{5pt}    
    \begin{tabular}{c | r@{.}l r@{.}l r@{.}l | r@{.}l r@{.}l}
        $(i)$ & \multicolumn{2}{c}{${(a_\mu^{3,3})}^{\mathrm{ID}}$} & \multicolumn{2}{c}{$\frac{1}{3}{(a_\mu^{8,8})}^{\mathrm{ID}}$} & \multicolumn{2}{c|}{$\frac{4}{9}{(a_\mu^{c,c})}^{\mathrm{ID}}$} & \multicolumn{2}{c}{$\frac{1}{9}{(a_\mu^{s,s})}^{\mathrm{ID}}$} & \multicolumn{2}{c}{${(a_\mu^{\mathrm{disc}})}^{\mathrm{ID}}$} \\
        \noalign{\smallskip}\hline\noalign{\smallskip}
        (4a)    & $-$64&91(16)(21) & $-$16&606(48)(55) & $-$1&311(21)(23)  & $-$9&809(19)(25)  & 0&415(34)(65) \\
        (4b)    & 29&11(7)(10) & 7&306(23)(25) & 0&460(7)(9)  & 4&275(8)(11)  & $-$0&203(16)(30) \\
        (4a\&b) & $-$35&80(8)(11) & $-$9&300(25)(30) & $-$0&851(13)(14)  & $-$5&534(10)(14)  & 0&212(18)(36) \\
    \end{tabular}
    \caption{Same as Tab.~\ref{tab:isoQCD_SD} but for the ID window.}
    \label{tab:isoQCD_ID}
\end{table}

\subsubsection{Long-distance window}
\label{sec:LD_result}

\begin{figure}[t]
    \centering
    \begin{subfigure}[b]{0.49\textwidth}
        \centering
        \includegraphics[width=\textwidth]{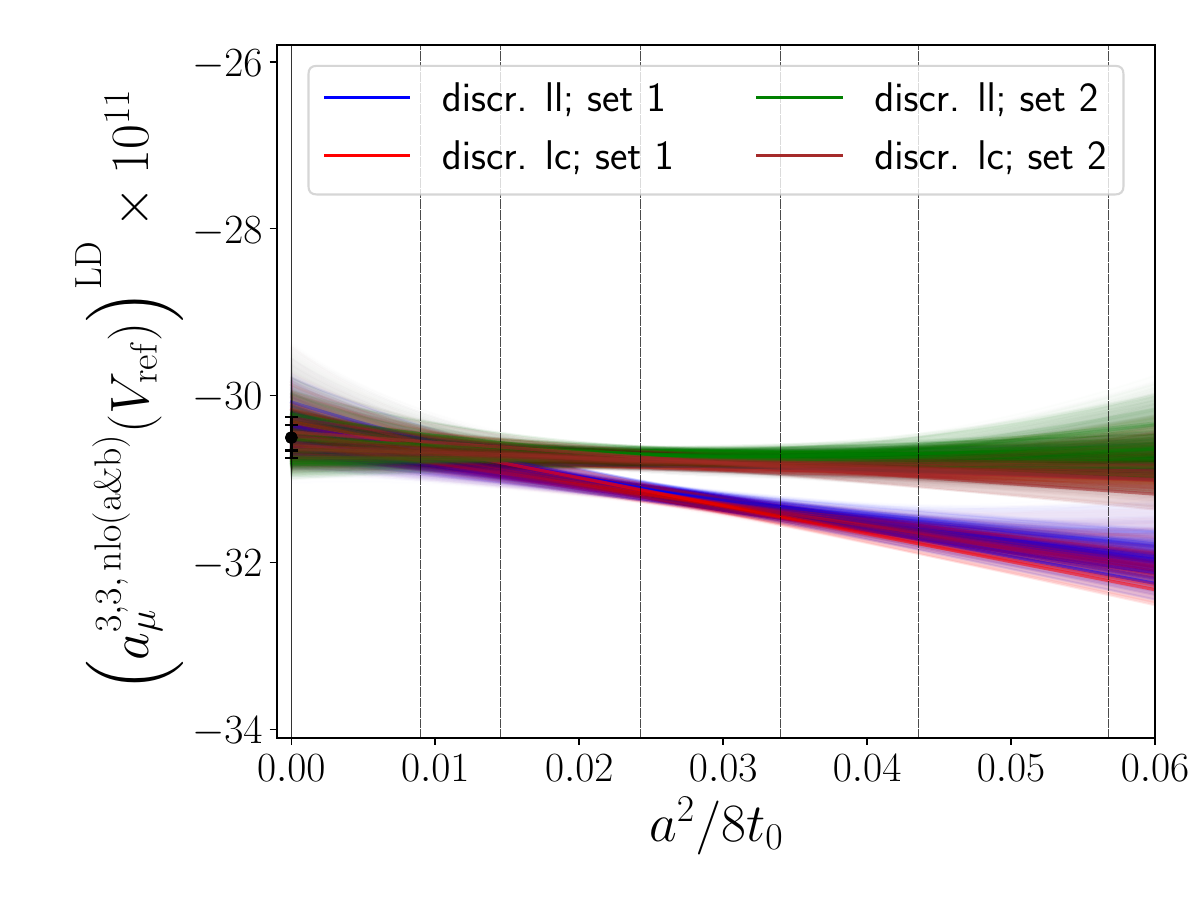}
        \caption{}
        \label{fig:LD_cont}
    \end{subfigure}
    \hfill
    \begin{subfigure}[b]{0.49\textwidth}
        \centering
        \includegraphics[width=\textwidth]{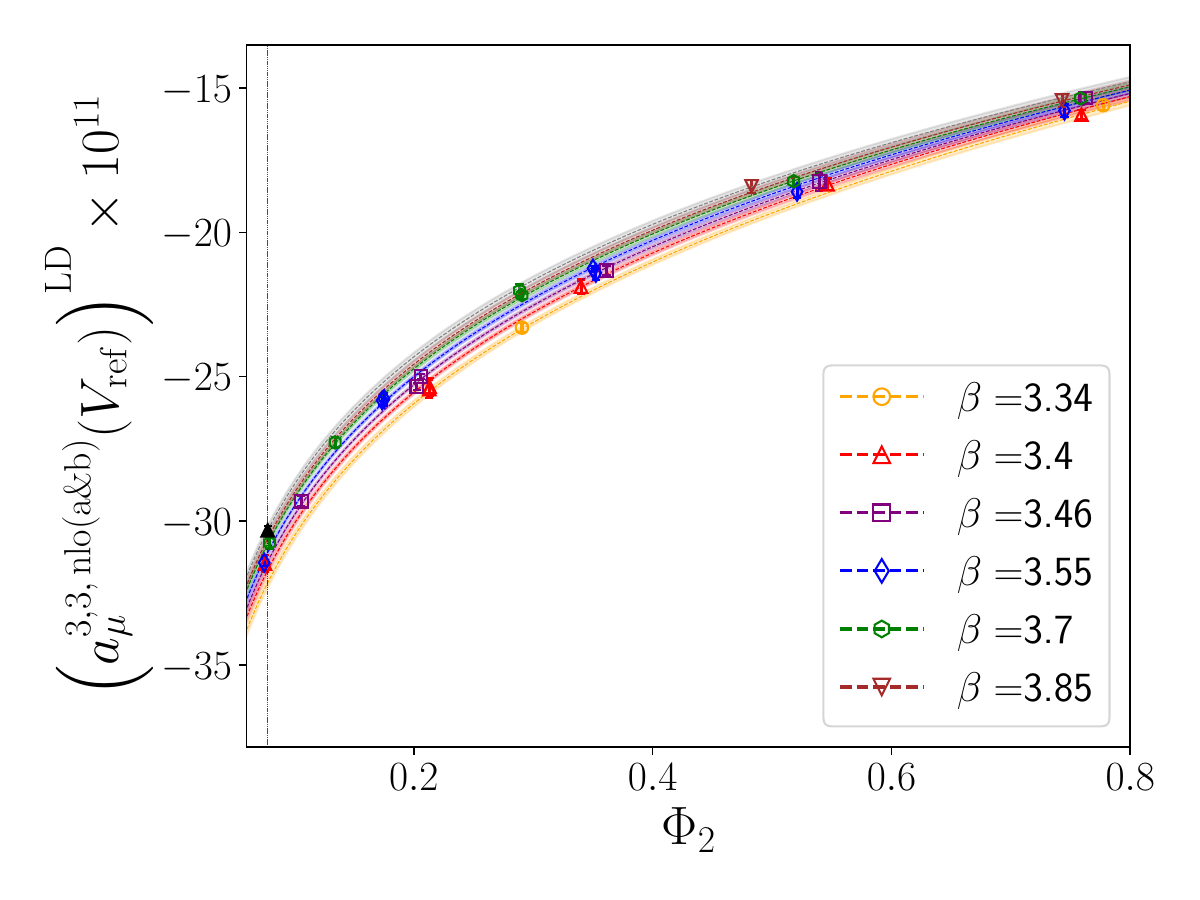}
        \caption{}
        \label{fig:LD_chircont}
    \end{subfigure}
    \caption{Same as Fig.~\ref{fig:ID_extrap} but for the LD window. The best fit example in the right-hand panel corresponds to the local-local (ll) discretization and improvement set~1, for which we find $\chi^2/\chi^2_{\mathrm{exp}}\approx0.8$.}
  \label{fig:LD_extrap}
\end{figure}

As described in previous sections, the LD window---particularly in the isovector channel---dominates the overall uncertainty of our estimates. It is therefore crucial to maintain good control over the extrapolation of this contribution. As expected, we observe a very strong chiral dependence (see Fig.~\ref{fig:LD_chircont}); in particular, terms proportional to $\ln \Phi_2$ dominate in the model average. By contrast, discretization effects are comparatively mild (see Fig.~\ref{fig:LD_cont}). Nevertheless, we allow for higher-order lattice-spacing terms in the fits, as we find that fits containing terms scaling as $a^3$ receive a non-negligible weight.

In Tab.~\ref{tab:isoQCD_LD}, we present the partial results for the LD window in the isoQCD world, obtained after unblinding according to Eq.~\eqref{eq:blinding}.
\begin{table}[H]
    \centering
    \small
    \renewcommand{\arraystretch}{1.1}
    \setlength{\tabcolsep}{5pt}
    \begin{tabular}{c | r@{.}l r@{.}l r@{.}l | r@{.}l r@{.}l}
        $(i)$ & \multicolumn{2}{c}{${(a_\mu^{3,3})}^{\mathrm{LD}}$} & \multicolumn{2}{c}{${\frac{1}{3}(a_\mu^{8,8})}^{\mathrm{LD}}$} & \multicolumn{2}{c|}{${\frac{4}{9}(a_\mu^{c,c})}^{\mathrm{LD}}$} & \multicolumn{2}{c}{${\frac{1}{9}(a_\mu^{s,s})}^{\mathrm{LD}}$} & \multicolumn{2}{c}{${(a_\mu^{\mathrm{disc}})}^{\mathrm{LD}}$} \\
        \noalign{\smallskip}\hline\noalign{\smallskip}
        (4a)    & $-$88&81(58)(70) & $-$10&86(29)(44) & $-$0&00465(12)(10)  & $-$4&616(44)(65)  & 3&62(27)(45) \\
        (4b)    & 57&81(42)(56) & 6&60(21)(28) & 0&002129(57)(49)  & 2&695(35)(45)  & $-$2&52(20)(29) \\
        (4a\&b) & $-$30&98(15)(20) & $-$4&30(8)(12) & $-$0&002525(63)(55)  & $-$1&906(13)(21)  & 1&05(7)(13) \\
    \end{tabular}
    \caption{Same as Tab.~\ref{tab:isoQCD_SD} but for the LD window.}
    \label{tab:isoQCD_LD}
\end{table}

\subsubsection{NLOa \& NLOb isoQCD results}
\label{sec:NLOa_NLOb_result}

We summarize the individual results from Secs.~\ref{sec:SD_result},~\ref{sec:ID_result}, and~\ref{sec:LD_result} in Tab.~\ref{tab:isoQCD_result}. We break down the results into the different time windows, as well as into the isospin and flavor components. All systematic uncertainties are combined in quadrature. We expect correlations between different systematic effects to be negligible, as evidenced by the distinct extrapolation behaviors observed for different windows and channels. Each window and channel exhibits different sensitivities to lattice-spacing and chiral extrapolation effects, supporting the assumption of independence.
\begin{table}[H]
    \centering
    \small
    \renewcommand{\arraystretch}{1.1}
    \setlength{\tabcolsep}{5pt}
    \begin{tabular}{c | r@{.}l r@{.}l r@{.}l}
        $(i)$ & \multicolumn{2}{c}{${(a_\mu^{\mathrm{hvp}})}^{\mathrm{SD}}$} & \multicolumn{2}{c}{${(a_\mu^{\mathrm{hvp}})}^{\mathrm{ID}}$} & \multicolumn{2}{c}{${(a_\mu^{\mathrm{hvp}})}^{\mathrm{LD}}$} \\
        \noalign{\smallskip}\hline\noalign{\smallskip}
        (4a)    & $-$34&777(98)(160) & $-$82&83(20)(22) & $-$99&68(71)(83) \\
        (4b)    & 10&723(24)(47) & 36&88(9)(11) & 64&42(51)(63) \\
        (4a\&b) & $-$24&053(73)(110) & $-$45&95(11)(12) & $-$35&28(20)(23) \\
    \end{tabular}
    \break\break\break
    \begin{tabular}{c | r@{.}l r@{.}l r@{.}l | r@{.}l r@{.}l}
        $(i)$ & \multicolumn{2}{c}{$a_\mu^{3,3}$} & \multicolumn{2}{c}{$\frac{1}{3}a_\mu^{8,8}$} & \multicolumn{2}{c|}{$\frac{4}{9}a_\mu^{c,c}$} & \multicolumn{2}{c}{$\frac{1}{9}a_\mu^{s,s}$} & \multicolumn{2}{c}{$a_\mu^{\mathrm{disc}}$}  \\
        \noalign{\smallskip}\hline\noalign{\smallskip}
        (4a)    & $-$174&95(66)(74) & $-$34&35(30)(44) & $-$7&983(95)(119) & $-$18&951(53)(74)  & 4&04(28)(46) \\
        (4b)    & 93&66(46)(57) & 16&08(21)(29) & 2&277(24)(34) & 8&392(38)(47)  & $-$2&72(20)(30) \\
        (4a\&b) & $-$81&27(21)(24) & $-$18&31(9)(13) & $-$5&703(71)(83) & $-$10&544(21)(30)  & 1&26(8)(14) \\
    \end{tabular}
    \caption{Final isoQCD estimates for diagram sets NLOa, NLOb, and their combination. We decompose the results into the different time windows, as well as into the isospin and flavor components. The light quark contribution can be obtained from the isovector channel via $(5/9)\, a_\mu^{l,l} = (10/9)\, a_\mu^{3,3}$. All results are given in units of $10^{-11}$. }
    \label{tab:isoQCD_result}
\end{table}
\noindent In Eq.~\eqref{eq:NLOab_result} we quote the final isoQCD results for each diagram set.
\begin{equation}
    \begin{aligned}
        {\left(a_\mu^{\mathrm{hvp,\,nlo(a)}}\right)}^{\mathrm{isoQCD}}\, &=\, -217.28(81)(87)(31)[1.23]\, \times\, 10^{-11}\, , \\
        {\left(a_\mu^{\mathrm{hvp,\,nlo(b)}}\right)}^{\mathrm{isoQCD}}\, &=\, \phantom{-}112.02(55)(64)(25)[88]\, \times\, 10^{-11}\, , \\
        {\left(a_\mu^{\mathrm{hvp,\,nlo(a\&b)}}\right)}^{\mathrm{isoQCD}}\, &=\, -105.29(28)(28)(5)[40]\, \times\, 10^{-11}\, ,
    \end{aligned}
    \label{eq:NLOab_result}
\end{equation}
where the uncertainties in brackets represent the statistical error of the lattice data, the systematic uncertainty of the model average, extracted from the weighted spread of the results and the uncertainty we associate to the FV shift applied on the continuum respectively. In square brackets, we quote the total quadratic combination of these uncertainties. So far, we do not quote any uncertainty coming from the physical value of our scale-setting quantity ($t_0^{\mathrm{ph}}$), although we include it in our final estimate (see Sec.~\ref{sec:finalresult}).

We emphasize that the chiral-continuum extrapolations for NLOa, NLOb, and NLOa\&b are performed independently, with only the combined NLOa\&b result entering the final NLO HVP determination. The agreement between the independently extrapolated NLOa\&b central value and the sum of the separately extrapolated NLOa and NLOb central values—evident in Tables~\ref{tab:isoQCD_SDpre} to~\ref{tab:isoQCD_LD} and Eq.~\eqref{eq:NLOab_result}—provides a non-trivial consistency check of our analysis.

\begin{figure}[t]
    \centering
    \includegraphics[width=0.625\textwidth]{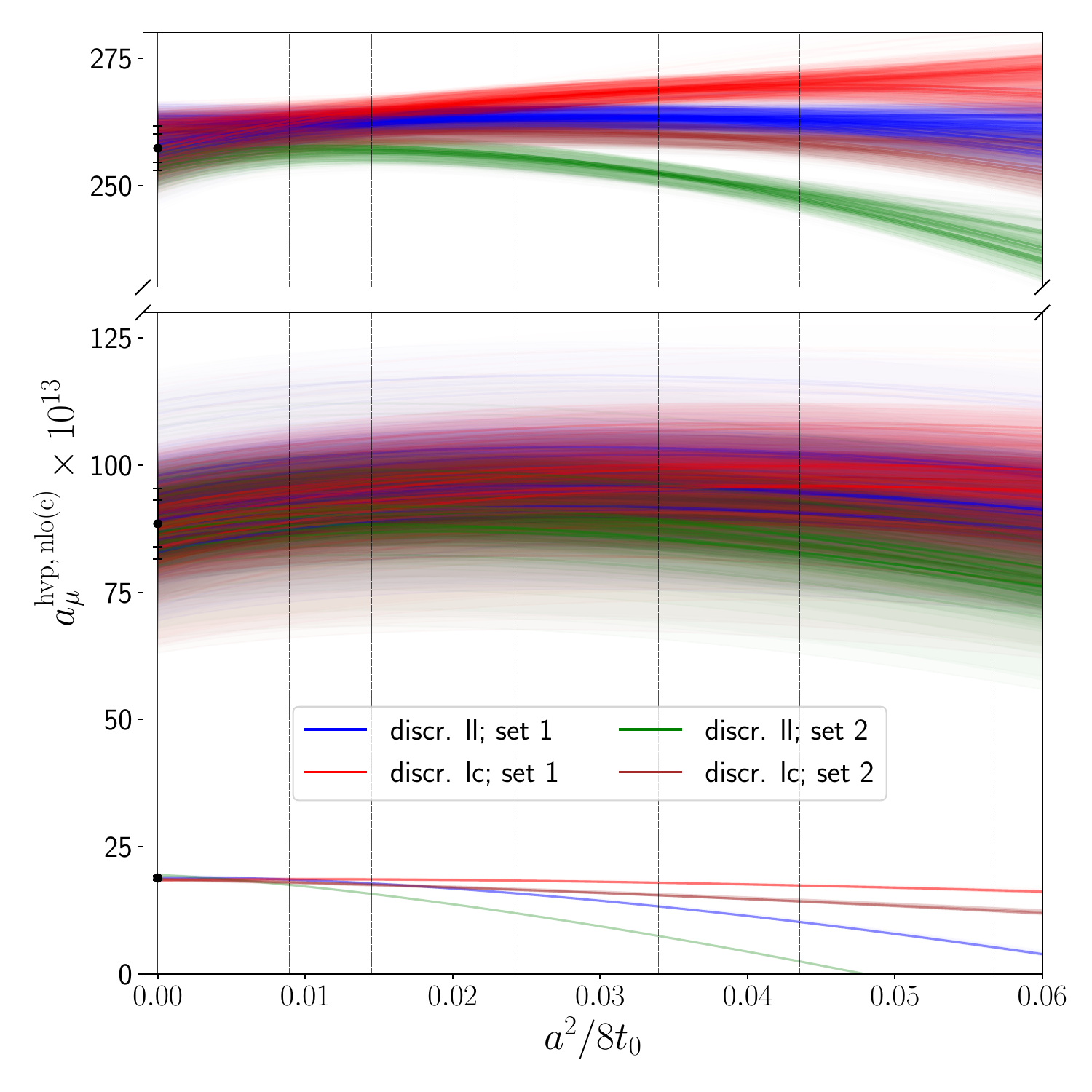}
    \caption{Continuum extrapolation of the three most contributing pieces to our NLOc estimate. From top to bottom, we show the extrapolations for $a_\mu^{3,3-3,3}$, $(2/3)\, a_\mu^{3,3-8,8}$, and $(8/9)\, a_\mu^{3,3-c,c}$.}
  \label{fig:NLOc_extrap}
\end{figure}

\subsection{NLOc results in isoQCD}
\label{sec:isoQCD_results_NLOc}

For diagram NLOc, the final estimate is obtained without any time-window splitting. All relevant contributions, together with their corresponding charge factors, are summarized in Tab.~\ref{tab:isoQCD_NLOc}. The result is dominated by the isovector channel, with $a_\mu^{3,3-3,3}$ providing the largest individual contribution. The total uncertainty, however, is dominated by the isovector-isoscalar crossed term $a_\mu^{3,3-8,8}$, where the convolution of the noisy isoscalar channel with the isovector correlator limits our precision to 8\% for this contribution. Nevertheless, after combining all contributions, we achieve a 2.4\% precision for the final isoQCD estimate~\eqref{eq:NLOc_result}, which is sufficient for the final NLO HVP determination. Figure~\ref{fig:NLOc_extrap} shows the continuum extrapolation with the three most relevant contributions displayed separately.
\begin{table}[H]
    \centering
    \small
    \renewcommand{\arraystretch}{1.1}
    \begin{tabular}{c c c c c c}
        $a_\mu^{3,3-3,3}$ & $\frac{2}{3}a_\mu^{3,3-8,8}$ & $\frac{8}{9}a_\mu^{3,3-c,c}$ & $\frac{1}{9}a_\mu^{8,8-8,8}$ & $\frac{8}{27}a_\mu^{8,8-c,c}$ & $\frac{16}{81}a_\mu^{c,c-c,c}$ \\
        \noalign{\smallskip}\hline\noalign{\smallskip}
        256.6(2.7)(3.5) & 88.2(4.6)(5.2) & 19.22(26)(38) & 9.23(38)(43) & 3.88(18)(39) & 0.386(8)(12) \\
    \end{tabular}
    \caption{IsoQCD estimates for all NLOc contributions. All values are given in units of $10^{-13}$.}
    \label{tab:isoQCD_NLOc}
\end{table}
\noindent This leads to the final isoQCD estimate for the NLOc diagram:
\begin{equation}
    {\left(a_\mu^{\mathrm{hvp,\,nlo(c)}}\right)}^{\mathrm{isoQCD}}\, =\, 378(6)(6)[9]\, \times\, 10^{-13}\, .
    \label{eq:NLOc_result}
\end{equation}
\subsection{Isospin-breaking effects}
\label{sec:isobreak}

All ensembles used in this computation are based on pure QCD simulations with degenerate light quark masses. In order to relate to the SM and compare our results with data-driven determinations of the NLO diagrams, we must therefore convert our isoQCD results to the physical theory by including isospin-breaking effects. 

In our previous works~\cite{Ce:2022kxy,Kuberski:2024bcj,Djukanovic:2024cmq}, we employed the RM123 method~\cite{Rome123} to compute a subset of these corrections directly in the TMR formulation~\cite{Risch:2019xio,Risch:2021hty} from lattice data. Moreover, to obtain a more accurate result at large distances~\cite{Djukanovic:2024cmq}, we extended these results by adding further contributions computed using the covariant coordinate-space (CCS)~\cite{Meyer:2017hjv,Chao:2022ycy} approach (see Refs.~\cite{Parrino:2025afq,Volodymyr3Erb_SIB}). However, CCS kernels for the NLO diagrams are currently unavailable, and generating the required data lies beyond the scope of this work. For this reason, we adopt an alternative strategy and estimate the required corrections using a spacelike representation.
\begin{equation}
    a_\mu^{\mathrm{hvp},\,(i)} = {\left(\frac{\alpha}{\pi}\right)}^{1+N_i} \int_0^1 dx\, \mathcal{K}^{(i)}(x)\, \left(4\pi\alpha\, \Pi^{(\gamma,\gamma)}(t(x))\right), 
    \qquad 
    t(x) = \frac{x^2 m_\mu^2}{1-x}\, .
    \label{eq:spacelike_def}
\end{equation}
The expression in Eq.~\eqref{eq:spacelike_def} applies to $(i)=(2)$, $(4a)$, and $(4b)$, with
\begin{equation}
    \mathcal{K}^{(i)}(x) = \frac{1}{m_\mu^2}\, t^\prime(x)\, \hat{f}^{(i)}\!\left(\frac{x^2}{x-1}\right)
    = \frac{(x-2)x}{{(x-1)}^2}\, \hat{f}^{(i)}\!\left(\frac{x^2}{x-1}\right)\, .
\end{equation}
Isospin-breaking effects naturally separate into two contributions according to their physical origin. For any HVP observable $\mathcal{O}$ we classify them into electromagnetic ($\Delta^\gamma \mathcal{O}$) and strong ($\Delta^{\delta m}\mathcal{O}$) IB corrections, with the total correction given by their sum,
\begin{equation}
    \Delta^{\mathrm{IB}}\mathcal{O}\, =\, \Delta^\gamma \mathcal{O}\, +\, \Delta^{\delta m}\mathcal{O}\, .
\end{equation}
The HVP scalar function entering Eq.~\eqref{eq:spacelike_def} is approximated using phenomenological models. In this work, we employ vector-meson dominance (VMD) for the electromagnetic effects~\cite{Volodymyr1,Volodymyr2} together with a model for the $(3,8)$ channel~\cite{Volodymyr3Erb_SIB} for the strong-IB effects. We assign conservative uncertainties to account for the limitations of these phenomenological approaches, as discussed below. The resulting IB effects turn out to be particularly small for the NLO contributions, although they are individually relevant for each subdiagram.

The electromagnetic corrections arise from the different electric charges of the up and down quarks. They are dominated by the shift from the neutral to the charged pion mass, with subleading contributions from the $\pi^+\pi^-\gamma$ and $\pi^0\gamma$ channels. We model these effects within the VMD framework following the procedure of Refs.~\cite{Volodymyr1,Volodymyr2}. The resulting correction to the HVP scalar can be written as
\begin{equation}
    \Delta^\gamma \Pi \approx 
    \left( 
        \Pi ^{(\pi^+\pi^-)}_{m_\pi=m_{\pi^\pm}}
        -
        \Pi^{(\pi^+\pi^-)}_{m_\pi=m_{\pi^0}}
    \right)
    + 
    \Pi^{(\pi^+\pi^-\gamma)}
    + 
    \Pi^{(\pi^0\gamma)}\, .
    \label{eq:IB_em}
\end{equation}
The strong IB effects originate from the mass difference between the up and down quarks. These effects are well approximated by the isospin component $(3,8)$ of the vacuum polarization~\cite{Volodymyr3Erb_SIB},
\begin{equation}
    \Delta^{\delta m}\Pi = 2\Pi^{(3,8)} + \mathrm{O}\left(\delta m^2\right)\, .
    \label{eq:IB_strong}
\end{equation}
Equations~\eqref{eq:IB_em} and~\eqref{eq:IB_strong} are inserted into Eq.~\eqref{eq:spacelike_def} in place of $\Pi^{(\gamma,\gamma)}$ in order to compute the IB corrections to the LO, NLOa, and NLOb diagrams, using the appropriate kernel $\mathcal{K}^{(i)}$. For NLOc, the leading correction is given by
\begin{equation}
    \Delta^{\mathrm{IB}}a_\mu^{\mathrm{hvp,\,nlo(c)}} \approx 32\pi\alpha^3\int_0^\infty dx \mathcal{K}^{(2)}(x)\Pi^{(\gamma,\gamma)}(t(x))\Delta^{\mathrm{IB}}\Pi(t(x))\, ,
    \label{eq:NLOc_IB}
\end{equation}
where $\mathcal{K}^{(2)} = 1 - x$. To obtain the HVP scalar function from our lattice data, we employ the Padé approximant of Ref.~\cite{Conigli:2025qvh} to model $\Pi^{(\gamma,\gamma)}$ in Eq.~\eqref{eq:NLOc_IB}. As a consistency check of our modeling, we estimate the total IB correction to the LO HVP contribution as
\begin{equation}
    \Delta^{\mathrm{IB}} a_\mu^{\mathrm{hvp,\,lo}}
    \approx -4.1 \times 10^{-10}\, ,
\end{equation}
which agrees with the lattice determination of Ref.~\cite{Djukanovic:2024cmq}. The consistency between these two estimates supports the validity of our modelling.

In Tab.~\ref{tab:isobreak}, we collect our estimates of the electromagnetic and strong IB contributions to the NLO diagrams. We assign a conservative $50\%$ uncertainty to each contribution and treat them as uncorrelated when forming the total IB correction.
\begin{table}[t]
    \centering
    \small
    \renewcommand{\arraystretch}{1.1}
    \begin{tabular}{c | r@{.}l r@{.}l | r@{.}l}
        diag & \multicolumn{2}{c}{$\Delta^\gamma a_\mu^{\mathrm{hvp}}$} & \multicolumn{2}{c|}{$\Delta^{\delta m}a_\mu^{\mathrm{hvp}}$} & \multicolumn{2}{c}{$\Delta^{\mathrm{IB}}a_\mu^{\mathrm{hvp}}$} \\
        \noalign{\smallskip}\hline\noalign{\smallskip}
        NLOa    &  1&43(72) & $-$0&75(37) &   0&69(81) \\
        NLOb    & $-$1&06(53) &  0&45(23) &  $-$0&61(58) \\
        NLOa\&b &  0&38(19) & $-$0&29(15) &   0&08(23) \\
        NLOc    & $-$0&055(28) & 0&028(14) & $-$0&027(31) \\
    \end{tabular}
    \caption{Electromagnetic and strong IB effects, along with their sum, in units of $10^{-11}$. A conservative $50\%$ uncertainty is assigned to both $\Delta^\gamma a_\mu^{\mathrm{hvp}}$ and $\Delta^{\delta m}a_\mu^{\mathrm{hvp}}$.}
    \label{tab:isobreak}
\end{table}

The cancellation between the NLOa and NLOb kernels at large Euclidean times corresponds to a cancellation at small $x$ in Eq.~\eqref{eq:spacelike_def}, where IB effects are concentrated. For the combined NLOa\&b contribution, this yields a fortuitously small central value and, more importantly, a substantially reduced uncertainty relative to the individual NLOa and NLOb contributions.

We obtain the total IB correction to the NLO HVP by summing the NLOa\&b and NLOc contributions, which we treat as 100\% correlated since both arise from the same phenomenological models:
\begin{equation}
\Delta^{\mathrm{IB}}a_\mu^{\mathrm{hvp,\,nlo}}
    = 0.06(27) \times 10^{-11}\, .
\end{equation}
\subsection{Further small contributions}
\label{sec:smallcontr}

In this section, we estimate several subleading contributions required to complete the connection between our isoQCD calculations in Secs.~\ref{sec:isoQCD_result} and~\ref{sec:isoQCD_results_NLOc} and the physical theory. The bottom quark and charm quark-disconnected contributions are computed using standard lattice techniques and perturbative methods. The tau-loop correction to NLOb is evaluated directly on one ensemble near the physical point. Finally, charm sea-quark effects are estimated using phenomenological models with conservative uncertainties.

\subsubsection{Bottom quark contribution}
\label{sec:bottom_quark}

Following~\cite{rivera2026}, we employ the time-like integration to estimate the contribution from the bottom quark to $(i)=(4a)$ and $(4b)$. In this representation, the quantity of interest can be written as
\begin{equation}
    a_\mu^{b,b,\,(i)} = {\left(\frac{\alpha m_\mu}{6\pi m_b}\right)}^2{\left(\frac{\alpha}{\pi}\right)}^{N_i}\int_{1}^\infty \frac{dz}{z^2} K^{(i)}\left(\frac{4m_b^2}{m_\mu^2}z\right)R_b(z)\, ,
    \label{eq:bottom_def}
\end{equation}
In this context, the large mass of the bottom quark is particularly advantageous. Above the threshold $s \geq s_{\mathrm{th}} = 4m_b^2$, the kernel function $K^{(i)}(s/m_\mu^2)$ entering the dispersion integral becomes essentially constant, with power-suppressed corrections that rapidly vanish at high energies. As a result, the contribution to the integral is dominated by the leading, kernel-independent term, which reduces the problem to simple integral moments of the hadronic spectral function $R_q(s)$. These moments are directly related, via the optical theorem, to the vector current correlator computed in perturbative QCD, which is known up to $\mathrm{O}(\alpha_s^3)$~\cite{Erler:2016atg,Erler:2022mzd}. Importantly, at this scale the strong coupling is small, $\alpha_s(\hat{m}_b^2)/\pi\approx0.07$, ensuring excellent convergence of the perturbative series and allowing these moments to be determined with very high precision. In Eq.~\eqref{eq:bottom}, we quote our estimates for the bottom quark contributions to NLOa and NLOb, which are added as exact contributions to our isoQCD results.
\begin{equation}
    \begin{aligned}
        \frac{1}{9}\, a_\mu^{b,b,\,\mathrm{nlo(a)}}\, &\approx \, -0.23 \times10^{-11}\, ,\\
        \frac{1}{9}\, a_\mu^{b,b,\,\mathrm{nlo(b)}}\, &\approx \, \phantom{-}0.048 \times10^{-11}\, .
    \end{aligned}
    \label{eq:bottom}
\end{equation}
The dominant bottom quark contribution to NLOc arises from the $a_\mu^{3,3-b,b}$ term. Based on simple kinematic arguments, we expect $a_\mu^{3,3-b,b} \sim {(m_c/m_b)}^2\, a_\mu^{3,3-c,c}$. We consider a 100\% error for this estimate, which is almost insignificant when compared to the results for NLOc in~\eqref{eq:NLOc_result}.
\begin{equation}
    \frac{2}{9}\, a_\mu^{3,3-b,b,\,\mathrm{nlo(c)}} = 0.4(4) \times 10^{-13}\, ,
\end{equation}
As a cross-check, similar estimates for the bottom quark contribution to NLOa and NLOb in~\eqref{eq:bottom} can be obtained by rescaling the results $(4/9)\, a_\mu^{c,c}$ from Tab.~\ref{tab:isoQCD_result} by the factor $\frac{Q_b^2}{Q_c^2}\frac{m_c^2}{m_b^2}$.

\subsubsection{Charm quark-disconnected contributions}
\label{sec:disccharm}

Based on previous observations~\cite{Kuberski:2024bcj}, we expect the contributions from charm quark-disconnected diagrams to be very small. These contributions arise from the $G^{(c,c)}_{\mathrm{disc}}$ and $G^{(c,8)}_{\mathrm{disc}}$ terms in Eq.~\eqref{eq:decomp}. We make use of conserved vector currents at source and sink to evaluate these correlation functions. Since they are rather noisy at large Euclidean times and, due to the large charm quark mass, are expected to contribute predominantly to the short-distance region of the integrand, we compute them only for the SD window. That is, we approximate $a_\mu \approx {(a_\mu)}^{\mathrm{SD}}$ for these terms. We refer to Appendix~C in Ref.~\cite{Ce:2022eix} for more information on the treatment of the quark-disconnected diagrams.

In Tab.~\ref{tab:charm_disc}, we present a low-resolution estimate of these contributions. All results are far smaller than the precision achieved for the total NLOa\&b result in Eq.~\eqref{eq:NLOab_result}, and are therefore numerically irrelevant for the final estimates.
\begin{table}[t]
    \centering
    \small
    \renewcommand{\arraystretch}{1.1}
    \begin{tabular}{c | r@{.}l r@{.}l}
        diag & \multicolumn{2}{c}{$\frac{4}{9}a_\mu^{c,c(\mathrm{disc})}$} & \multicolumn{2}{c}{$\frac{2}{3\sqrt{3}}a_\mu^{c,8(\mathrm{disc})}$} \\
        \noalign{\smallskip}\hline\noalign{\smallskip}
        NLOa    & 21&6(5.0)(5.3)$\times10^{-15}$ & $-$2&4(1.6)(1.7)$\times10^{-15}$ \\
        NLOb    & $-$6&4(1.7)(1.4)$\times10^{-15}$ & 0&87(57)(58)$\times10^{-15}$ \\
        NLOa\&b & 15&1(3.4)(3.9)$\times10^{-15}$ & $-$1&5(1.0)(1.2)$\times10^{-15}$ \\
    \end{tabular}
    \caption{Charm quark-disconnected contributions to the main diagram sets.}
    \label{tab:charm_disc}
\end{table}

\subsubsection{Charm quark quenching}
\label{sec:seacharm}

The gauge configurations used in this work do not include dynamical charm quarks in the sea, so effects arising from charm quark vacuum loops are not accounted for. To estimate a systematic uncertainty to correct for this quenching effect, we follow the approach presented in Appendix~D of Ref.~\cite{Ce:2022kxy}, where a simplified phenomenological model for the contribution of the $D^0\bar{D}^0$, $D^+D^-$, and $D_s^+D_s^-$ channels to the $R$-ratio associated with $(u,d,s)$ valence quarks is employed. We also evaluate the difference $\Pi_v^{(N_{\mathrm{f}}=4)}-\Pi_v^{(N_{\mathrm{f}}=3)}$ explicitly using perturbative QCD. These estimates predict about a one per-mille effect for the NLOa and NLOb contributions separately, while the relative effect is slightly larger for the combined NLOa\&b contribution.

However, the long-distance window presents additional concerns. First, the phenomenological models become increasingly unreliable at low energies where non-perturbative effects dominate. Second, the long-distance contribution is particularly sensitive to the scale-setting observable $t_0^{\mathrm{ph}}$, whose determination in Ref.~\cite{t0madrid} was itself performed with $N_{\mathrm{f}}=2+1$ ensembles lacking dynamical charm. To account for these compounded low-energy uncertainties, we add in quadrature an additional uncertainty equal to 0.2\% of the long-distance window contribution for each diagram set. The final contribution to the error budget of each diagram set from the charm quark quenching reads
\begin{equation}
    \begin{aligned}
        \delta^{\mathrm{c-sea}}\left(a_\mu^{\mathrm{hvp,\,nlo(a)}}\right) &= 0.34\times10^{-11} \,,\\
        \delta^{\mathrm{c-sea}}\left(a_\mu^{\mathrm{hvp,\,nlo(b)}}\right) &= 0.16\times10^{-11}\,,\\
        \delta^{\mathrm{c-sea}}\left(a_\mu^{\mathrm{hvp,\,nlo(a\&b)}}\right) &= 0.18\times10^{-11}\,.
    \end{aligned}
\end{equation}
\subsubsection{Tau-loop effects to NLOb}
\label{sec:NLOb_tau}

In Sec.~\ref{sec:timekernels}, we assumed---without detailed justification---that the only relevant contribution to NLOb arises from the electron loop, and that effects from tau loops can be neglected. Formally, taking $M_\tau \equiv m_\tau/m_\mu \gg 1$, the corresponding kernel function is strongly suppressed, scaling as $f^{(4b)}_\tau \sim 1/M_\tau^2$.

To estimate the numerical size of this contribution, we use ensemble E250, our closest ensemble to the physical point for which data in all relevant channels are available. We approximate $a_\mu^{\mathrm{hvp,\,nlo(b;\tau)}} \approx (\alpha/\pi)^3 \sum_t \tilde{f}^{(4b)}_\tau(\hat{t})\, G_{\mathrm{E250}}^{(\gamma,\gamma)}(t)$ and assign a conservative $50\%$ uncertainty to account for possible discretization effects and pion mass mistuning. We obtain
\begin{equation}
    a_\mu^{\mathrm{hvp,\,nlo(b;\tau)}} = 0.06(3)\times10^{-11}\, .
\end{equation}

\subsection{Total NLO HVP contribution}
\label{sec:finalresult}

We summarize here the results obtained in Secs.~\ref{sec:isoQCD_result} and~\ref{sec:isoQCD_results_NLOc}, together with the isospin-breaking estimates of Sec.~\ref{sec:isobreak} and the subleading contributions discussed in Sec.~\ref{sec:smallcontr}. Equation~\eqref{eq:final_result_diag} collects the final results for each diagram set, while Eq.~\eqref{eq:final_result} gives our final estimate for the NLO HVP, obtained from the addition of the NLOa\&b and NLOc results. The uncertainties presented between brackets correspond to the statistical error, the systematic uncertainty derived from the spread, the uncertainty arising from $t_0^{\mathrm{ph}}$, and the uncertainty associated to the IB shift. For NLOa, NLOb and their combination, where the charm quark sea effects have been estimated, and we correct to a reference volume, we also quote the uncertainty associated to the respective corrections applied on the continuum. The total uncertainty quoted in square brackets corresponds to their quadratic combination.
\begin{equation}
    \begin{aligned}
        a_\mu^{\mathrm{hvp,\,nlo(a)}} & = -216.83(81)(87)(99)(81)(34)(31)[1.81] \times 10^{-11} \, , \\
        a_\mu^{\mathrm{hvp,\,nlo(b)}} & = \phantom{-}111.51(55)(64)(64)(58)(16)(25)[1.24] \times 10^{-11} \, , \\
        a_\mu^{\mathrm{hvp,\,nlo(a\&b)}} & = -105.33(28)(28)(35)(24)(18)(5)[61] \times 10^{-11} \, , \\
        a_\mu^{\mathrm{hvp,\,nlo(c)}} & = \phantom{-00}3.75(6)(6)(3)(3)[10] \times 10^{-11} \, , \\
    \end{aligned}
    \label{eq:final_result_diag}
\end{equation}
\begin{equation}
    \phantom{-}a_\mu^{\mathrm{hvp,\,nlo}} = -101.57(26)(29)(31)(27)(18)(5)[59] \times 10^{-11} \, .
    \label{eq:final_result}
\end{equation}
\begin{figure}[t]
    \centering
    \includegraphics[width=\textwidth]{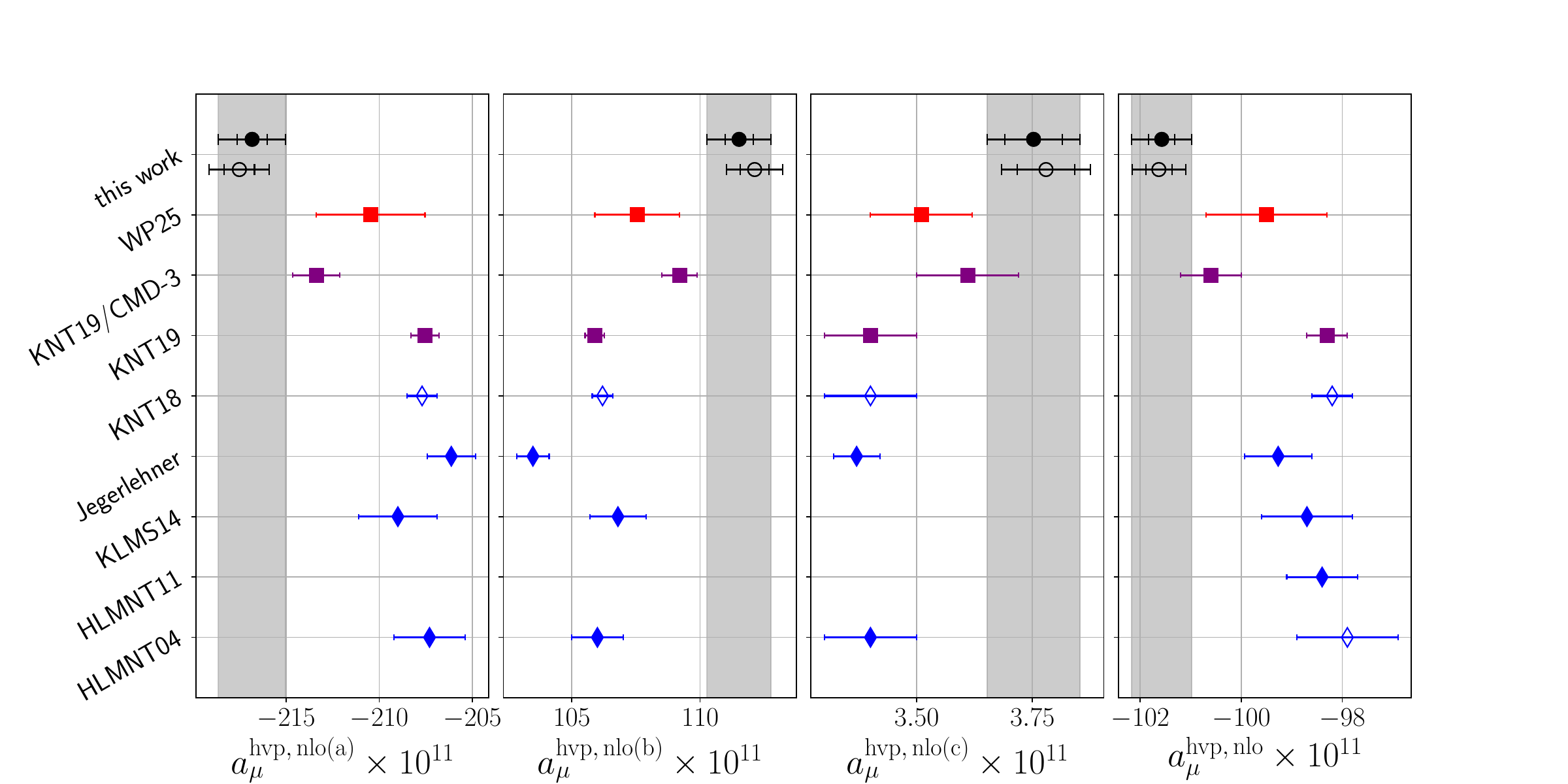}
    \caption{Comparison of the NLO diagram sets and their sum between our results and data-driven determinations. Filled and empty black points denote the final and isoQCD results of this work, respectively. In red squares, the WP25~\cite{WP25} average is shown, together with the KNT19~\cite{Keshavarzi:2019abf} and KNT19/CMD-3~\cite{DiLuzio:2024sps} inputs used in its calculation, in this case shown in purple squares. Blue rhombuses show older data-driven results from Refs.~\cite{Hagiwara:2006jt,HLMNT11,KLMS14,Jegerlehner:2017gek,KNT18}.}
    \label{fig:results}
\end{figure}

Figure~\ref{fig:results} compares the results of this work (black filled point and shaded band) with a variety of data-driven determinations. The empty black point represents our isoQCD determination prior to the inclusion of isospin-breaking effects. The red squares show the current WP25~\cite{WP25} estimate\footnote{Reference~\cite{WP25} quotes only the combined NLO HVP result $a_\mu^{\mathrm{hvp,nlo}} = -9.96(13) \times 10^{-10}$, without providing values for individual diagram sets. The WP25 average is based on KNT19 ($-9.83(4) \times 10^{-10}$) and a variant incorporating CMD-3 data ($-10.08(6) \times 10^{-10}$). An updated analysis correcting an issue in the NLOc contribution yields revised values of $-10.06(6) \times 10^{-10}$ for KNT19/CMD-3 and $-9.95(12) \times 10^{-10}$ for the average. We thank A. Keshavarzi for providing the results for the individual
diagrams shown in Fig.~\ref{fig:results} and their updated values.}, which accounts for the tension between different data-driven evaluations shown as purple squares: KNT19~\cite{Keshavarzi:2019abf} and the variant incorporating CMD-3 data for the two-pion channel~\cite{DiLuzio:2024sps}. Older data-driven results are shown in blue~\cite{Hagiwara:2006jt,HLMNT11,KLMS14,Jegerlehner:2017gek,KNT18}.

\begin{figure}[h!]
    \centering
    \includegraphics[width=\textwidth]{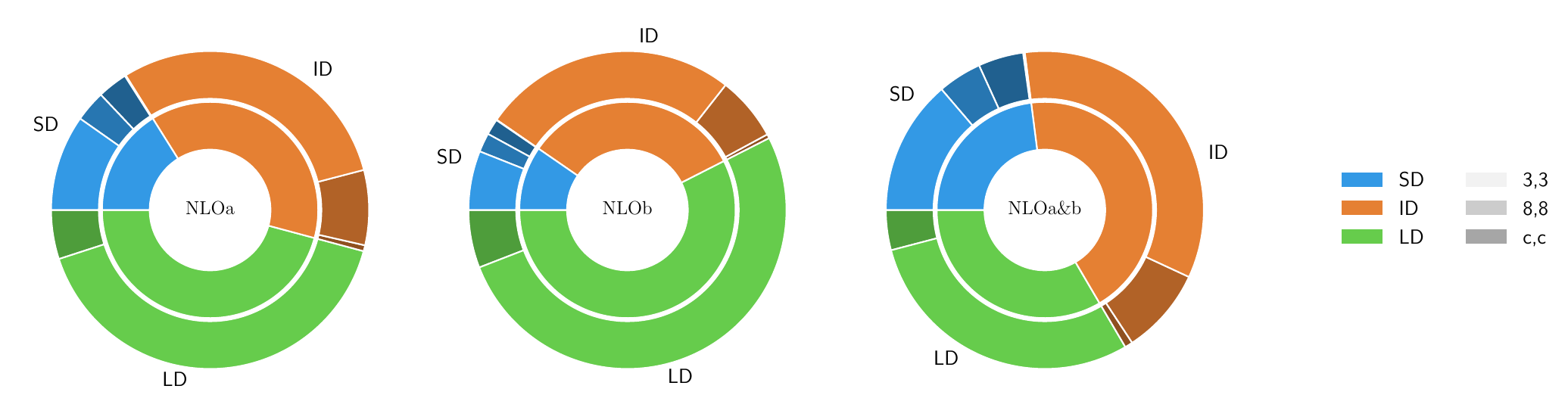}
    \caption{Relative contribution to the central value of diagram sets NLOa, NLOb, and their combination. Blue, orange, and green correspond to the SD, ID, and LD windows, respectively. Shaded areas denote the contributions of the different isospin components.}
    \label{fig:CV_piechart}
    \centering
    \includegraphics[width=\textwidth]{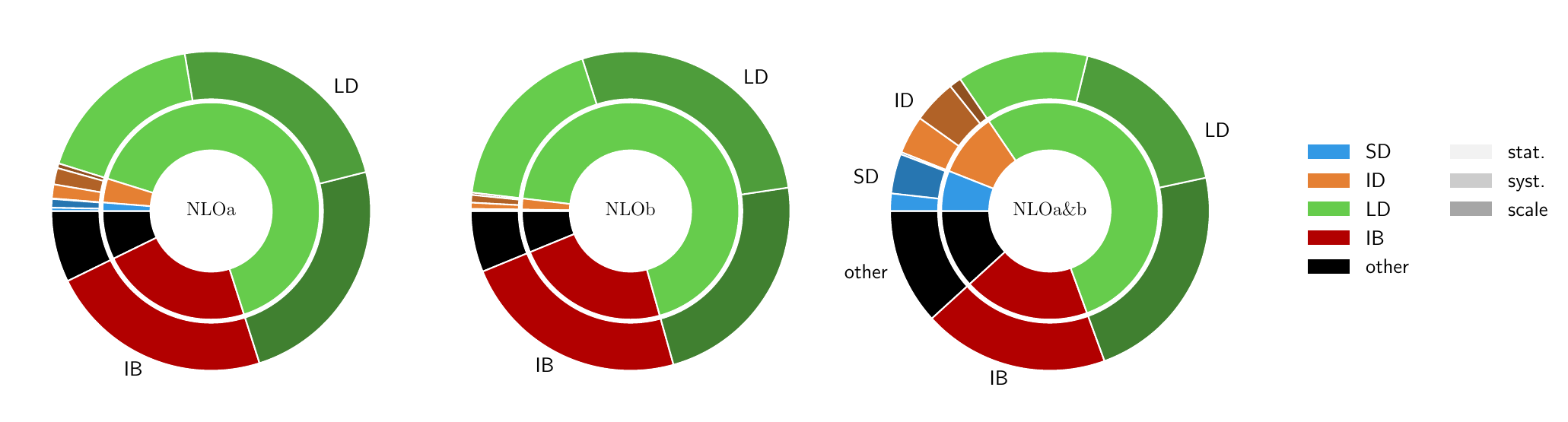}
    \caption{Relative contribution to the variance of diagram sets NLOa, NLOb, and their combination. Blue, orange, and green correspond to the SD, ID, and LD windows, respectively, dark red denotes the contribution from IB effects, and black summarizes the other uncertainty sources, like the FVC in the continuum and the charm sea-quark effects. Shading indicates the breakdown into statistical uncertainties, model-average systematics, and scale setting.}
    \label{fig:Var_piechart}
    \centering
    \begin{subfigure}{0.49\textwidth}
        \centering
        \includegraphics[height=4.5cm]{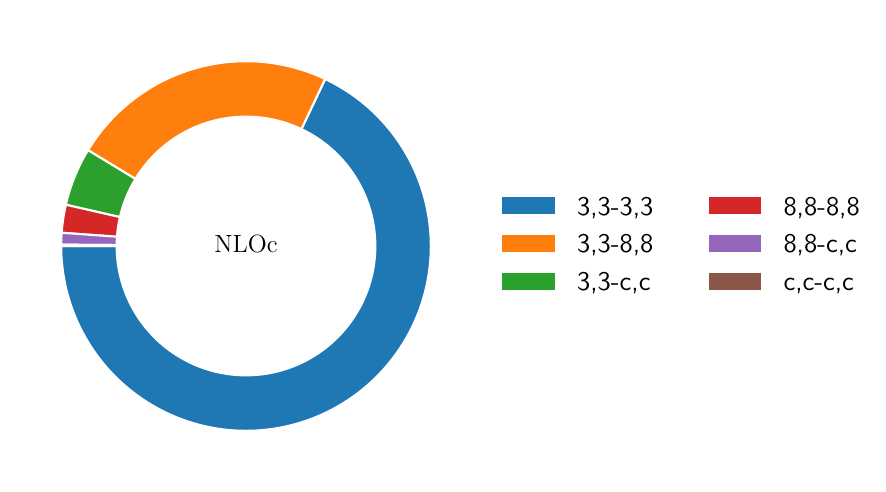}
    \end{subfigure}
    \hfill
    \begin{subfigure}{0.49\textwidth}
        \centering
        \includegraphics[height=4.5cm]{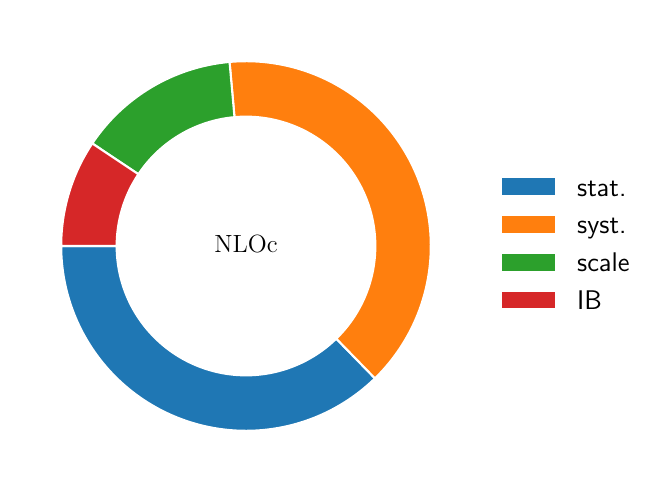}
    \end{subfigure}
    \caption{Relative contributions to the central value (left) and variance (right) of the NLOc diagram.}
    \label{fig:CV_VAR_NLOc_piechart}
\end{figure}

Figures~\ref{fig:CV_piechart} and~\ref{fig:Var_piechart} illustrate the relative contributions to the final central value and variance of NLOa, NLOb, and combination NLOa\&b. Figure~\ref{fig:CV_piechart} highlights the substantial reduction in the sensitivity to the long-distance window achieved when combining NLOa and NLOb. Figure~\ref{fig:Var_piechart} shows how this translates into a reduced total variance and a more balanced distribution among the different sources of uncertainty. The corresponding breakdown for the NLOc diagram is shown in Fig.~\ref{fig:CV_VAR_NLOc_piechart}.

\section{Conclusion}
\label{sec:conclusions}

The tensions surrounding the evaluation of the LO HVP contribution via the data-driven dispersive approach have yet to be explained, and provide a strong motivation for performing a state-of-the-art lattice calculation of the NLO correction that can be combined consistently with the corresponding calculation of the LO part, which is what we have achieved with this paper.

Our result for $a_\mu^{\mathrm{hvp,\,nlo}}$, given in Eq.~\eqref{eq:final_result}, has a total precision of 0.6\% and lies consistently below data-driven evaluations based on hadronic cross section measurements taken prior to the CMD-3 result (see the rightmost panel of Fig.~\ref{fig:results}). This shift between lattice and data-driven evaluations of $a_\mu^{\mathrm{hvp,\,nlo}}$ is consistent with the analogous shift observed for $a_\mu^{\mathrm{hvp,\,lo}}$.

This level of precision is comparable to that achieved by data-driven $R$-ratio approaches, rendering our result sufficiently precise for a meaningful comparison with experiment. Crucially, our lattice determination is free from the inconsistencies among the experimental data sets afflicting the dispersive method. The first-principles nature of our calculation thus provides an independent cross-check of the Standard Model prediction, unencumbered by these experimental ambiguities.

One important ingredient for achieving a level of precision which significantly exceeds that of the LO HVP contribution is the strong cancellation of the long-distance contribution of the vector current correlation function between the NLOa and NLOb diagram sets. As a result, the TMR integrand for the total NLO contribution is far less affected by statistical noise and by finite-volume and isospin-breaking corrections (and their uncertainties) which mostly arise from the long-distance regime. Most importantly, the integrand is less sensitive to uncertainties arising from the scale-setting procedure, which in our case is a dominant source of uncertainty and is also predominant in the long-distance regime. This is illustrated by the pie charts shown in Figs.~\ref{fig:CV_piechart} and~\ref{fig:Var_piechart}.

Looking ahead, we note that our results can be straightforwardly updated as determinations of fundamental QCD scales improve. In Appendix~\ref{app:derivatives}, we provide the dimensionless scheme dependencies $S/\mathcal{O} \times \partial\mathcal{O}/\partial S$ for all observables with respect to $\sqrt{t_0}$, $m_\pi$, $m_K$, and $m_{D_s}$. These derivatives enable practitioners to convert our results to alternative scale-setting schemes or to incorporate future refinements---particularly in the determination of $t_0$, which currently contributes $\approx0.3 \times 10^{-11}$ to our total uncertainty---without requiring a re-analysis of the underlying lattice correlators.

Now that a consistent and precise determination of the NLO HVP contribution has been achieved, the focus of future lattice studies returns to the LO contribution, which continues to dominate the total uncertainty of the SM prediction for $a_\mu$. The sub-percent precision reached for the NLO contribution ensures it will not become a limiting factor as lattice determinations of the LO HVP continue to improve.

\section*{Acknowledgments}
\label{sec:acknowledgments}

We thank Gilberto Colangelo, Martin Hoferichter, Alex Keshavarzi, Stefano Laporta and Georg von Hippel for valuable discussions. A.B. is grateful to Dalibor Djukanovic for discussions and crosschecks regarding finite-size corrections. A.B. is also grateful to Pere Masjuan and Antonio Rivera, for their help in the perturbative estimation of the bottom quark effects.
We thank Andrew Hanlon, Ben H\"{o}rz, Nolan Miller, Daniel Mohler, Colin Morningstar and Srijit Paul for the collaboration on the data generation and analysis for the spectral reconstruction.
We thank Volodymyr Biloshytskyi for providing auxiliary data used in the estimate of isospin-breaking effects.
We are grateful to our colleagues in the CLS initiative for sharing ensembles.
Calculations for this project were performed on the HPC clusters Clover and HIMster-II at the Helmholtz Institute Mainz and Mogon-II and Mogon-NHR at Johannes Gutenberg-Universität (JGU) Mainz, as well as on the GCS Supercomputers JUQUEEN and JUWELS at J\"ulich Supercomputing Centre (JSC), HAZELHEN and HAWK at the H\"ochstleistungsrechenzentrum Stuttgart (HLRS), and SuperMUC at the Leibniz Supercomputing Centre (LRZ).
The authors gratefully acknowledge the support of the Gauss Centre for Supercomputing (GCS) and the John von Neumann-Institut für Computing (NIC) by providing computing time via the projects HMZ21, HMZ23 and HINTSPEC at JSC, as well as projects GCS-HQCD and GCS-MCF300 at HLRS and LRZ. We also gratefully acknowledge the scientific support and HPC resources provided by NHR-SW of Johannes Gutenberg-Universität Mainz (project NHR-Gitter).
This work has been supported by Deutsche Forschungsgemeinschaft (German Research Foundation, DFG) through the Collaborative Research Center 1660 ``Hadrons and Nuclei as Discovery Tools'', through the research unit FOR 5327 ``Photon-photon interactions in the Standard Model and beyond exploiting the discovery potential from MESA to the LHC'' (Project No.\ 458854507), under grant HI~2048/1-2 (Project No.\ 399400745), and through the Cluster of Excellence ``Precision Physics, Fundamental Interactions and Structure of Matter'' (PRISMA++ EXC2118/2), funded within the German Excellence strategy (Project No.\ 390831469). This project has received funding from the European Union's Horizon Europe research and innovation programme under the Marie Sk\l{}odowska-Curie grant agreement No.\ 101106243.

\newpage

\appendix

\section{Closed form solution for NLOc}
\label{app:timekern_NLOc}

An analytic form for the time-kernel of diagram NLOc can be obtained by solving Eq.~\eqref{eq:timekernel_i}, where $\hat{f}^{(4c)}=\hat{f}^{(2)}$ and $m_{4c}=2$.
\begin{equation}
    \begin{aligned}
        &\frac{m_\mu^4}{32\pi^4}\Tilde{f}^{(4c)}\left(\hat{t}, \hat{\tau}\right) = \frac{\hat{\tau} ^2 \hat{t}^2}{4}+\frac{\hat{t}^2}{\hat{\tau} ^2}+\frac{\hat{\tau} ^2}{\hat{t}^2}-\frac{1}{2} \left(\hat{t}^2+\hat{\tau} ^2\right) +\frac{1}{6} -2 (1+\gamma_E)\\
        & + 2 \hat{t}^2 (\ln \hat{\tau} +\gamma_E )  + 2 \hat{\tau} ^2 (\ln \hat{t}+\gamma_E) + 2 \left(\hat{t}^2-1\right) \ln \hat{t} + 2 \left(\hat{\tau} ^2-1\right) \ln \hat{\tau} \\
        & + \left[1-(\hat{t}+\hat{\tau} )^2\right] \ln (\hat{t}+\hat{\tau} ) + \left[1-(\hat{t}-\hat{\tau} )^2\right] \ln | \hat{t}-\hat{\tau} |\\
        & + \left(\frac{\hat{t}^2}{6}-2\right) K_0(2 t) + \left(\frac{\hat{\tau}^2}{6}-2\right) K_0(2 \tau )+\left(1-\frac{1}{12} (\hat{t}+\hat{\tau} )^2\right) K_0(2 (\hat{t}+\hat{\tau} ))\\
        & + \left(1-\frac{1}{12} (\hat{t}-\hat{\tau} )^2\right) K_0(2 | \hat{t}-\hat{\tau} | )-\left(\frac{2 \hat{t}^2}{\hat{\tau} }+\frac{\hat{\tau} }{12}\right) K_1(2 \hat{\tau} )-\left(\frac{2 \hat{\tau}^2}{\hat{t}}+\frac{\hat{t}}{12}\right) K_1(2 \hat{t}) \\
        & + \frac{1}{24} | \hat{t}-\hat{\tau} |  K_1(2 | \hat{t}-\hat{\tau} | )+\frac{1}{24} (\hat{t}+\hat{\tau} ) K_1(2 (\hat{t}+\hat{\tau} )) \\
        & + \left(\frac{\hat{t}^2}{12}+\frac{\hat{\tau}^2}{4}-\frac{15}{16}\right) G_{1,3}^{2,1}\left(\hat{t}^2|
        \begin{array}{c}
         \frac{3}{2} \\
         0,1,\frac{1}{2} \\
        \end{array}
        \right)+\left(\frac{\hat{\tau}^2}{12}+\frac{\hat{t}^2}{4}-\frac{15}{16}\right) G_{1,3}^{2,1}\left(\hat{\tau}^2|
        \begin{array}{c}
         \frac{3}{2} \\
         0,1,\frac{1}{2} \\
        \end{array}
        \right) \\
        & + \left(\frac{15}{32}-\frac{1}{24} (\hat{t}+\hat{\tau} )^2\right) G_{1,3}^{2,1}\left((\hat{t}+\hat{\tau} )^2|
        \begin{array}{c}
         \frac{3}{2} \\
         0,1,\frac{1}{2} \\
        \end{array}
        \right) \\
        & + \left(\frac{15}{32}-\frac{1}{24} (\hat{t}-\hat{\tau} )^2\right) G_{1,3}^{2,1}\left((\hat{t}-\hat{\tau} )^2|
        \begin{array}{c}
         \frac{3}{2} \\
         0,1,\frac{1}{2} \\
        \end{array}
        \right)\, .
    \end{aligned}
    \label{eq:diagram_c_exact}
\end{equation}

\section{Time-kernel coefficient tables}
\label{sec:tables}

\subsection{NLOb}
\begin{table}[h!]
    \scriptsize
    \centering
    \renewcommand{\arraystretch}{1.4}
    \begin{tabular}{c|c c c}
        n & $m = 0$ & $m = 4$ & $m = 6$\\
        \noalign{\smallskip}\hline\noalign{\smallskip}
        $4$ & $0$ & $0$ & $0$ \\
        $6$ & $-\frac{2}{3}$ & $0$ & $0$ \\
        $8$ & $-4$ & $0$ & $0$ \\
        $10$ & $-\frac{56}{3}$ & $4$ & $0$ \\
        $12$ & $-80$ & $24$ & $\frac{16}{3}$ \\
        $14$ & $-330$ & $112$ & $32$ \\
        $16$ & $-\frac{4004}{3}$ & $480$ & $\frac{448}{3}$ \\
        $18$ & $-\frac{16016}{3}$ & $1980$ & $640$ \\
        $20$ & $-21216$ & $8008$ & $2640$ \\
        $22$ & $-83980$ & $32032$ & $\frac{32032}{3}$ \\
        $24$ & $-\frac{994840}{3}$ & $127296$ & $\frac{128128}{3}$ \\
        $26$ & $-1307504$ & $503880$ & $169728$ \\
        $28$ & $-\frac{15452320}{3}$ & $1989680$ & $671840$ \\
        $30$ & $-20281170$ & $7845024$ & $\frac{7958720}{3}$ \\
    \end{tabular}
    \qquad\qquad
    \begin{tabular}{c|c c c c}
        n & $m = 0$ & $\mathbf{m = 3}$ & $m = 4$ & $m = 6$\\
        \noalign{\smallskip}\hline\noalign{\smallskip}
        $4$ & $0$ & $-\frac{2}{3}$ & $\frac{2}{3}$ & $-\frac{4}{9}$ \\
        $6$ & $\frac{1}{6}$ & $0$ & $0$ & $\frac{8}{9}$ \\
        $8$ & $1$ & $0$ & $0$ & $0$ \\
        $10$ & $\frac{14}{3}$ & $0$ & $-1$ & $0$ \\
        $12$ & $20$ & $0$ & $-6$ & $-\frac{4}{3}$ \\
        $14$ & $\frac{165}{2}$ & $0$ & $-28$ & $-8$ \\
        $16$ & $\frac{1001}{3}$ & $0$ & $-120$ & $-\frac{112}{3}$ \\
        $18$ & $\frac{4004}{3}$ & $0$ & $-495$ & $-160$ \\
        $20$ & $5304$ & $0$ & $-2002$ & $-660$ \\
        $22$ & $20995$ & $0$ & $-8008$ & $-\frac{8008}{3}$ \\
        $24$ & $\frac{248710}{3}$ & $0$ & $-31824$ & $-\frac{32032}{3}$ \\
        $26$ & $326876$ & $0$ & $-125970$ & $-42432$ \\
        $28$ & $\frac{3863080}{3}$ & $0$ & $-497420$ & $-167960$ \\
        $30$ & $\frac{10140585}{2}$ & $0$ & $-1961256$ & $-\frac{1989680}{3}$ \\
    \end{tabular}
    \caption{$f_{nm}$ (on the left) and $h_{nm}$ (on the right) coefficients. Notice that $h_{43}$ is the only non-zero "odd coefficient". All coefficients not explicitly shown are zero.}
    \label{tab:NLOb_coefficients3}
\end{table}

\newgeometry{left=4.5cm,right=-2.0cm,top=4.0cm,bottom=0.0cm}
\begin{landscape}
    \centering
    \begin{table}[h!]
    \scriptsize
    \centering
    \renewcommand{\arraystretch}{1.4}
    \begin{tabular}{c|c c c c}
        n & $m = 0$ & $m = 2$ & $m = 4$ & $m = 6$\\
        \noalign{\smallskip}\hline\noalign{\smallskip}
        $4$ & $-\frac{2}{9} \ln M-\frac{1}{18}$ & $1$ & $4 + 4 \ln^2M$ & $-\frac{46}{27}+\frac{28}{9}\ln M-\frac{1}{3} 8 \ln ^2M$ \\
        $6$ & $\frac{169}{90}\ln M-\frac{36931}{10800}$ & $-\frac{2}{3}$ & $-2 -2 \ln M$ & $\frac{80}{27}+\frac{56}{9}\ln M+\frac{16}{3}\ln ^2M$ \\
        $8$ & $\frac{1604}{105}\ln M-\frac{210047}{8820}$ & $\frac{704}{105}$ & $-\frac{10}{9} +\frac{2}{3}\ln M$ & $-\frac{-38}{9}-\frac{8}{3}\ln M$ \\
        $10$ & $\frac{11018}{135}\ln M-\frac{21513067}{170100}$ & $\frac{5344}{105}$ & $\frac{17550779}{396900}-\frac{4756}{315}\ln M$ & $-\frac{4}{9} + \frac{16}{9}\ln M $ \\
        $12$ & $\frac{29412}{77}\ln M-\frac{2894965393}{4802490}$ & $\frac{130858}{495}$ & $\frac{1206809563}{4002075}-\frac{127228}{1155}\ln M$ & $\frac{1883184914}{36018675}-\frac{228584}{10395}\ln M$ \\
        $14$ & $\frac{612359}{364}\ln M-\frac{2138512657033}{787026240}$ & $\frac{1218348}{1001}$ & $\frac{127013964109}{82818450}-\frac{3616148}{6435}\ln M$ & $\frac{99638442364}{289864575}-\frac{7043536}{45045}\ln M$ \\
        $16$ & $\frac{967681}{135}\ln M-\frac{82365133883}{6949800}$ & $\frac{965085}{182}$ & $\frac{761763635183}{108216108}-\frac{2560820}{1001}\ln M$ & $\frac{428362589497}{248455350}-\frac{15209336}{19305}\ln M$ \\
        $18$ & $\frac{13741442}{459}\ln M-\frac{101648836935029}{2008492200}$ & $\frac{17151277}{765}$ & $\frac{58307355653567}{1895421528}-\frac{17107145}{1547}\ln M$ & $\frac{184237817350667}{23455841409}-\frac{544829680}{153153}\ln M$ \\
        $20$ & $\frac{2708773132}{21945}\ln M-\frac{272062917990438103}{1277158182300}$ & $\frac{1353220726}{14535}$ & $\frac{3951352873426328}{30211070175}-\frac{675640394}{14535}\ln M$ & $\frac{8767893549452644}{256592689353}-\frac{1347926756}{88179}\ln M$ \\
        $22$ & $\frac{349608740}{693}\ln M-\frac{143168111013691151}{161325244080}$ & $\frac{2795440492}{7315}$ & $\frac{462494576627921483}{845909964900}-\frac{2793108812}{14535}\ln M$ & $\frac{91904978008606424}{634432473675}-\frac{2789228936}{43605}\ln M$ \\
        $24$ & $\frac{16542939190}{8073}\ln M-\frac{39648833260883466677}{10806172437060}$ & $\frac{8268943735}{5313}$ & $\frac{254493286584114300817}{112602779739450}-\frac{132237346072}{168245}\ln M$ & $\frac{202849476857291708837}{335614778574075}-\frac{264260177584}{1002915}\ln M$ \\
        $26$ & $\frac{169660886992}{20475}\ln M-\frac{2066775856731885303313}{137034795397500}$ & $\frac{84815201126}{13455}$ & $\frac{1648232909159109674719}{177793862746500}-\frac{84789942526}{26565}\ln M$ & $\frac{10537105313665625555402}{4222604240229375}-\frac{2711963087072}{2523675}\ln M$ \\
        $28$ & $\frac{208288358800}{6237}\ln M-\frac{2760330939401667649181}{44724543113250}$ & $\frac{520668477976}{20475}$ & $\frac{1094517880631620904114}{28945104742125}-\frac{520577023756}{40365}\ln M$ & $\frac{12274302884650696941196}{1200108573538875}-\frac{1040850944312}{239085}\ln M$ \\
        $30$ & $\frac{1027838381233}{7656}\ln M-\frac{19232398525323098271692683}{76420755826272000}$ & $\frac{6166722206768}{60291}$ & $\frac{3793236255849090686929301}{24695627770563750}-\frac{30830570730448}{593775}\ln M$ & $\frac{3045196077582565137399176}{73028499264381375}-\frac{61650532771376}{3511755}\ln M$ \\
    \end{tabular}
    \caption{$e_{nm}(\ln M)$ coefficients. All coefficients not explicitly shown are zero.}
    \label{tab:NLOb_coefficients1}
\end{table}
    \vspace{0.3cm}
    
\begin{table}[h!]
    \scriptsize
    \centering
    \renewcommand{\arraystretch}{1.4}
    \begin{tabular}{c|c c c c}
        n & $m = 0$ & $m = 2$ & $m = 4$ & $m = 6$\\
        \noalign{\smallskip}\hline\noalign{\smallskip}
        $4$ & $0$ & $0$ & $0$ & $0$ \\
        $6$ & $\frac{97}{45}-\frac{4}{3}\ln M$ & $0$ & $0$ & $0$ \\
        $8$ & $\frac{1583}{105}-8 \ln M$ & $-4$ & $0$ & $0$ \\
        $10$ & $\frac{10562}{135}-\frac{112}{3}\ln M$ & $-24$ & $8 \ln M-\frac{8011}{315}$ & $0$ \\
        $12$ & $\frac{251684}{693}-160 \ln M$ & $-112$ & $48 \ln M-\frac{185902}{1155}$ & $\frac{32}{3}\ln M-\frac{307124}{10395}$ \\
        $14$ & $\frac{871433}{546}-660 \ln M$ & $-480$ & $224 \ln M-\frac{5047292}{6435}$ & $64 \ln M-\frac{8412904}{45045}$ \\
        $16$ & $\frac{1835959}{270}-\frac{8008 }{3}\ln M$ & $-1980$ & $960 \
        \ln M-\frac{10466956}{3003}$ & $\frac{896}{3}\ln M-\frac{17570552}{19305}$ \\
        $18$ & $\frac{21741802}{765}-\frac{32032}{3}\ln M$ & $-8008$ & \
        $3960 \ln M-\frac{45885963}{3094}$ & $1280 \ln M-\frac{206607056}{51051}$ \\
        $20$ & $\frac{171608860}{1463}-42432 \ln M$ & $-32032$ & $16016 \
        \ln M-\frac{59715011}{969}$ & $5280 \ln M-\frac{506619670}{29393}$ \\
        $22$ & $\frac{221733355}{462}-167960 \ln M$ & $-127296$ & $64064 \
        \ln M-\frac{81545956}{323}$ & $\frac{64064}{3}\ln M-\frac{208425692}{2907}$ \\
        $24$ & $\frac{5251828975}{2691}-\frac{1989680}{3}\ln M$ & \
        $-503880$ & $254592 \ln M-\frac{34489442472}{33649}$ & \
        $\frac{256256}{3}\ln M-\frac{19661441456}{66861}$ \\
        $26$ & $\frac{53918626534}{6825}-2615008 \ln M$ & $-1989680$ & \
        $1007760 \ln M-\frac{36626670967}{8855}$ & $339456 \ln M-\frac{1005477461344}{841225}$ \\
        $28$ & $\frac{993906911612}{31185}-\frac{30904640}{3}\ln M$ & \
        $-7845024$ & $3979360 \ln M-\frac{670965894406}{40365}$ & $1343680 \
        \ln M-\frac{1154533778012}{239085}$ \\
        $30$ & $\frac{7363774861147}{57420}-40562340 \ln M$ & $-30904640$ \
        & $15690048 \ln M-\frac{39547820883388}{593775}$ & $\frac{15917440 \
       }{3} \ln M-\frac{68226557140376}{3511755}$ \\
    \end{tabular}
    \caption{$g_{nm}(\ln M)$ coefficients. All coefficients not explicitly shown are zero.}
    \label{tab:NLOb_coefficients2}
\end{table}
\end{landscape}
\restoregeometry

\subsection{NLOc}

\begin{table}[H]
\scriptsize
\centering
\renewcommand{\arraystretch}{1.4}
\begin{tabular}{c|cccc}
n& $i_n$& $j_n$& $k_n$& $l_n$\\
\noalign{\smallskip}\hline\noalign{\smallskip}
$4$ & $\frac{1}{3}$ & $0$ & $0$ & $0$ \\
$6$ & $-\frac{169}{60}$ & $2$ & $-\frac{1}{3}$ & $0$ \\
$8$ & $-\frac{802}{35}$ & $12$ & $\frac{352}{105}$ & $-2$ \\
$10$ & $-\frac{5509}{45}$ & $56$ & $\frac{2672}{105}$ & $-12$ \\
$12$ & $-\frac{44118}{77}$ & $240$ & $\frac{65429}{495}$ & $-56$ \\
$14$ & $-\frac{1837077}{728}$ & $990$ & $\frac{609174}{1001}$ & $-240$ \\
$16$ & $-\frac{967681}{90}$ & $4004$ & $\frac{965085}{364}$ & $-990$ \\
$18$ & $-\frac{6870721}{153}$ & $16016$ & $\frac{17151277}{1530}$ & $-4004$ \\
$20$ & $-\frac{1354386566}{7315}$ & $63648$ & $\frac{676610363}{14535}$ & $-16016$ \\
$22$ & $-\frac{174804370}{231}$ & $251940$ & $\frac{1397720246}{7315}$ & $-63648$ \\
$24$ & $-\frac{8271469595}{2691}$ & $994840$ & $\frac{8268943735}{10626}$ & $-251940$ \\
$26$ & $-\frac{84830443496}{6825}$ & $3922512$ & $\frac{42407600563}{13455}$ & $-994840$ \\
$28$ & $-\frac{104144179400}{2079}$ & $15452320$ & $\frac{260334238988}{20475}$ & $-3922512$ \\
$30$ & $-\frac{1027838381233}{5104}$ & $60843510$ & $\frac{3083361103384}{60291}$ & $-15452320$ \\
\end{tabular}
\caption{Coefficients $i_n$, $j_n$, $k_n$ and $l_n$. All coefficients not explicitly shown are zero.}
\label{tab:NLOc_small}
\end{table}

\section{pQCD computation of the subtracted piece}
\label{app:SDsub_NLOa}

The origin of the logarithmically enhanced terms in the SD window can be traced back to taking the naive continuum limit of the lower integration bound in the TMR integration. This can be understood by focusing on the behavior of the integrand in the neighborhood of $t\sim0$. At small distances, the $\mathrm{O}(a)$-improved electromagnetic correlator behaves as
$G(t)\sim 1/t^3\left(A + (a/t)^2 B + \mathrm{O}(a^3)\right)$,
where $A$ and $B$ are dimensionless coefficients whose explicit values are irrelevant for this argument. Then, using a realistic kernel function $K(t)$ and up to an irrelevant integration constant, the naive $a\to0$ limit of $\int K(t)G(t)$ reads
\begin{equation*}
    \begin{aligned}
        K(t)\sim t^4\, :&\quad \int dt\, K(t)G(t)\bigg|_{t=a} \sim \frac{A}{2}\, a^2 + B\, a^2\ln a + \mathrm{O}(a^3)\, , \\
        K(t)\sim t^4\ln t\, :&\quad \int dt\, K(t)G(t)\bigg|_{t=a} \sim -\frac{A}{4}\, a^2 + \frac{A}{2}\, a^2\ln a + \frac{B}{2}\, a^2\ln^2 a + \mathrm{O}(a^3)\, .
    \end{aligned}
\end{equation*}
As described in Sec.~\ref{sec:SD_sub}, effects arising from these logarithmically enhanced terms are very difficult to constrain in a continuum extrapolation and must therefore be subtracted. We expect this subtraction to be computable within perturbation theory. For the simple case $K(t)\sim t^4$, this is straightforward: the $C_4$ coefficient factors out of the integral~\eqref{eq:b_tilde}, and the computation reduces to evaluating $b^{(3,3)}(Q^2) \sim \Pi(Q^2) - \Pi(Q^2/4)$. In the massless theory, $\Pi(Q^2)$ is known up to $\mathrm{O}(\alpha_s^5)$.

The remaining challenge is to compute the second term in Eq.~\eqref{eq:b_tilde_NLOa}. To this end, we make use of the Chetyrkin expansion of the two-point function in the Euclidean coordinate-space representation~\cite{Chetyrkin_pQCD}, which is known up to $\mathrm{O}(\alpha_s^4)$ in the $\widetilde{\mathrm{MS}}$-scheme. In this representation, the two-point function at small distances is given by
\begin{equation}
    \Tilde{\Pi}_{\mu \nu}^{v}(X) = \frac{6}{\pi^4 (X^2)^3} \left[ \left( \frac{\delta_{\mu \nu}}{2} - \frac{X_\mu X_\nu}{X^2} \right) \Tilde{C}^v + \delta_{\mu \nu} \Tilde{D}^v \right]\, ,
\end{equation}
with $\Tilde{C}^v = \sum_{n=0}^{\infty} \Tilde{C}^{(n),v} \Tilde{a}_s^n$ and $\Tilde{D}^v = \sum_{n=0}^{\infty} \Tilde{D}^{(n),v} \Tilde{a}_s^n$, where the coefficients $\Tilde{C}^{(n),v}$ and $\Tilde{D}^{(n),v}$ are known up to $n=4$ ($\Tilde{C}^{(0),v}=\Tilde{C}^{(1),v}=1$ and $\Tilde{D}^{(0),v}=\Tilde{D}^{(1),v}=0$).

This allows for a short-distance description of the light quark correlator,
\begin{equation}
    G^{(l,l)}(t) = \frac{1}{3}4\pi\int dx  \frac{6}{\pi^4 (t^2+x^2)^3} \left[ \left( \frac{3}{2} - \frac{x^2}{t^2+x^2} \right) \Tilde{C}^v + 3 \Tilde{D}^v \right]\,,
\end{equation}
where $\Tilde{a}_s = \Tilde{a}_s\left(1/\sqrt{t^2+x^2}\right)$. 

The $\widetilde{\mathrm{MS}}$- and $\overline{\mathrm{MS}}$-schemes are related by the following perturbative relation:
\begin{equation}
    \begin{aligned}
            \tilde{a}_s(\mu) = & a_s(\mu) \left\{ 1 - a_s(\mu) l \, \beta_0 + a_s^2(\mu) l \left( \beta_0^2 l - \beta_1 \right) + a_s^3(\mu) l \left( -\beta_0^3 l^2 + \frac{5}{2} \beta_0 \beta_1 l - \beta_2 \right) \right.\\
            & \left. + a_s^4(\mu) l \left[ \beta_0^4 l^3 - \frac{13}{3} \beta_0^2 \beta_1 l^2 + 3 \left( \frac{\beta_1^2}{2} + \beta_0 \beta_2 \right) l - \beta_3 \right]
    + \mathcal{O}(a_s^5 l^5) \right\}\,,
    \end{aligned}
\end{equation}
with $l\equiv2(\ln2 + \gamma_E)$.
\begin{figure}[t]
    \centering
    \begin{subfigure}{0.49\textwidth}
        \centering
        \includegraphics[height=4.5cm]{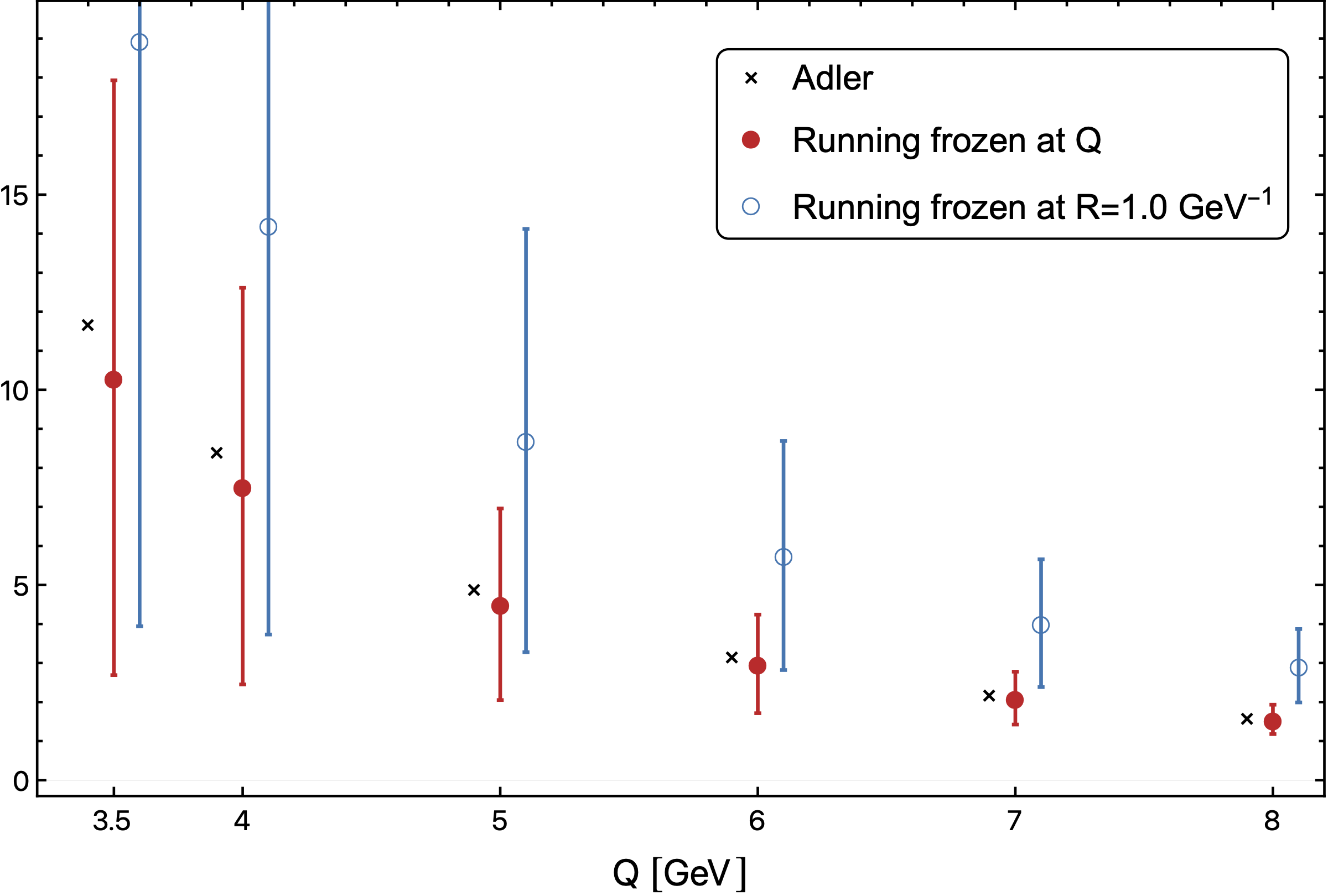}
        \caption{}
        \label{fig:SDsub_O1_crosscheck}
    \end{subfigure}
    \hfill
    \begin{subfigure}{0.49\textwidth}
        \centering
        \includegraphics[height=4.5cm]{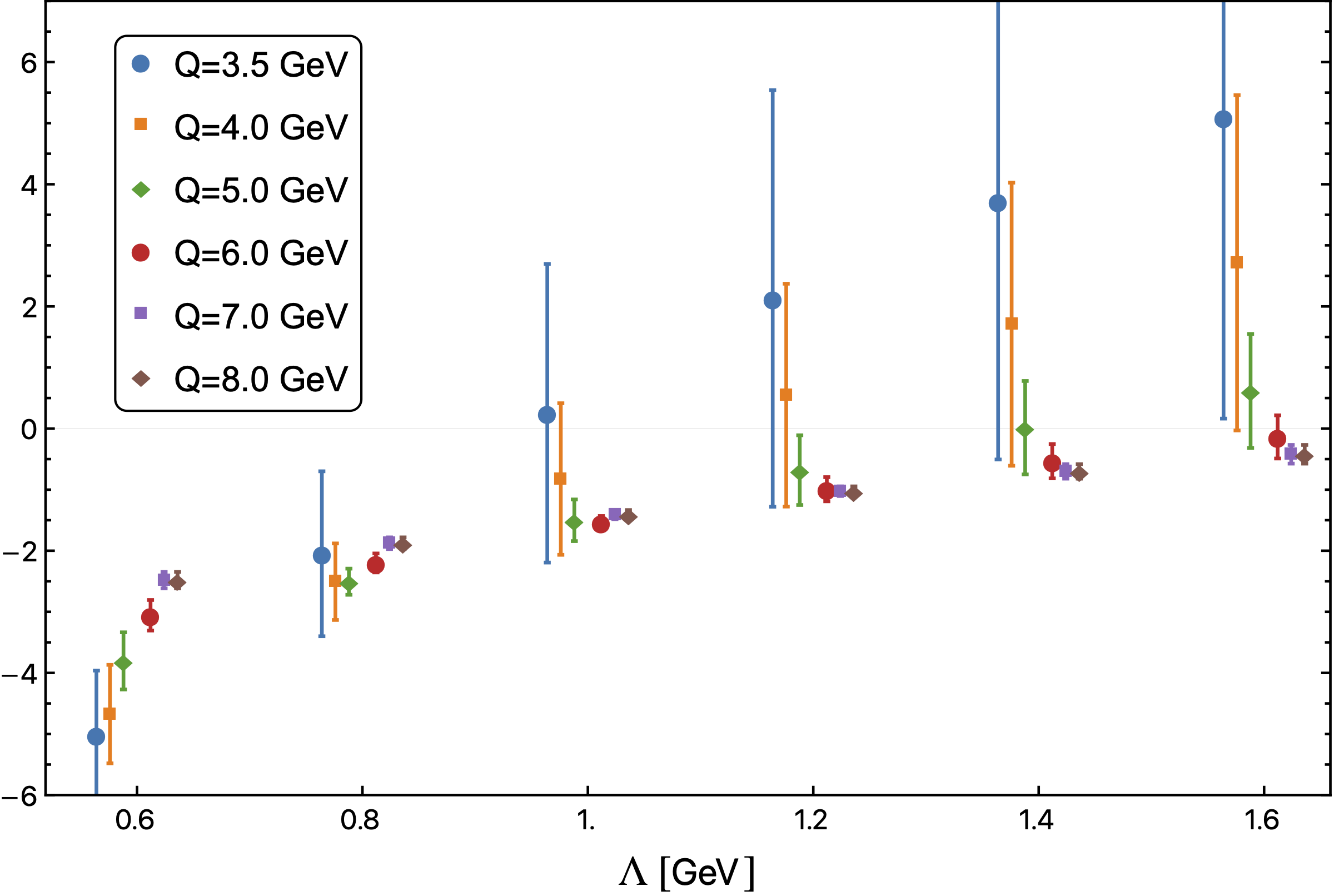}
        \caption{}
        \label{fig:SDsub_O1_logint}
    \end{subfigure}
    \caption{We show the first perturbative correction ($\mathrm{O}(\alpha_s^1)$) of integral~\eqref{eq:pQCD_id} (left) and~\eqref{eq:pQCD_log} (right) for all explored values of $Q$ and $\Lambda$. In both cases, the y-axis is given in units of $\frac{1}{2} \left(\frac{16 m_\mu \alpha}{3 Q^2}\right)^2\times 10^{-10}$. On the left panel, we compare the exact computation of $b^{(3,3)}(Q^2)$ using different methods. In black crosses, the Adler function is used to estimate this quantity. The method that we will follow in this work where the running of $\alpha_s$ is frozen for $r\geq1/Q$ is shown by red filled circles. And the empty blue circles show an alternative approach where we freeze all points at $r\geq R$ for $R = 1.0\, \mathrm{GeV}^{-1}$.}
    \label{fig:SDsub_O1}
\end{figure}

We can now write the integrals of interest in the coordinate-space representation as an expansion in the strong coupling at small distances, in both schemes. In both cases, the four-dimensional integration can be reduced to an integral over the radial coordinate,
\begin{equation}
    \begin{aligned}
        \int_0^\infty dt\ G^{(l,l)}(t) \sin^4\frac{Q t}{4} &= \int_0^\infty dr\ \frac{4}{\pi^2 r^3} \left[
        (\tilde{C}^v + 6 \tilde{D}^v) \left( \frac{3}{32} - \frac{J_1\left(\frac{Q r}{2}\right)}{2 Q r} + \frac{J_1(Q r)}{16 Q r} \right) \right. \\
        & \left. + \tilde{C}^v \left( 
        \frac{3}{64} 
        - \frac{2 J_2\left( \frac{Q r}{2} \right)}{Q^2 r^2} 
        + \frac{J_2(Q r)}{8 Q^2 r^2} 
        + \frac{J_3\left( \frac{Q r}{2} \right)}{Q r} 
        - \frac{J_3(Q r)}{8 Q r}
        \right)
        \right]\,,
    \end{aligned}
    \label{eq:pQCD_id}
\end{equation}
\begin{equation}
    \int_0^\infty dt\ G^{(l,l)}(t)\ln \Lambda t \sin^4\frac{Q t}{4} = \int_0^\infty dr\ \frac{4}{\pi^2 r^3} \left[ (\tilde{C}^v + 6 \tilde{D}^v) \mathcal{L}_0(r;Q,\Lambda) + \tilde{C}^v \mathcal{L}_2(r;Q,\Lambda)\right]\,,
    \label{eq:pQCD_log}
\end{equation}
where
\begin{equation}
    \mathcal{L}_n(r;Q,\Lambda) = \int_0^1 dc \sqrt{1-c^2}\ln(\Lambda rc) c^n \sin^4\frac{Q r c}{4}\, .
\end{equation}
At tree level, $\alpha_s$ is set to zero and the integration can be carried out analytically. For sufficiently large values of $Q$, this contribution is expected to capture the dominant part of the integral.
\begin{equation}
    \int_0^\infty dt\left\{\begin{matrix}1 \\ \ln\Lambda t\end{matrix}\right\}G^{(3,3)}_{tl}(t)\sin^4\frac{Q t}{4} = \left\{\begin{matrix}\frac{Q^2}{64\pi^2}\ln 2 \\ \frac{Q^2}{128\pi^2}\ln 2\big(3-2\gamma_E-\ln\frac{Q^2}{2\Lambda}\big)\end{matrix}\right\}\, .
\end{equation}
To compute higher-order contributions in pQCD, we must account for the running of the coupling with the radial distance, i.e.\ $\alpha_s=\alpha_s(1/r)$. For large values of $r$, $\alpha_s(1/r)$ becomes too large and the perturbative expansion in Eqs.~\eqref{eq:pQCD_id} and~\eqref{eq:pQCD_log} breaks down. To address this issue, we freeze the running coupling at the characteristic energy scale, which we expect to be of order $Q$. To be conservative, all contributions from the region where the expansion is expected to break, i.e.\ $r\gtrsim1.0\,\mathrm{GeV}^{-1}$, are assigned a $100\%$ uncertainty.
\begin{figure}[t]
        \centering
    \includegraphics[height=5.5cm]{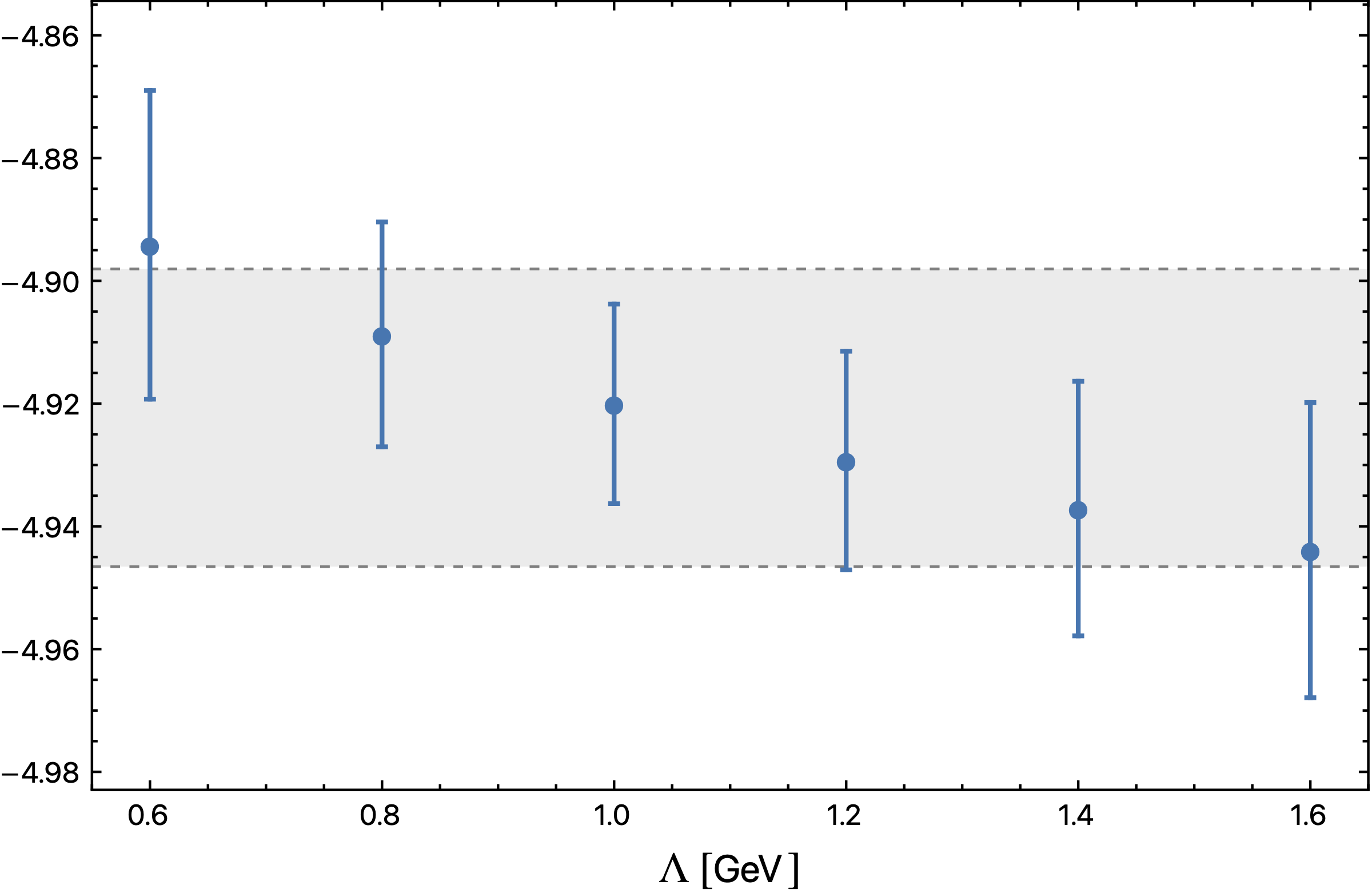}
    \caption{We show our estimation of $\tilde{b}^{(3,3),\,\mathrm{nlo(a)}}(5.0\, \mathrm{GeV})$ following Eq.~\eqref{eq:b_tilde_NLOa} for the different explored values of $\Lambda$. The shaded area corresponds to the final estimate where we compensate for the trend by including the spread of the points in the estimate of the final uncertainty.}
    \label{fig:lambda_stabil}
\end{figure}

We use the integral in Eq.~\eqref{eq:pQCD_id} to assess the accuracy of this procedure, since this quantity can be computed to high precision using its momentum-space representation in Eq.~\eqref{eq:b_momentum} and the well-controlled expansion of the Adler function. In Fig.~\ref{fig:SDsub_O1_crosscheck}, we compare the $\mathrm{O}(\alpha_s^1)$ contribution obtained with both approaches (and a third approach where $\alpha_s$ is not frozen until we reach $R\approx1.0\,\mathrm{GeV}^{-1}$), making use of the conversion between the $\widetilde{\mathrm{MS}}$ and $\overline{\mathrm{MS}}$ schemes. As can be seen, the agreement is excellent. Figure~\ref{fig:SDsub_O1_logint} shows the corresponding result for the integral in Eq.~\eqref{eq:pQCD_log} for the explored values of $Q$ and $\Lambda$.

The final evaluation of Eq.~\eqref{eq:pQCD_log} is carried out up to $\mathrm{O}(\tilde{\alpha}_s^2)$. Including higher orders would not improve the overall precision, since the uncertainty arising from the region $r\geq1.0\,\mathrm{GeV}^{-1}$ dominates over the truncation error. We expect $\tilde{b}^{(3,3)}(Q^2)$ in Eq.~\eqref{eq:b_tilde} (Eq.~\eqref{eq:b_tilde_NLOa} for NLOa) to be independent of $\Lambda$. In Fig.~\ref{fig:lambda_stabil}, we demonstrate the stability of the full result under variations of $\Lambda$. The most precise point is obtained for $\Lambda=1.0\, \mathrm{GeV}$. We observe a clear trend for other values of $\Lambda$. We compensate for this effect by taking a weighted average where the spread is included in the final uncertainty.

\section{Dimensionless scheme dependencies}
\label{app:derivatives}

To allow for an easy conversion to other schemes, we provide in Tab.~\ref{tab:dim_dep} the dimensionless scheme dependencies. For all studied channels and diagram sets, we compute the following dimensionless quantity:
\begin{equation}
    \frac{S}{\mathcal{O}}\pdv{\mathcal{O}}{S}\ ,
\label{eq:dimless_def}
\end{equation}
where $\mathcal{O}$ is the observable and $S$ is the quantity used to define the scheme. In this work, $S=\sqrt{t_0}$, $m_\pi$, and $m_K$ for all channels, with $m_{D_s}$ additionally included for charm quark quantities. As expected, the pion mass dependence dominates in the isovector channel, while the kaon mass dominates in the isoscalar channel. A strong dependence on the gradient-flow scale $\sqrt{t_0}$ is observed across all channels.

\begin{table}[t]
    \centering
    \small
    \renewcommand{\arraystretch}{1.2}
    \begin{tabular}{c | r@{.}l r@{.}l r@{.}l r@{.}l}
        \multicolumn{9}{c}{NLOa} \\
        \noalign{\smallskip}\hline\noalign{\smallskip}
        & \multicolumn{8}{c}{S} \\
        \noalign{\smallskip}\hline\noalign{\smallskip}
        $\mathcal{O}$ & \multicolumn{2}{c}{$\sqrt{t_0}$} & \multicolumn{2}{c}{$m_\pi$} & \multicolumn{2}{c}{$m_K$} & \multicolumn{2}{c}{$m_{D_s}$} \\
        \noalign{\smallskip}\hline\noalign{\smallskip}
        $a_\mu^{3,3}$ & 0&93 & $-$0&34 & $-$0&09 & \multicolumn{2}{c}{-} \\
        $a_\mu^{8,8}$ & 0&79 & $-$0&00 & $-$0&66 & \multicolumn{2}{c}{-} \\
        $a_\mu^{c,c}$ & 1&79 & $-$0&01 & 0&12 & $-$2&08 \\
        \noalign{\smallskip}\hline\noalign{\smallskip}
        $a_\mu^{\mathrm{hvp}}$ & 0&94 & $-$0&27 & $-$0&17 & $-$0&08 \\
    \end{tabular}
    \qquad
    \begin{tabular}{c | r@{.}l r@{.}l r@{.}l r@{.}l}
        \multicolumn{9}{c}{NLOb} \\
        \noalign{\smallskip}\hline\noalign{\smallskip}
        & \multicolumn{8}{c}{S} \\
        \noalign{\smallskip}\hline\noalign{\smallskip}
        $\mathcal{O}$ & \multicolumn{2}{c}{$\sqrt{t_0}$} & \multicolumn{2}{c}{$m_\pi$} & \multicolumn{2}{c}{$m_K$} & \multicolumn{2}{c}{$m_{D_s}$} \\
        \noalign{\smallskip}\hline\noalign{\smallskip}
        $a_\mu^{3,3}$ & 1&18 & $-$0&46 & $-$0&10 & \multicolumn{2}{c}{-} \\
        $a_\mu^{8,8}$ & 1&07 & $-$0&01 & $-$0&78 & \multicolumn{2}{c}{-} \\
        $a_\mu^{c,c}$ & 2&10 & $-$0&02 & 0&14 & $-$2&43 \\
        \noalign{\smallskip}\hline\noalign{\smallskip}
        $a_\mu^{\mathrm{hvp}}$ & 1&18 & $-$0&38 & $-$0&19 & $-$0&05 \\
    \end{tabular}
    \break\break\break
    \begin{tabular}{c | r@{.}l r@{.}l r@{.}l r@{.}l}
        \multicolumn{9}{c}{NLOc} \\
        \noalign{\smallskip}\hline\noalign{\smallskip}
        & \multicolumn{8}{c}{S} \\
        \noalign{\smallskip}\hline\noalign{\smallskip}
        $\mathcal{O}$ & \multicolumn{2}{c}{$\sqrt{t_0}$} & \multicolumn{2}{c}{$m_\pi$} & \multicolumn{2}{c}{$m_K$} & \multicolumn{2}{c}{$m_{D_s}$} \\
        \noalign{\smallskip}\hline\noalign{\smallskip}
        $a_\mu^{3,3-3,3}$ & 2&09 & $-$0&76 & $-$0&14 & \multicolumn{2}{c}{-} \\
        $a_\mu^{3,3-8,8}$ & 2&18 & $-$0&23 & $-$0&70 & \multicolumn{2}{c}{-} \\
        $a_\mu^{3,3-c,c}$ & 1&70 & $-$0&31 & $-$1&17 & $-$2&51 \\
        $a_\mu^{8,8-8,8}$ & 1&40 & $-$0&09 & $-$1&66 & \multicolumn{2}{c}{-} \\
        $a_\mu^{8,8-c,c}$ & 1&43 & $-$0&05 & $-$1&82 & $-$2&39 \\
        $a_\mu^{c,c-c,c}$ & 3&38 & $-$0&05 & $-$0&09 & $-$2&10 \\
        \noalign{\smallskip}\hline\noalign{\smallskip}
        $a_\mu^{\mathrm{hvp}}$ & 2&07 & $-$0&59 & $-$0&38 & $-$0&16 \\
    \end{tabular}
    \qquad
    \begin{tabular}{c | r@{.}l r@{.}l r@{.}l r@{.}l}
        \multicolumn{9}{c}{NLOa\&b} \\
        \noalign{\smallskip}\hline\noalign{\smallskip}
        & \multicolumn{8}{c}{S} \\
        \noalign{\smallskip}\hline\noalign{\smallskip}
        $\mathcal{O}$ & \multicolumn{2}{c}{$\sqrt{t_0}$} & \multicolumn{2}{c}{$m_\pi$} & \multicolumn{2}{c}{$m_K$} & \multicolumn{2}{c}{$m_{D_s}$} \\
        \noalign{\smallskip}\hline\noalign{\smallskip}
        $a_\mu^{3,3}$ & 0&64 & $-$0&19 & $-$0&08 & \multicolumn{2}{c}{-} \\
        $a_\mu^{8,8}$ & 0&55 & 0&00 & $-$0&55 & \multicolumn{2}{c}{-} \\
        $a_\mu^{c,c}$ & 1&67 & $-$0&01 & 0&12 & $-$1&95 \\
        \noalign{\smallskip}\hline\noalign{\smallskip}
        $a_\mu^{\mathrm{hvp}}$ & 0&68 & $-$0&15 & $-$0&15 & $-$0&11 \\
    \end{tabular}
    \break\break\break
    \begin{tabular}{c | r@{.}l r@{.}l r@{.}l r@{.}l}
        \multicolumn{9}{c}{NLO} \\
        \noalign{\smallskip}\hline\noalign{\smallskip}
        & \multicolumn{8}{c}{S} \\
        \noalign{\smallskip}\hline\noalign{\smallskip}
        $\mathcal{O}$ & \multicolumn{2}{c}{$\sqrt{t_0}$} & \multicolumn{2}{c}{$m_\pi$} & \multicolumn{2}{c}{$m_K$} & \multicolumn{2}{c}{$m_{D_s}$} \\
        \noalign{\smallskip}\hline\noalign{\smallskip}
        $a_\mu^{\mathrm{hvp}}$ & 0&63 & -0&13 & -0&14 & -0&10 \\
    \end{tabular}
    \caption{Following Eq.~\eqref{eq:dimless_def}, we show the dimensionless scheme dependencies for all studied channels and diagram sets (observable $\mathcal{O}$) with respect to the scheme quantities $S$. The central values are $\sqrt{t_0}=0.1440\, \mathrm{fm}$, $m_\pi=134.9768\,\mathrm{MeV}$, $m_K=495.011\,\mathrm{MeV}$, and $m_{D_s}=1968.47\,\mathrm{MeV}$.}
    \label{tab:dim_dep}
\end{table}

\newpage

\bibliographystyle{JHEP}
\bibliography{ref}

\end{document}